\documentclass{article}%
\usepackage{amsfonts}
\usepackage{amssymb}
\usepackage{amsmath}
\usepackage{graphicx}%
\providecommand{\U}[1]{\protect\rule{.1in}{.1in}}

\newtheorem{theorem}{Theorem}

\newtheorem{corollary}{Corollary}

\newtheorem{lemma}[theorem]{Lemma}

\newtheorem{proposition}{Proposition}

\begin{document}

\title{The Quantum Logic of Direct-Sum Decompositions}
\author{David Ellerman\\University of California Riverside}
\maketitle

\begin{abstract}
Since the pioneering work of Birkhoff and von Neumann, quantum logic has been
interpreted as the logic of (closed) subspaces of a Hilbert space. There is a
progression from the usual Boolean logic of subsets to the "quantum logic" of
subspaces of a general vector space--which is then specialized to the closed
subspaces of a Hilbert space. But there is a "dual" progression. The notion of
a partition (or quotient set or equivalence relation) is dual (in a
category-theoretic sense) to the notion of a subset. Hence the Boolean logic
of subsets has a dual logic of partitions. Then the dual progression is from
that logic of partitions to the quantum logic of direct-sum decompositions
(i.e., the vector space version of a set partition) of a general vector
space--which can then be specialized to the direct-sum decompositions of a
Hilbert space. This allows the logic to express measurement by any
self-adjoint operators rather than just the projection operators associated
with subspaces. In this introductory paper, the focus is on the quantum logic
of direct-sum decompositions of a finite-dimensional vector space (including
such a Hilbert space). The primary special case examined is finite vector
spaces over $%
\mathbb{Z}
_{2}$ where the pedagogical model of quantum mechanics over sets (QM/Sets) is
formulated. In the Appendix, the combinatorics of direct-sum decompositions of
finite vector spaces over $GF\left(  q\right)  $ is analyzed with computations
for the case of QM/Sets where $q=2$.

\end{abstract}
\tableofcontents

\section{Introduction}

This paper is an introduction to quantum logic based on direct-sum
decompositions rather than on subspaces of vector spaces. This allows the
logic to express measurement by any self-adjoint operators rather than just
the projection operators corresponding to subspaces. A \textit{direct-sum
decomposition} (DSD) of a vector space $V$ over a base field $\mathbb{K}$ is a
set of (nonzero) subspaces $\left\{  V_{i}\right\}  _{i\in I}$ that are
\textit{disjoint} (i.e., their pair-wise intersections are the zero space $0$)
and that span the space so that each vector $v\in V$ has a unique expression
$v=\sum_{i\in I}v_{i}$ with each $v_{i}\in V_{i}$ (with only a finite number
of $v_{i}$'s nonzero). For introductory purposes, it is best to assume $V$ is
finite dimensional (although many of the proofs are more general).

Each self-adjoint operator, and in general diagonalizable operator, has
eigenspaces that form a direct-sum decomposition of the vector space but the
notion of a direct-sum decomposition makes sense over arbitrary vector spaces
independently of an operator. For instance, in the pedagogical model of
"quantum mechanics over sets" or QM/Sets (\cite{ell:qmoversets};
\cite{ell:mammoth}),the vector space is $%
\mathbb{Z}
_{2}^{n}$ so the only operators (always assumed diagonalizable) are projection
operators $P:%
\mathbb{Z}
_{2}^{n}\rightarrow%
\mathbb{Z}
_{2}^{n}$. But given a set $U=\left\{  v_{1},...,v_{n}\right\}  $ of basis
vectors for $%
\mathbb{Z}
_{2}^{n}$, any real-valued "random variable" or function $f:U\rightarrow%
\mathbb{R}
$ determines a DSD $\left\{  \wp\left(  f^{-1}\left(  r\right)  \right)
\right\}  _{r\in f\left(  U\right)  }$ of $%
\mathbb{Z}
_{2}^{n}$ (where $\wp()$ is the power-set and $f\left(  U\right)  $ is the
image or "spectrum" of "eigenvalues" of the numerical attribute $f$). Thus the
concept of a direct-sum decomposition of a vector space allows one to capture
many of the relevant properties of such a real-valued "observable" even though
it does not take values in the base field (which is only $%
\mathbb{Z}
_{2}$ in QM/Sets). It is only as the base field is increased up to the complex
numbers that all real-valued observables can be "internalized" as self-adjoint operators.

The genesis of the usual quantum logic of projection operators or subspaces
can be seen as starting with the Boolean logic of subsets in the Boolean
lattice $\wp\left(  U\right)  $ of subsets of a universe set $U$ and taking
the vector space version of a "subset"--which is a subspace. That yields the
lattice of subspaces of an arbitrary vector space--which can then be
specialized to the lattice of (closed) subspaces of a Hilbert space for the
strictly quantum mechanical case. In category theory, the notion of a
subobject, such as a subset or subspace, has a dual notion of a quotient
object. Thus the dual concept to a subset is the concept of a quotient set,
equivalence relation, or partition on a set $U$. That gives rise to the idea
of the logic of partitions (\cite{ell:partitions}; \cite{ell:intropartlogic})
instead of the Boolean logic of subsets. The origin and topic of this paper is
the vector space version of a partition, namely a direct-sum decomposition
(NB: not a quotient space). The analogue of a partial Boolean algebra
\cite{kochen:hv-in-qm} is then a "partial partition algebra" of DSDs on an
arbitrary vector space (our topic here)--which can then be specialized to a
Hilbert space for the strictly quantum mechanical interpretation or
specialized to a vector space over $%
\mathbb{Z}
_{2}$ for pedagogical purposes.%

\begin{center}
\includegraphics[
height=1.4883in,
width=5.8098in
]%
{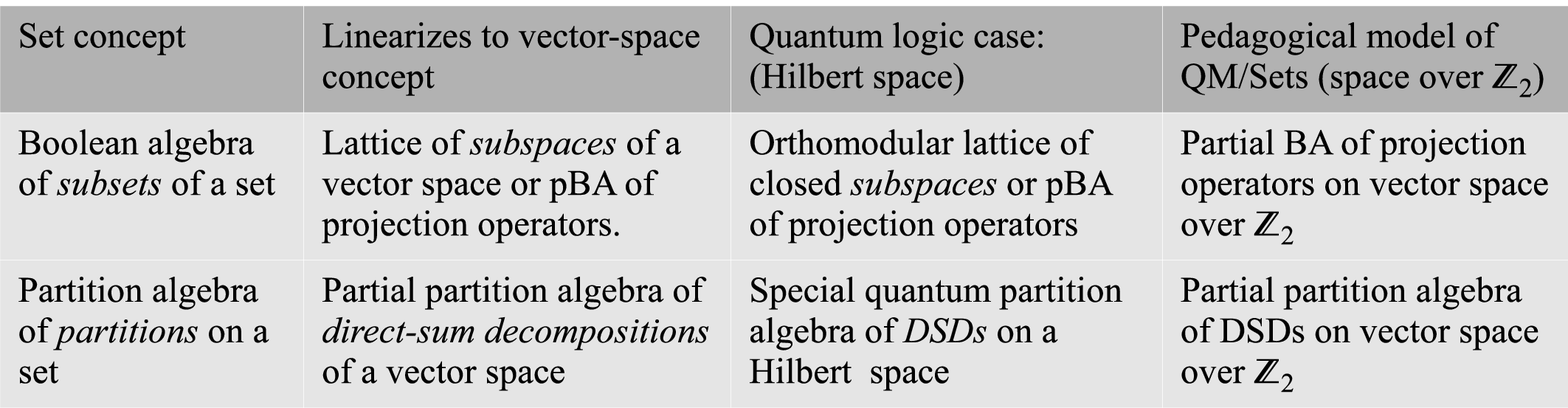}%
\end{center}

Figure 1: Progressions from sets to vector spaces starting with dual concepts
of subset and partition.

There is a natural partial order ("refinement" as with partitions on sets) on
the DSDs of a vector space $V$ and there is a minimum element $\mathbf{0}%
=\left\{  V\right\}  $, the \textit{indiscrete} DSD (nicknamed the "blob")
which consists of the whole space $V$. A DSD is "atomic" in the partial order
if there is no DSD between it and the minimum DSD $\mathbf{0}$, and the atomic
DSDs are the binary ones consisting of just two subspaces. Each atomic DSD
determines a pair of projection operators, and the indiscrete DSD also
determines a pair of projection operators, namely the zero operator $\hat{0}$
and the identity operator $I$. Conversely, each projection operator
$P:V\rightarrow V$ on an arbitrary vector space $V$ (other than the identity
or zero operator) determines an atomic DSD consisting of the image of $P$ and
the image of $I-P$, while the identity and zero operators determine the
indiscrete DSD. In that sense, the quantum logic of DSDs extends the quantum
logic of projection operators associated with atomic DSDs. In the quantum
logic of (closed) subspaces, only measurement of projection operators
(associated with atomic DSDs and the blob) can be represented, so the quantum
logic of direct-sum decompositions allows the representation of the
measurement of any self-adjoint operators.

\section{The partial partition algebra of direct-sum decompositions}

\subsection{Compatibility of DSDs}

Let $V$ be a finite dimensional vector space over a field $\mathbb{K}$. A
\textit{direct sum decomposition }(DSD) of $V$ is a set of subspaces $\left\{
V_{i}\right\}  _{i\in I}$ such that $V_{i}\cap\sum_{i^{\prime}\neq
i}V_{i^{\prime}}=0$ (the zero space) for $i\in I$ and which span the space:
written $\oplus_{i\in I}V_{i}=V$. Let $DSD\left(  V\right)  $ be the set of
DSDs of $V$.

In the algebra of partitions on a set, the operations of join, meet, and
implication are always defined, but in the context of "vector space
partitions," i.e., DSDs, we need to define a notion of compatibility. If we
were dealing with operators (and their associated DSDs of eigenspaces), then
compatibility would be defined by commutativity. But we are dealing with DSDs
directly with no assumption that they are the eigenspace DSDs of operators.

Given two DSDs $\pi=\left\{  V_{i}\right\}  _{i\in I}$ and $\sigma=\left\{
W_{j}\right\}  _{j\in J}$, their \textit{proto-join} is the set of non-zero
subspaces $\left\{  V_{i}\cap W_{j}\right\}  _{\left(  i,j\right)  \in I\times
J}$. If the two DSDs $\pi$ and $\sigma$ were defined as the eigenspace DSDs of
two operators, then the space spanned by the proto-join would be the space
spanned by the simultaneous eigenvectors of the two operators, and that space
would be the kernel of the commutator of the two operators. If the two
operators commuted, then their commutator would be the zero operator whose
kernel is the whole space so the proto-join would span the whole space. Hence
the natural definition of compatibility without any mention of operators is:

\begin{center}
$\pi$ and $\sigma$ are \textit{compatible}, written $\pi\leftrightarrow\sigma
$, if the proto-join spans the whole space $V$.
\end{center}

The \textit{indiscrete DSD} $\mathbf{0}=\left\{  V\right\}  $ (the "blob") is
compatible with all DSDs, i.e., $\mathbf{0}\leftrightarrow\pi$ for any $\pi$.

\subsection{The join of compatible DSDs}

When two DSDs are compatible, the proto-join is the\textit{ join: }

\begin{center}
$\pi\vee\sigma=\left\{  V_{i}\cap W_{j}\right\}  _{\left(  i,j\right)  \in
I\times J}$

Join of DSDs when $\pi\leftrightarrow\sigma$.
\end{center}

\noindent The binary relation of compatibility on DSDs is reflexive and
symmetric. The indiscrete DSD acts as the identity for the join:
$\mathbf{0}\vee\pi=\pi$ for any DSD $\pi$.

In a set of mutually compatible DSDs, we need to show that the join operation
preserves compatibility. If $\pi\leftrightarrow\sigma$, it is trivial that
$\left(  \pi\vee\sigma\right)  \leftrightarrow\pi$ and $\left(  \pi\vee
\sigma\right)  \leftrightarrow\sigma$, but for a third DSD $\tau$ with
$\pi\leftrightarrow\tau$ and $\sigma\leftrightarrow\tau$, does $\left(
\pi\vee\sigma\right)  \leftrightarrow\tau$?

\begin{lemma}
Let the DSDs $\pi=\left\{  V_{i}\right\}  _{i\in I}$ and $\sigma=\left\{
W_{j}\right\}  _{j\in J}$ be compatible so that $\pi\vee\sigma=\left\{
V_{i}\cap W_{j}\right\}  _{\left(  i,j\right)  \in I\times J}$ is a DSD and
thus any $v\in V$ has a unique expression $v=\sum_{\left(  i,j\right)  \in
I\times J}v_{ij}$ where $v_{ij}\in V_{i}\cap W_{j}$. Let $v_{i}=\sum_{j\in
J}v_{ij}\in V_{i}$ so that $v=\sum_{i\in I}v_{i}$. If $v\in V_{i}$, then
$v=v_{i}$.
\end{lemma}

Proof: Let $\widehat{v_{i}}=\sum_{i^{\prime}\in I,i^{\prime}\neq
i}v_{i^{\prime}}$ so that $v=v_{i}+\widehat{v_{i}}$. Then $v-v_{i}%
=\widehat{v_{i}}\in V_{i}$. If $\widehat{v_{i}}\neq0$, then $\widehat{v_{i}}$
itself and $\sum_{i^{\prime}\neq i,j}v_{i^{\prime}j}$ are two different
expressions for $\widehat{v_{i}}$ of vectors in a direct sum, so
$\widehat{v_{i}}=0$. $\square$

\begin{theorem}
\noindent Given three DSDs, $\pi=\left\{  V_{i}\right\}  _{i\in I}$,
$\sigma=\left\{  W_{j}\right\}  _{j\in J}$, and $\tau=\left\{  X_{k}\right\}
_{k\in K}$ that are mutually compatible, i.e., $\pi\leftrightarrow\sigma$,
$\pi\leftrightarrow\tau$, and $\sigma\leftrightarrow\tau$, then $\left(
\pi\vee\sigma\right)  \leftrightarrow\tau$.
\end{theorem}

Proof: We need to prove $\pi\vee\sigma=\left\{  V_{i}\cap W_{j}\right\}
_{\left(  i,j\right)  \in I\times J}$ is compatible with $\tau=\left\{
X_{k}\right\}  _{k\in K}$, i.e., that $\oplus_{\left(  i,j,k\right)  \in
I\times J\times K}\left(  V_{i}\cap W_{j}\cap X_{k}\right)  =V$. Consider any
nonzero $v\in V$ where since $\pi\leftrightarrow\sigma$, $\exists v_{ij}\in
V_{i}\cap W_{j}$ such that $v=\sum_{\left(  i,j\right)  \in I\times J}v_{ij}$.

\qquad Now since $\pi\leftrightarrow\tau$, for each nonzero $v_{ij}\in
V_{i}\cap W_{j}$, $\exists v_{ij,i^{\prime}k}\in V_{i^{\prime}}\cap X_{k}$
such that $v_{ij}=\sum_{\left(  i^{\prime},k\right)  \in I\times
K}v_{ij,i^{\prime}k}$. But since $v_{ij}\in V_{i}$, by the Lemma, only
$v_{ij,ik}$ is nonzero, so $v_{ij}=\sum_{k\in K}v_{ij,ik}$.

\qquad Symmetrically, since $\sigma\leftrightarrow\tau$, for each $v_{ij}\in
V_{i}\cap W_{j}$, $\exists v_{ij,j^{\prime}k}\in W_{j^{\prime}}\cap X_{k}$
such that $v_{ij}=\sum_{\left(  j^{\prime},k\right)  \in J\times
K}v_{ij,j^{\prime}k}$. But since $v_{ij}\in W_{j}$, by the Lemma, only
$v_{ij,jk}$ is nonzero, so $v_{ij}=\sum_{k\in K}v_{ij,jk}$.

\qquad Now since $\left\{  X_{k}\right\}  _{k\in K}$ is a DSD, there is a
unique expression for each nonzero $v_{ij}=\sum_{k\in K}v_{ijk}$ where
$v_{ijk}\in X_{k}$. Hence by uniqueness: $v_{ijk}=v_{ij,ik}=v_{ij,jk}$. But
since $v_{ij,ik}\in V_{i}$ and $v_{ij,jk}\in W_{j}$ and $v_{ij,ik}%
=v_{ijk}=v_{ij,jk}$, we have $v_{ijk}\in V_{i}\cap W_{j}\cap X_{k}$. Thus

$v=\sum_{\left(  i,j\right)  \in I\times J}v_{ij}=\sum_{\left(  i,j\right)
\in I\times J}\sum_{k\in K}v_{ijk}=\sum_{\left(  i,j,k\right)  \in I\times
J\times K}v_{ijk}$. Since $v$ was arbitrary,

\begin{center}
$\oplus_{\left(  i,j,k\right)  \in I\times J\times K}\left(  V_{i}\cap
W_{j}\cap X_{k}\right)  =V$. $\square$
\end{center}

\subsection{The meet of two DSDs}

For any two DSDs $\pi$ and $\sigma$, the \textit{meet} $\pi\wedge\sigma$ is
the DSD whose subspaces are direct sums of subspaces from $\pi$ and the direct
sum of subspaces from $\sigma$ and are minimal subspaces in that regard. That
is, $\left\{  Y_{l}\right\}  _{l\in L}$ is the meet if there is a set
partition $\left\{  I_{l}\right\}  _{l\in L}$ on $I$ and a set partition
$\left\{  J_{l}\right\}  _{l\in L}$ such that:

\begin{center}
$Y_{l}=\oplus_{i\in I_{l}}V_{i}=\oplus_{j\in J_{l}}W_{j}$
\end{center}

\noindent and that holds for no more refined partitions on the index sets. If
$\pi\leftrightarrow\tau$ and $\sigma\leftrightarrow\tau$, then it is trivial
that $\left(  \pi\wedge\sigma\right)  \leftrightarrow\tau$.

As in the movie of the same name,%

\begin{center}
\includegraphics[
height=2.0306in,
width=1.9346in
]%
{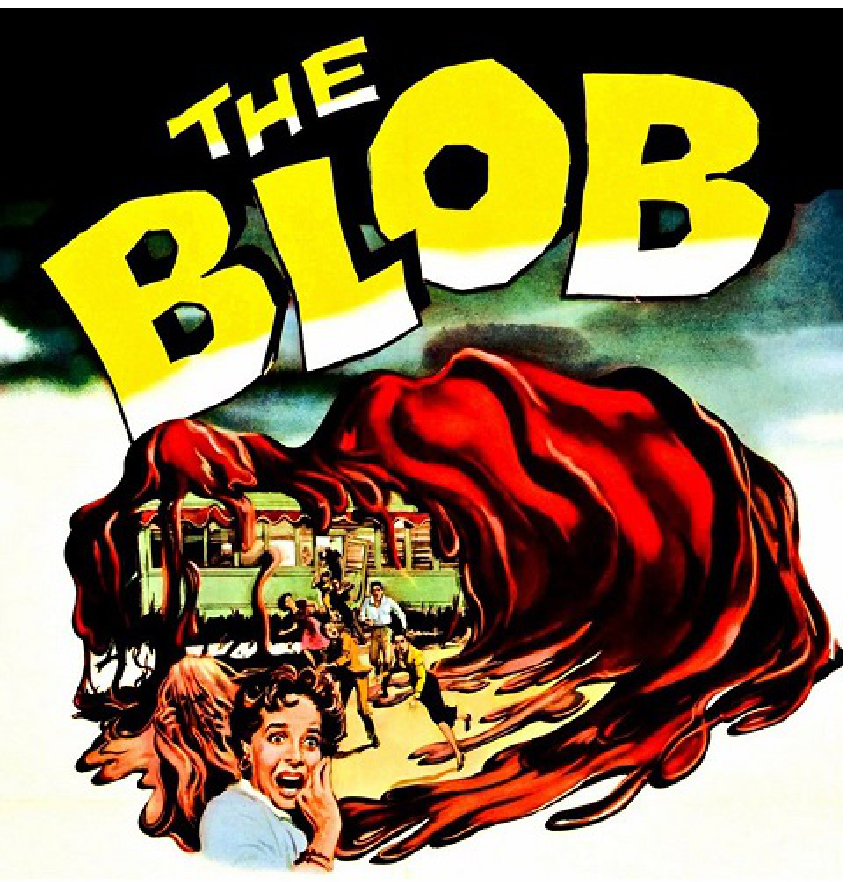}%
\end{center}

\begin{center}
Figure 2: "The blob absorbs everything it meets": $\mathbf{0}\wedge
\pi=\mathbf{0}$.
\end{center}

\subsection{The refinement partial order on DSDs}

The \textit{partial order} on the DSDs of $V$ is defined as for set partitions
but with subspaces replacing subsets:

\begin{center}
$\pi$ \textit{refines} $\sigma$, written $\sigma\preceq\pi$, if for every
$V_{i}\in\pi$, $\exists W_{j}\in\sigma$ such that $V_{i}\subseteq W_{j}$.
\end{center}

\noindent If $\sigma\preceq\pi$ holds, then each $W_{j}=\oplus\left\{
V_{i}:V_{i}\subseteq W_{j}\right\}  $ so $\pi\leftrightarrow\sigma$ and
$\pi\vee\sigma=\pi$ as well as $\pi\wedge\sigma=\sigma$ as expected.

\begin{proposition}
Where it exists, the join $\pi\vee\sigma$ is the least upper bound of $\pi$
and $\sigma$.
\end{proposition}

Proof: If $\pi,\sigma\preceq\tau$, i.e., $\pi$ and $\sigma$ have a common
upper bound $\tau=\left\{  X_{k}\right\}  _{k\in K}$, then $V_{i}%
=\oplus\left\{  X_{k}:X_{k}\subseteq V_{i}\right\}  $ and $W_{j}%
=\oplus\left\{  X_{k}:X_{k}\subseteq W_{j}\right\}  $. Given a nonzero $v\in
V_{i}\cap W_{j}$, it can be expressed uniquely as $v=\sum_{X_{k}\subseteq
V_{i}}v_{ik}$ where $v_{ik}\in X_{k}$ and as $v=\sum_{X_{k}\subseteq W_{j}%
}v_{jk}$ where $v_{jk}\in X_{k}$. But since $\tau$ is a DSD, there is a unique
expression $v=\sum_{k\in K}v_{k}$ so $v_{ik}=v_{jk}=v_{k}$ where $v_{k}\in
V_{i}\cap W_{j}\cap X_{k}$ where $X_{k}\subseteq V_{i}$ and $X_{i}\subseteq
W_{j}$ so $X_{k}\subseteq V_{i}\cap W_{j}$ and thus $\pi\vee\sigma\preceq\tau
$. Hence $\pi\vee\sigma$ is the least upper bound of $\pi$ and $\sigma$.
$\square$

\begin{corollary}
If $\pi$ and $\sigma$ have a common upper bound, i.e., $\pi,\sigma\preceq\tau
$, then $\pi\leftrightarrow\sigma$.
\end{corollary}

Two DSDs $\pi$ and $\sigma$ need not have a common upper bound so $DSD\left(
V\right)  $ is not a join-semilattice.

\begin{lemma}
Given a DSD $\pi=\left\{  V_{i}\right\}  _{i\in I}$, let $X=\oplus_{i\in
I_{X}}V_{i}$ and $Y=\oplus_{i\in I_{Y}}V_{i}$ both be direct sums of some
$V_{i}$'s. If $X\cap Y$ is nonzero, then it is also a direct sum of some
$V_{i}$'s.
\end{lemma}

Proof: Consider a nonzero $v\in X\cap Y$ so there is a unique expression
$v=\sum_{i\in I_{X}}v_{i,X}$ where $v_{i,X}\in V_{i}\subseteq X$ and a unique
expression $v=\sum_{i\in I_{Y}}v_{i,Y}$ where $v_{i,Y}\in V_{i}\subseteq Y$.
Since $\pi$ is a DSD, there is also a unique expression $v=\sum_{i\in I}v_{i}$
so for each nonzero $v_{i}$. $v_{i}=v_{i,X}=v_{i,Y}\in V_{i}\cap X\cap Y$.
Thus for any such $i$, $V_{i}$ is a common direct summand to $X$ and $Y$, so
$V_{i}\subseteq X\cap Y$. Thus every nonzero element $v\in X\cap Y$ is in a
direct sum of $V_{i}$'s for $V_{i}\subseteq X\cap Y$ and thus $X\cap Y$ is the
direct sum of $V_{i}$ that are common direct summands of $X$ and $Y$.
$\square$

\begin{proposition}
The meet $\pi\wedge\sigma$ is the greatest lower bound of $\pi$ and $\sigma$.
\end{proposition}

Proof:\textit{ }If $\tau\preceq\pi,\sigma$ then each $X_{k}=\oplus\left\{
V_{i}:V_{i}\subseteq X_{k}\right\}  =\oplus\left\{  W_{j}:W_{j}\subseteq
X_{k}\right\}  $. By the construction of $\pi\wedge\sigma$, there is a set
partition $\left\{  I_{l}\right\}  _{l\in L}$ on $I$ and a set partition
$\left\{  J_{l}\right\}  _{l\in L}$ on $J$ such that the subspaces in the meet
$\pi\wedge\sigma=\left\{  Y_{l}\right\}  $ are:

\begin{center}
$Y_{l}=\oplus_{i\in I_{l}}V_{i}=\oplus_{j\in J_{l}}W_{j}$,
\end{center}

\noindent and where no subsets of $I$ smaller than $I_{l}$ and subsets of $J$
smaller than $J_{l}$ have that property. Since each $V_{i}$ is contained in
some $X_{k}$, if $i\in I_{l}$, then $V_{i}\subseteq Y_{l}\cap X_{k}$. Since
both $Y_{l}$ and $X_{k}$ are direct sums of some $V_{i}$, then by the Lemma
the nonzero subspace $Y_{l}\cap X_{k}$ is also a direct sum of some $V_{i}$'s.
Symmetrically, since the same $Y_{l}$ and $X_{k}$ are direct sums of some
$W_{j}$'s, then by the Lemma the nonzero subspace $Y_{l}\cap X_{k}$ is also a
direct sum of some $W_{j}$'s. But since $Y_{l}$ is the smallest direct sum of
both $V_{i}$'s and $W_{j}$'s, $Y_{l}\cap X_{k}=Y_{l}$, i.e., $Y_{l}\subseteq
X_{k}$, and thus $\pi\wedge\sigma$ is the greatest (in the refinement partial
ordering) lower bound on $\pi$ and $\sigma$. $\square$

As the blob is compatible with all DSDs, it is the minimum element in the
ordering: $\mathbf{0}\preceq\pi$ for any $\pi$. Hence any two DSDs $\pi$ and
$\sigma$ always have a common lower bound, so they always have a meet
$\pi\wedge\sigma$ , i.e., $DSD\left(  V\right)  $ is a meet-semilattice. Thus
the \textit{partial partition algebra} $DSD\left(  V\right)  $ could also be
called the \textit{meet-semi-lattice of DSDs} on a vector space $V$.

The binary DSDs $\alpha=\left\{  A_{1},A_{2}\right\}  $ are the atoms of the
meet-semi-lattice $DSD\left(  V\right)  $. A meet-semi-lattice is said to be
\textit{atomistic} if every element is the join of the atoms below it.

\begin{proposition}
The meet-semi-lattice $DSD(V)$ is \textit{atomistic.}
\end{proposition}

Proof: Consider a non-blob DSD $\pi=\left\{  V_{i}\right\}  _{i\in I}$. If
$\alpha=\left\{  A_{1},A_{2}\right\}  \preceq\pi=\left\{  V_{i}\right\}
_{i\in I}$, then $A_{k}=\oplus\left\{  V_{i}:V_{i}\subseteq A_{k}\right\}  $
for $k=1,2$. Thus for any other atom $\alpha^{\prime}=\left\{  A_{1}^{\prime
},A_{2}^{\prime}\right\}  \preceq\pi$, the join $\alpha\vee\alpha^{\prime}$ is
defined and $\alpha\vee\alpha^{\prime}\preceq\pi$, and each nonzero subspace
$A_{k}\cap A_{k^{\prime}}^{\prime}$ is the direct sum of some $V_{i}$'s. If a
join of atoms had a subspace $V_{i_{1}}\oplus V_{i_{2}}$, $i_{1},i_{2}\in I$,
then the join with the atom $\left\{  V_{i_{1}},\oplus_{i^{\prime}\neq
i_{1},i^{\prime}\in I}V_{i^{\prime}}\right\}  $ would split apart $V_{i_{1}%
}\oplus V_{i_{2}}$, so the join of the atoms below $\pi$ gives $\pi$.$\square$

$DSD\left(  V\right)  $ is the \textit{quantum partition logic} determined by
$V$. To be more specifically "quantum", $V$ could be a finite-dimensional
Hilbert space.

\section{Partition logics in a partial partition algebra}

\subsection{The implication DSD in partition logics}

Just as a partial Boolean algebra is made up of overlapping Boolean algebras,
so a partial partition algebra is made up of overlapping partition logics or
algebras. There is no maximum DSD, only maximal DSDs. Each maximal element in
the partial ordering is a \textit{discrete }(or "non-degenerate") DSD
$\omega=\left\{  U_{k}\right\}  _{k\in K}$ of one-dimensional subspaces (rays)
of $V$ (so $\left\vert K\right\vert $ is the dimension of $V$). A
\textit{partition logic} is determined by the set of DSDs $\prod\left(
\omega\right)  =\left\{  \pi:\pi\preceq\omega\right\}  =\left[  \mathbf{0}%
,\omega\right]  \subseteq DSD\left(  V\right)  $ compatible with a maximal
element $\omega$ with the induced ordering and operations (which is analogous
to the way in a partial Boolean algebra, a complete set of one-dimensional
subspaces determines a Boolean algebra).%

\begin{center}
\includegraphics[
height=2.495in,
width=2.6507in
]%
{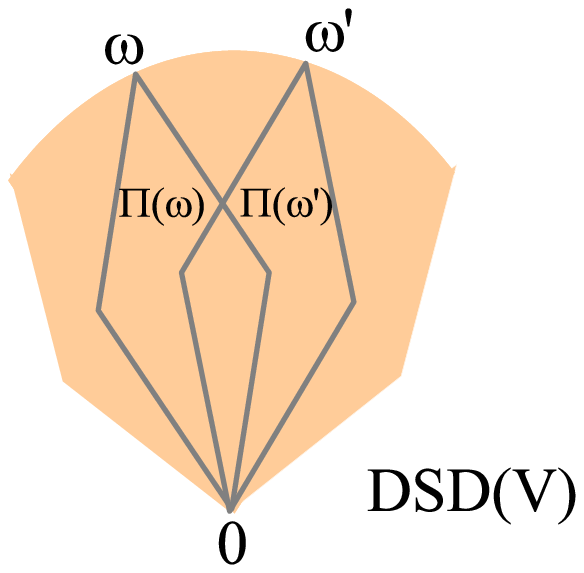}%
\end{center}

\begin{center}
Figure 3: Partial Partition Logic or Meet-Semi-Lattice of DSDs of $V$

with partition logics $\prod\left(  \omega\right)  $ and $\prod\left(
\omega^{\prime}\right)  $.
\end{center}

For any $\pi\in\prod\left(  \omega\right)  $, $\pi\preceq\omega$ so $\omega$
is (by construction) the maximum or \textit{top} DSD in $\prod\left(
\omega\right)  $ and thus might be symbolized as the discrete DSD
$\mathbf{1}_{\omega}$ ($=\omega$). Each subspace $V_{i}\in\pi\preceq\omega$
has $V_{i}=\oplus\left\{  U_{k}:U_{k}\subseteq V_{i},k\in K\right\}  $ so
$\omega$ absorbs what it joins and is the unit element for meets within
$\prod\left(  \omega\right)  $:

\begin{center}
$\pi\vee\omega=\omega$ and $\pi\wedge\omega=\pi$.
\end{center}

All the DSDs $\pi$ and $\sigma$ compatible with $\omega$, i.e., $\pi,\sigma
\in\prod\left(  \omega\right)  $, are compatible with each other since they
have a common upper bound. Explicitly, each $\pi=\left\{  V_{i}\right\}
_{i\in I}$ in $\prod\left(  \omega\right)  $ determines a set partition
$\pi\left(  \omega\right)  $ on $K$ (the index set for $\omega$), and thus
$\left\vert \prod\left(  \omega\right)  \right\vert =B\left(  \left\vert
K\right\vert \right)  =B\left(  \dim\left(  V\right)  \right)  $, the Bell
number for the dimension of $V$. Each $U_{k}$ would be contained in some block
of the set-partition join $\pi\left(  \omega\right)  \vee\sigma\left(
\omega\right)  $ and thus those corresponding subspaces $V_{i}\cap
W_{j}=\oplus\left\{  U_{k}:U_{k}\subseteq V_{i}\cap W_{j},k\in K\right\}  $
would span $V$ so $\pi\leftrightarrow\sigma$.

In order to be properly called a "logic," each partition lattice $\prod\left(
\omega\right)  $ has a naturally defined implication inherited from the logic
of set partitions (so "partition logic" refers to a partition lattice plus the
implication operation). For $\sigma,\pi\in\prod\left(  \omega\right)  $, and
for each $V_{i}\in\pi$, the corresponding subspaces of the
\textit{implication} $\sigma\Rightarrow\pi$ are:

\begin{center}
$\left\{
\begin{array}
[c]{l}%
\text{all }U_{k}\subseteq V_{i}\text{ if }\exists W_{j}\in\sigma\text{ such
that }V_{i}\subseteq W_{j}\text{ }\\
V_{i}\text{ if not.}%
\end{array}
\right.  $ .
\end{center}

Since each $V_{i}=\oplus\left\{  U_{k}:U_{k}\subseteq V_{i}\right\}  $, the
implication $\sigma\Rightarrow\pi$ is still a DSD in spite of some of the
$V_{i}\in\pi$ being "discretized" into the $U_{k}$ contained in it. In the
implication DSD $\sigma\Rightarrow\pi$, each $V_{i}\in\pi$ either remains
whole like a mini-zero-blob $\mathbf{0}_{V_{i}}=\left\{  V_{i}\right\}  $ on
the space $V_{i}$ (if $V_{i}$ is not contained in any $W_{j}\in\sigma$) or it
is discretized into the "atoms" $U_{k}\subseteq V_{i}$ which in effect assigns
a "$\mathbf{1}$" to $V_{i}$ if $\exists W_{j}$ such that $V_{i}\subseteq
W_{j}$. In other words, the implication $\sigma\Rightarrow\pi$ acts like an
indicator or characteristic function assigning a $1$ or $0$ to each $V_{i}$
depending on whether or not $\exists W_{j}$ such that $V_{i}\subseteq W_{j}$.
Thus trivially:

\begin{center}
$\sigma\Rightarrow\pi=\mathbf{1}_{\omega}$ iff $\sigma\preceq\pi$.
\end{center}

If we just take $\omega=\left\{  U_{k}\right\}  _{k\in K}$ as a set of
entities (forgetting about any vector space structure), then each DSD
$\sigma=\left\{  W_{j}\right\}  _{j\in J}$ in $\prod\left(  \omega\right)
=\left[  \mathbf{0},\omega\right]  $ defines a set partition on $\omega
=\left\{  U_{k}\right\}  _{k\in K}$ where each subspace $W_{j}$ determines a
block $\left\{  U_{k}:U_{k}\subseteq W_{j}\right\}  $.

Indeed, given any DSD $\pi=\left\{  V_{i}\right\}  _{i\in I}$, each subspace
$W_{j}$ of $\sigma\in\left[  \mathbf{0},\pi\right]  $ determines a block
$\left\{  V_{i}:V_{i}\subseteq W_{j}\right\}  $ so $\sigma$ defines a set
partition on $\pi$. Thus the interval $\left[  \mathbf{0},\pi\right]  $ is
isomorphic to the set-based partition logic (join, meet, and implication
operations) on that set $\pi$ \cite{ell:intropartlogic}. As a partition
lattice, $\left[  \mathbf{0},\pi\right]  $ has many of well-known properties
(\cite{ore:eq-rel}; \cite{birkhoff:lt}; \cite[Chapter IV, section
4]{grat:lattice-theory}). However, the late development of partition logic was
in part retarded by the practice of referring to the lattice of equivalence
relations as the "lattice of partitions" where the partial order however
"corresponds to set inclusion for the corresponding equivalence relations"
\cite[p. 251]{grat:lattice-theory} so instead of being refinement it is
actually "reverse refinement" \cite[p.30]{kung;rota}. The partial order on the
partition lattice $\prod\left(  \omega\right)  $ (as defined here) corresponds
to set inclusion of the binary relations that are the complements of
equivalence relations and are called \textit{partition relations}
\cite{ell:intropartlogic} or \textit{apartness relations}. In the lattice of
equivalence relations, the top is the biggest (indiscrete) equivalence
relation (where everything is identified) and the bottom is the smallest
(discrete) equivalence relation where each element is identified only with
itself--whereas the partition lattice $\prod\left(  \omega\right)  $ uses the
opposite partial order.\footnote{Instead of the usual duality relation within
a Boolean algebra, there is a duality relation between the logic of partitions
and the "logic" of equivalence relations \cite{ell:intropartlogic}.} Either
way the lattice is complete and relatively complemented but not distributive.
But the reversed order reverses the join and meet, the top and bottom, and the
atoms and coatoms.

\subsection{DSDs, CSCOs, and measurement}

Given a self-adjoint operator $L$, the projections $P_{\lambda}$ can be
reconstructed from the DSD $\pi=\left\{  V_{\lambda}\right\}  _{\lambda}$of
eigenspaces and then the operator can reconstructed--given the
eigenvalues--from the spectral decomposition $L=\sum_{\lambda}\lambda
P_{\lambda}$. What information about self-adjoint operators is lost by dealing
only with their DSDs of eigenspaces? The information about which eigenvalues
for eigenvectors are the same or different is retained by the distinct
eigenspaces in the DSD. It is only the specific numerical values of the
eigenvalues that is lost, and those numerical values are of little importance
in QM. Any transformation into other real numbers that is one-to-one (thus
avoiding "accidental" degeneracy) would do as well. Thus we can say that the
essentials of the measurement process in QM can be translated into the
language of the quantum logic of direct-sum decompositions. Kolmogorov
referred to the set partition given by the inverse-image of a random variable
as the "experiment" \cite[p. 6]{kolmogorov:foundations}, so it is natural to
abstractly represent the direct-sum decomposition of eigenspaces given by a
self-adjoint operator as the "measurement."

Thus unlike the quantum logic of subspaces, the quantum logic of direct-sum
decompositions can directly represent the process of measurement for any
self-adjoint operators (rather than just projection operators). Given a state
$\psi$ and a self-adjoint operator $L:V\rightarrow V$ on a finite dimensional
Hilbert space, the operator determines the DSD $\pi=\left\{  V_{\lambda
}\right\}  _{\lambda}$of eigenspaces for the eigenvalues $\lambda$. The
measurement operation uses the eigenspace DSD to decompose $\psi$ into the
unique parts given by the projections $P_{\lambda}\left(  \psi\right)  $ into
the eigenspaces $V_{\lambda}$, where $P_{\lambda}\left(  \psi\right)  $ is the
outcome of the projective measurement with probability $\Pr\left(
\lambda|\psi\right)  =\left\Vert P_{\lambda}\left(  \psi\right)  \right\Vert
^{2}/\left\Vert \psi\right\Vert ^{2}$.

The eigenspace DSD $\pi=\left\{  V_{\lambda}\right\}  _{\lambda}$ of $L$ is
refined by one or more maximal eigenvector DSDs, $\pi=\left\{  V_{\lambda
}\right\}  _{\lambda}\preceq\omega=\left\{  U_{k}\right\}  _{k\in K}$. For
each $\omega$, there is a set partition $\left\{  B_{\lambda}\right\}
_{\lambda}$ on the index set $K$ such that $V_{\lambda}$ is the direct sum of
the $U_{k}$ for $k\in B_{\lambda}$, i.e., $V_{\lambda}=\oplus_{k\in
B_{\lambda}}U_{k}$.

If some of the $V_{\lambda}$ have dimension larger than one ("degeneracy"),
then more measurements by commuting operators will be necessary to further
decompose down to single eigenvectors. If two operators commute, that means
that their eigenspace DSDs are compatible. Given another self-adjoint operator
$M:V\rightarrow V$ commuting with $L$, its eigenspace DSD $\sigma=\left\{
W_{\mu}\right\}  _{\mu}$ (for eigenvalues $\mu$ of $M$) is compatible with
$\pi=\left\{  V_{\lambda}\right\}  _{\lambda}$ and thus has a join DSD
$\pi\vee\sigma$ in $DSD\left(  V\right)  $ which is also in $\prod\left(
\omega\right)  $ for one or more maximal $\omega$ each representing an
orthonormal basis of simultaneous eigenvectors. The combined measurement by
the two commuting operators is just the single measurement using the join DSD
$\pi\vee\sigma$.

Dirac's notion of a Complete Set of Commuting Operators (CSCO)
\cite{dirac:principles} translates into the language of the quantum logic of
DSDs as a set of compatible DSDs whose join is a maximal DSD $\omega$ in
$DSD\left(  V\right)  $ and thus is the maximum DSD $\mathbf{1}_{\omega}$ in
$\prod\left(  \omega\right)  $. As the join, that DSD $\omega$ refines each of
the compatible DSDs. The combined measurement of the CSCO of commuting
operators is the single (non-degenerate) measurement by the maximal DSD that
is the join of their eigenspace DSDs.

In addition to being able to naturally represent measurement, the quantum
logic of DSDs in useful for quite different reasons. There is a pedagogical
model of quantum mechanics using vector spaces over $%
\mathbb{Z}
_{2}$, called "quantum mechanics over sets" (QM/Sets) (\cite{ell:qmoversets},
\cite{ell:mammoth}), whose probability calculus is a non-commutative version
of the classical Laplace-Boole finite probability theory with real-valued
random variables. Such real-valued random variables on a finite sample space
$U$ cannot be represented or "internalized" as operators on $%
\mathbb{Z}
_{2}^{\left\vert U\right\vert }$--but they \textit{can be represented} by DSDs
on $%
\mathbb{Z}
_{2}^{\left\vert U\right\vert }$. This allows the quantum logic of DSDs'
treatment of measurement in QM to be reproduced in an appropriate form in the
pedagogical model of QM/Sets, and that in turn allows simplified pedagogical
versions of quantum results such as the two-slit experiment, the indeterminacy
principle, Bell's Theorem, and so forth.

In the remainder of this introductory treatment of quantum partition logic, we
will focus on this pedagogical model of QM/Sets using vector spaces over $%
\mathbb{Z}
_{2}$--together with an Appendix on the combinatorics of DSDs over finite
vector spaces over $GF\left(  q\right)  $ since QM/Sets uses the special case
of $q=2$.

\section{Review of QM/Sets}

\subsection{Previous attempts to model QM over sets}

QM/Sets is a pedagogical or "toy" model of quantum mechanics over sets where
the quantum probability calculus is a non-commutative version of the ordinary
Laplace-Boole finite logical probability theory (\cite{laplace:probs},
\cite{boole:lot}) and where the usual vector spaces over $%
\mathbb{C}
$ for QM are replaced with vector spaces $%
\mathbb{Z}
_{2}^{n}$ over $%
\mathbb{Z}
_{2}$. Fix a basis for $%
\mathbb{Z}
_{2}^{n}$ [i.e., a maximal DSD in $DSD\left(
\mathbb{Z}
_{2}^{n}\right)  $] and that basis set is the sample space or outcome space
for the Laplace-Boole finite probability calculus. But there are many
incompatible basis sets for $%
\mathbb{Z}
_{2}^{n}$ so, in that sense, the probability calculus of QM/Sets is a
non-commutative version of the Laplace-Boole calculus.

Quantum mechanics over sets is a bare-bones (e.g., non-physical\footnote{In
full QM, the DeBroglie relations connect mathematical notions such as
frequency and wave-length to physical notions such as energy and momentum.
QM/sets is "non-physical" in the sense that it is a sets-version of the pure
mathematical framework of (finite-dimensional) QM without those direct
physical connections.}) "logical skeleton" of QM with appropriate versions of
spectral decomposition, the Dirac brackets, the norm, observable-attributes,
the Born rule, commutators, and density matrices all in the simple setting of
sets,\footnote{Given a basis set for $%
\mathbb{Z}
_{2}^{n}$, each vector is expressed as a subset of the basis set.} but that
nevertheless provides models of characteristically quantum results (e.g., a
QM/Sets version of the double-slit experiment \cite{ell:mammoth}). In that
manner, QM/Sets can serve not only as a pedagogical (or "toy") model of QM but
perhaps as an engine to better elucidate QM itself by representing the quantum
features in a simple setting.

There have been at least three previous attempts at developing a version of QM
where the base field of $%
\mathbb{C}
$ is replaced by $%
\mathbb{Z}
_{2}$ (\cite{schum:modal}, \cite{hansonsabry:dqt}, and \cite{tak:mutant}).
Since there are no inner products in vector spaces over a finite field, the
"trick" is how to define the brackets, the norm, and then the probability
algorithm. All these previous attempts use the aspect of full QM that the bras
are dual vectors so the brackets take their values in the base field of $%
\mathbb{Z}
_{2}$. For instance, the Schumacher-Westmoreland model does "not make use of
the idea of probability" \cite[p. 919]{schum:modal} and have instead only a
modal interpretation ($1=$ possibility and $0=$ impossibility). There is a
fourth category-theoretic model where the objects are sets
\cite{abram-coecke:catqm} but it also has the "brackets" taking only $0,1$
values and thus has only a modal or "possibilistic" interpretation.

\subsection{The Yoga of transporting vector space structures}

There is a method or "Yoga" to transport some structures from a vector space
$V$ over a field $\mathbb{K}$ to a vector space $V^{\prime}$ over a
\textit{different} field $\mathbb{K}^{\prime}$. Select a basis set $U$ for the
source space $V$ and then consider a structure on $V$ that can be
characterized in terms of the basis set $U$. Then apply the free vector space
over the field $\mathbb{K}^{\prime}$ construction to $U$ to generate the
target vector space $V^{\prime}$. Since the source structure was defined in
terms of the basis set $U$, it can be carried over or "transported" to
$V^{\prime}$ via \textit{its} basis set $U$.

This Yoga can be stated in rigorous terms using category theory
(\cite{maclane:cwm2}; \cite{awodey:cat-theory}). The construction of the free
vector space over a field $\mathbb{K}$ is a functor from the category $Sets$
of sets and functions to the category $Vect_{\mathbb{K}}$ of vector spaces
over $\mathbb{K}$ and linear transformations. The functor will only be used
here on finite sets where it takes a finite set $U$ to the vector space
$\mathbb{K}^{U}$. This paper is about direct-sum decompositions of a
finite-dimensional vector space $V$. A DSD a set $\left\{  V_{i}\right\}  $ of
disjoint subspaces so that the whole space $V$ is their direct sum, or, in
terms of category theory, $V$ is the coproduct $V=\oplus V_{i}$ of the
subspaces $\left\{  V_{i}\right\}  $. In the category $Sets$, a set $\left\{
B_{i}\right\}  $ of disjoint subsets of a set $U$ is a set partition of $U$ if
$\cup B_{i}=U$, or, in terms of category theory, $U$ is the coproduct of the
disjoint subsets $\left\{  B_{i}\right\}  $. The free vector space over
$\mathbb{K}$ functor is a left adjoint, "left adjoints preserve colimits"
\cite[p. 197]{awodey:cat-theory}, and coproducts are a special type of
colimit. Hence the free vector space functor carries a set partition
$\pi=\left\{  B_{i}\right\}  _{i=1,...,m}$ to the DSD $\left\{  V_{i}%
=\mathbb{K}^{B_{i}}\right\}  $ of $V=\mathbb{K}^{U}=\oplus\mathbb{K}^{B_{i}}$.

Now start with the structure of a DSD $\left\{  V_{i}\right\}  $ on $V\in
Vect_{\mathbb{K}}$. What we previously called "characterizing the structure in
terms of a basis set $U$" is rigorously interpreted to mean, in this case,
finding a basis $U$ and a partition $\left\{  B_{i}\right\}  $ on $U$ so that
the given DSD $\left\{  V_{i}\right\}  $ is the image of the free vector space
functor, i.e., $V=\mathbb{K}^{U}=\oplus\mathbb{K}^{B_{i}}=\oplus V_{i}$. But
then the free vector space functor over a different field $\mathbb{K}^{\prime
}$ can be applied to the same set partition $\left\{  B_{i}\right\}  $ of the
set $U$ to generate a DSD $\left\{  V_{i}^{\prime}=\mathbb{K}^{\prime B_{i}%
}\right\}  $ of $V^{\prime}=\mathbb{K}^{\prime U}$. That is how to rigorously
describe "transporting" a set-based structure on a vector $V$ over
$\mathbb{K}$ to a vector space $V^{\prime}$ over a different field
$\mathbb{K}^{\prime}$.

To show that any given DSD $\left\{  V_{i}\right\}  $ of $V$ is in the image
of the free vector space over $\mathbb{K}$ functor, pick basis set $B_{i}$ of
$V_{i}$. The sets $B_{i}$ are disjoint and since $\left\{  V_{i}\right\}  $ is
a DSD, the union $U=\cup B_{i}$ is a basis for $V$ so $V_{i}=\mathbb{K}%
^{B_{i}}$ and $V=\mathbb{K}^{U}=\oplus\mathbb{K}^{B_{i}}$.

This method is applied to the transporting of self-adjoint operators from $V=%
\mathbb{C}
^{n}$ to $V^{\prime}=%
\mathbb{Z}
_{2}^{n}$ that motivates QM/Sets. A self-adjoint operator $F:%
\mathbb{C}
^{n}\rightarrow%
\mathbb{C}
^{n}$ has a basis $U=\left\{  u_{1},...,u_{n}\right\}  $ of orthonormal
eigenvectors and it has real distinct eigenvalues $\left\{  \phi_{i}\right\}
_{j=1,...,m}$, so it defines the real eigenvalue function $f:U\rightarrow%
\mathbb{R}
$ where for $u_{j}\in U$, $f\left(  u_{j}\right)  $ is one of the distinct
eigenvalues $\left\{  \phi_{i}\right\}  _{i=1,...m}$. For each distinct
eigenvalue $\phi_{i}$, there is the eigenspace $V_{i}$ of its eigenvectors and
$\left\{  V_{i}\right\}  _{i=1,...,m}$ is a DSD on $V=%
\mathbb{C}
^{n}$. The inverse-image $\pi=\left\{  B_{i}=f^{-1}\left(  \phi_{i}\right)
\right\}  _{i=1,...,m}$ of the eigenvalue function $f:U\rightarrow%
\mathbb{R}
$ is a set partition on $U$.

Thus the set-based structure we have is the set $U$ with a partition $\left\{
B_{i}=f^{-1}\left(  \phi_{i}\right)  \right\}  _{i}$ on $U$ induced by a
real-value function $f:U\rightarrow%
\mathbb{R}
$ on $U$. That set-based structure is sufficient to reconstruct the DSD
$\left\{  V_{i}=%
\mathbb{C}
^{B_{i}}\right\}  _{i}$ on $V=%
\mathbb{C}
^{n}\cong%
\mathbb{C}
^{U}=\oplus%
\mathbb{C}
^{B_{i}}$ as well as the original operator $F$. The operator $F$ is defined on
the basis $U$ by $Fu_{j}=f\left(  u_{j}\right)  u_{j}$ for $j=1,...,n$. But it
might be helpful to go through the categorical construction. Scalar
multiplication in the vector space is given by the set function $%
\mathbb{C}
\times%
\mathbb{C}
^{n}\overset{\cdot}{\rightarrow}%
\mathbb{C}
^{n}$. There is the injection of the generators function $\iota:U\rightarrow%
\mathbb{C}
^{U}\cong%
\mathbb{C}
^{n}$ and there is the function $f:U\rightarrow%
\mathbb{R}
\subseteq%
\mathbb{C}
$ so by the universal mapping property (UMP) of the product $%
\mathbb{C}
\times%
\mathbb{C}
^{n}$, we have the factor map $<f,\iota>:U\rightarrow%
\mathbb{C}
\times%
\mathbb{C}
^{n}$ and thus the composition $f\cdot\iota:U\overset{<f,\iota
>}{\longrightarrow}%
\mathbb{C}
\times%
\mathbb{C}
^{n}\overset{\cdot}{\longrightarrow}%
\mathbb{C}
^{n}$. Then we use the UMP of the \textit{free} vector space over $%
\mathbb{C}
$ functor.

\begin{center}
$%
\begin{array}
[c]{ccc}%
U &  & \\
^{\iota}\downarrow^{{}} & \searrow^{\varphi} & \\%
\mathbb{C}
^{n} & \overset{\exists!F}{\longrightarrow} & W
\end{array}
$

UMP of free vector space functor
\end{center}

\noindent That UMP is that for any function $\varphi:U\rightarrow W$ from the
set $U$ to any vector space $W$ over $%
\mathbb{C}
$, there is a linear transformation $F:%
\mathbb{C}
^{n}=%
\mathbb{C}
^{U}\rightarrow W$ such that $F\iota=\varphi$. Taking $\varphi=f\cdot\iota$
and $W=%
\mathbb{C}
^{n}$, there is a unique linear operator $F$ on $%
\mathbb{C}
^{n}$ such that for any $u_{j}\in U$, $F\iota\left(  u_{j}\right)  $ is the
scalar multiple $f\left(  u_{j}\right)  \cdot\iota\left(  u\right)  _{j}$ or,
where we write $u_{j}=\iota\left(  u_{j}\right)  $ (and scalar multiplication
by juxtaposition), $Fu_{j}=f\left(  u_{j}\right)  u_{j}$ for $j=1,...,n$. That
process of going from the function $f:U\rightarrow%
\mathbb{R}
$ on a basis set $U$ of $%
\mathbb{C}
^{U}$ to an operator on $%
\mathbb{C}
^{U}$ might be called \textit{internalizing} the function $f:U\rightarrow%
\mathbb{R}
$ in $%
\mathbb{C}
^{U}$.

Given the set-based structure of a real-valued function $f:U\rightarrow%
\mathbb{R}
$, which determines the set partition $\left\{  f^{-1}\left(  \phi_{i}\right)
\right\}  _{i=1,...,m}$ on $U$, we then apply the free vector space over $%
\mathbb{Z}
_{2}$ functor to construct the vector space $%
\mathbb{Z}
_{2}^{U}$. That vector space is more familiar in the form of the powerset
$\wp\left(  U\right)  \cong%
\mathbb{Z}
_{2}^{U}$ since each function $U\rightarrow%
\mathbb{Z}
_{2}=\left\{  0,1\right\}  $ in $%
\mathbb{Z}
_{2}^{U}$ is the characteristic function $\chi_{S}$ of a subset $S\in
\wp\left(  U\right)  $. The free vector space functor $%
\mathbb{Z}
_{2}^{()}$ takes the coproduct $U=\cup_{i=1}^{m}f^{-1}\left(  \phi_{i}\right)
$ to the DSD $\left\{  \wp\left(  f^{-1}\left(  \phi_{i}\right)  \right)
\right\}  $ of $\wp\left(  U\right)  $. The attempt to internalize the real
function $f:U\rightarrow%
\mathbb{R}
$ would use the scalar multiplication function $%
\mathbb{Z}
_{2}\times%
\mathbb{Z}
_{2}^{U}\overset{\cdot}{\rightarrow}%
\mathbb{Z}
_{2}^{U}$ and would only work if $f$ took values in $%
\mathbb{Z}
_{2}=\left\{  0,1\right\}  \subseteq%
\mathbb{R}
$ in which case $f$ would be a characteristic function $\chi_{S}$ for some
subset $S\in\wp\left(  U\right)  $. In that special case, the internalized
operator would be the projection operator $P_{S}:%
\mathbb{Z}
_{2}^{U}\rightarrow%
\mathbb{Z}
_{2}^{U}$ which in terms of the basis $U$ has the action $P_{S}\left(
T\right)  =S\cap T$ taking any subset $T\in\wp\left(  U\right)  $ to $S\cap
T\in\wp\left(  S\right)  $.

Hence outside of characteristic functions, the real-valued functions
$f:U\rightarrow%
\mathbb{R}
$ cannot be internalized as operators on $%
\mathbb{Z}
_{2}^{U}$. But that is fine since the idea of the model QM/Sets is that given
a basis $U$ of $%
\mathbb{Z}
_{2}^{n}$, the quantum probability calculus will just be the classical finite
probability calculus with the outcome set or sample space $U$ where
$f:U\rightarrow%
\mathbb{R}
$ is a real-valued random variable. We have illustrated the transporting of
set-based structures on $%
\mathbb{C}
^{n}$ to $%
\mathbb{Z}
_{2}^{n}$ using a basis set $U$, but in the stand-alone model QM/Sets, we cut
the umbilical cord to $%
\mathbb{C}
^{n}$ and work with any other basis $U^{\prime}$ of $%
\mathbb{Z}
_{2}^{n}$ and real-valued random variables $g:U^{\prime}\rightarrow%
\mathbb{R}
$ on \textit{that} sample space.

Other structures can be transported across the bridge from $%
\mathbb{C}
^{n}$ to $%
\mathbb{Z}
_{2}^{n}$. QM/Sets differs from the other four attempts to define some toy
version of QM on sets by the treatment of the Dirac brackets. Starting with
our orthonormal basis $U$ on a finite-dimensional Hilbert space $%
\mathbb{C}
^{n}$ (where the bracket is the inner product), we need to define the
transported brackets applied to two subsets $S,T\subseteq U$ in $\wp\left(
U\right)  $. The two subsets define the vectors $\psi_{S}=\sum_{u\in
S}\left\vert u\right\rangle $ and $\psi_{T}=\sum_{u\in T}\left\vert
u\right\rangle $ in $%
\mathbb{C}
^{n}$ which have the bracket value $\left\langle \psi_{S}|\psi_{T}%
\right\rangle =\left\vert S\cap T\right\vert $. Since that value is defined
just in terms of the subsets $S,T\subseteq U$ as the cardinality of their
overlap, that value can be transported to $\wp\left(  U\right)  $ as the
real-valued basis-dependent brackets $\left\langle S|_{U}T\right\rangle
=\left\vert S\cap T\right\vert $ (see below).

\subsection{Laplace-Boole finite probability theory}

Since our purpose is conceptual rather than mathematical, we will stick to the
simplest case of finite probability theory with a finite sample space or
outcome space $U=\left\{  u_{1},...,u_{n}\right\}  $ of $n$ equiprobable
outcomes and to finite dimensional QM.\footnote{The mathematics can be
generalized to the case where each point $u_{j}$ in the sample space has a
probability $p_{j}$ (when it is a basis set with point probabilities that is
transported) but the simpler case of equiprobable points serves our conceptual
purposes.} The \textit{events} are the subsets $S\subseteq U$, and the
\textit{probability} of an event $S$ occurring in a trial is the ratio of the
cardinalities: $\Pr\left(  S\right)  =\frac{\left\vert S\right\vert
}{\left\vert U\right\vert }$. Given that a conditioning event $S\subseteq U$
occurs, the \textit{conditional probability} that $T\subseteq U$ occurs is:
$\Pr(T|S)=\frac{\Pr\left(  T\cap S\right)  }{\Pr(S)}=\frac{\left\vert T\cap
S\right\vert }{\left\vert S\right\vert }$. The ordinary probability
$\Pr\left(  T\right)  $ of an event $T$ can be taken as the conditional
probability with $U$ as the conditioning event so all probabilities can be
seen as conditional probabilities. Given a real-valued random variable
$f:U\rightarrow%
\mathbb{R}
$ on the elements of $U$, the \textit{probability of observing a value }%
$r$\textit{ given an event }$S$\textit{ }is the conditional probability of the
event $f^{-1}\left(  r\right)  $ given $S$:

\begin{center}
$\Pr\left(  r|S\right)  =\frac{\left\vert f^{-1}\left(  r\right)  \cap
S\right\vert }{\left\vert S\right\vert }$.
\end{center}

\noindent That is all the probability theory we will need here. Our first task
is to show how the mathematics of finite probability theory can be recast
using the mathematical notions of quantum mechanics with the base field of $%
\mathbb{Z}
_{2}$.

\subsection{Vector spaces over $%
\mathbb{Z}
_{2}$}

To show how classical Laplace-Boole finite probability theory can be recast as
a quantum probability calculus, we use finite dimensional vector spaces over $%
\mathbb{Z}
_{2}$. The power set $\wp\left(  U\right)  $ of $U=\left\{  u_{1}%
,...,u_{n}\right\}  $ is a vector space over $%
\mathbb{Z}
_{2}=\left\{  0,1\right\}  $, isomorphic to $%
\mathbb{Z}
_{2}^{n}$, where the vector addition $S+T$ is the \textit{symmetric
difference} of subsets. That is, for $S,T\subseteq U$,

\begin{center}
$S+T=\left(  S-T\right)  \cup\left(  T-S\right)  =S\cup T-S\cap T$
\end{center}

\noindent so the members of $S+T$ are the elements that are members of $S$ or
members of $T$ but not members of both.

\noindent The $U$\textit{-basis} in $\wp\left(  U\right)  $ is the set of
singletons $\left\{  u_{1}\right\}  ,\left\{  u_{2}\right\}  ,...,\left\{
u_{n}\right\}  $, i.e., the set $\left\{  \left\{  u_{j}\right\}  \right\}
_{j=1,...,n}$. In the context of $DSD\left(
\mathbb{Z}
_{2}^{n}\right)  $, that basis set would correspond to the maximal element
$\omega=\left\{  U_{j}\right\}  _{j=1,...,n}$ where $U_{j}$ is the
one-dimensional subspace $\left\{  \emptyset,\left\{  u_{j}\right\}  \right\}
$. A vector $S\in\wp\left(  U\right)  $ is specified in the $U$-basis as
$S=\sum_{u_{j}\in S}\left\{  u_{j}\right\}  $ and it is characterized by its $%
\mathbb{Z}
_{2}$-valued characteristic function $\chi_{S}:U\rightarrow%
\mathbb{Z}
_{2}\subseteq%
\mathbb{R}
$ of coefficients since $S=\sum_{u_{j}\in U}\chi_{S}\left(  u_{j}\right)
\left\{  u_{j}\right\}  $.

Consider the simple case of $U=\left\{  a,b,c\right\}  $ where the $U$-basis
is $\left\{  a\right\}  $, $\left\{  b\right\}  $, and $\left\{  c\right\}  $.
The three subsets $\left\{  a,b\right\}  $, $\left\{  b,c\right\}  $, and
$\left\{  a,b,c\right\}  $ also form a basis since:

$\left\{  b,c\right\}  +\left\{  a,b,c\right\}  =\left\{  a\right\}  $;

$\left\{  b,c\right\}  +\left\{  a,b\right\}  +\left\{  a,b,c\right\}
=\left\{  b\right\}  $; and

$\left\{  a,b\right\}  +\left\{  a,b,c\right\}  =\left\{  c\right\}  $.

\noindent These new basis vectors could be considered as the basis-singletons
in another equicardinal sample space $U^{\prime}=\left\{  a^{\prime}%
,b^{\prime},c^{\prime}\right\}  $ where $\left\{  a^{\prime}\right\}  $,
$\left\{  b^{\prime}\right\}  $, and $\left\{  c^{\prime}\right\}  $ refer to
the same abstract vector as $\left\{  a,b\right\}  $, $\left\{  b,c\right\}
$, and $\left\{  a,b,c\right\}  $ respectively.

In the following \textit{ket table}, each row is an abstract vector of $%
\mathbb{Z}
_{2}^{3}$ expressed in the $U$-basis, the $U^{\prime}$-basis, and a
$U^{\prime\prime}$-basis.

\begin{center}%
\begin{tabular}
[c]{|c|c|c|}\hline
$U=\left\{  a,b,c\right\}  $ & $U^{\prime}=\left\{  a^{\prime},b^{\prime
},c^{\prime}\right\}  $ & $U^{\prime\prime}=\left\{  a^{\prime\prime
},b^{\prime\prime},c^{\prime\prime}\right\}  $\\\hline\hline
$\left\{  a,b,c\right\}  $ & $\left\{  c^{\prime}\right\}  $ & $\left\{
a^{\prime\prime},b^{\prime\prime},c^{\prime\prime}\right\}  $\\\hline
$\left\{  a,b\right\}  $ & $\left\{  a^{\prime}\right\}  $ & $\left\{
b^{\prime\prime}\right\}  $\\\hline
$\left\{  b,c\right\}  $ & $\left\{  b^{\prime}\right\}  $ & $\left\{
b^{\prime\prime},c^{\prime\prime}\right\}  $\\\hline
$\left\{  a,c\right\}  $ & $\left\{  a^{\prime},b^{\prime}\right\}  $ &
$\left\{  c^{\prime\prime}\right\}  $\\\hline
$\left\{  a\right\}  $ & $\left\{  b^{\prime},c^{\prime}\right\}  $ &
$\left\{  a^{\prime\prime}\right\}  $\\\hline
$\left\{  b\right\}  $ & $\left\{  a^{\prime},b^{\prime},c^{\prime}\right\}  $
& $\left\{  a^{\prime\prime},b^{\prime\prime}\right\}  $\\\hline
$\left\{  c\right\}  $ & $\left\{  a^{\prime},c^{\prime}\right\}  $ &
$\left\{  a^{\prime\prime},c^{\prime\prime}\right\}  $\\\hline
$\emptyset$ & $\emptyset$ & $\emptyset$\\\hline
\end{tabular}

Ket table giving a vector space isomorphism: $%
\mathbb{Z}
_{2}^{3}\cong\wp\left(  U\right)  \cong\wp\left(  U^{\prime}\right)  \cong%
\wp\left(  U^{\prime\prime}\right)  $ where row = ket.
\end{center}

In the Dirac notation \cite{dirac:principles}, the\textit{ ket} $\left\vert
\left\{  a,c\right\}  \right\rangle $ represents the abstract vector that is
represented in the $U$-basis coordinates as $\left\{  a,c\right\}  $. A row of
the ket table gives the different representations of the \textit{same} ket in
the different bases, e.g., $\left\vert \left\{  a,c\right\}  \right\rangle
=\left\vert \left\{  a^{\prime},b^{\prime}\right\}  \right\rangle =\left\vert
\left\{  c^{\prime\prime}\right\}  \right\rangle $.

\subsection{The brackets and the norm}

In a Hilbert space, the inner product is used to define the brackets
$\left\langle v_{i}|v\right\rangle $ and the norm $\left\Vert v\right\Vert
=\sqrt{\left\langle v|v\right\rangle }$ but there are no inner products in
vector spaces over finite fields. The different attempts to develop a toy
model of QM over a finite field (\cite{schum:modal}, \cite{tak:mutant},
\cite{hansonsabry:dqt}) such as $%
\mathbb{Z}
_{2}$ differ from this model in how they address this problem. The treatment
of the Dirac brackets and norm defined here is distinguished by the fact that
the resulting probability calculus in QM/Sets is (a non-commutative version
of) classical finite probability theory (instead of just a modal calculus with
values $0$ and $1$).

For a singleton basis vector $\left\{  u_{j}\right\}  \subseteq U$, the
(basis-dependent) \textit{bra} $\left\langle \left\{  u_{j}\right\}
\right\vert _{U}:\wp\left(  U\right)  \rightarrow%
\mathbb{R}
$ is defined by the \textit{bracket}:

\begin{center}
$\left\langle \left\{  u\right\}  |_{U}S\right\rangle =\left\{
\begin{array}
[c]{c}%
1\text{ if }u\in S\\
0\text{ if }u\notin S
\end{array}
\right.  =\left\vert \left\{  u_{j}\right\}  \cap S\right\vert =\chi
_{S}\left(  u_{j}\right)  $.
\end{center}

\noindent Note that the bra and the bracket is defined in terms of the
$U$-basis and that is indicated by the $U$-subscript on the bra portion of the
bracket. Then for $u_{j},u_{k}\in U$, $\left\langle \left\{  u_{j}\right\}
|_{U}\left\{  u_{k}\right\}  \right\rangle =\chi_{\left\{  u_{k}\right\}
}\left(  u_{j}\right)  =\chi_{\left\{  u_{j}\right\}  }\left(  u_{k}\right)
=\delta_{jk}$ (the Kronecker delta function) which is the QM/Sets-version of
$\left\langle v_{j}|v_{k}\right\rangle =\delta_{jk}$ for an orthonormal basis
$\left\{  \left\vert v_{j}\right\rangle \right\}  $ of $%
\mathbb{C}
^{n}$. The bracket linearly extends \textit{in the natural numbers }$%
\mathbb{N}
\subseteq%
\mathbb{R}
$ to any two vectors $T,S\in\wp\left(  U\right)  $:\footnote{\noindent Here
$\left\langle T|_{U}S\right\rangle =\left\vert T\cap S\right\vert $ takes
values in the natural numbers $%
\mathbb{N}
$ outside the base field of $%
\mathbb{Z}
_{2}$ just like, say, the Hamming distance function $d_{H}\left(  T,S\right)
=\left\vert T+S\right\vert $ on vector spaces over $%
\mathbb{Z}
_{2}$ in coding theory. \cite{mceliece:coding} Thus the "size of overlap" bra
$\left\langle T\right\vert _{U}:\wp\left(  U\right)  \rightarrow%
\mathbb{N}
$ is not to be confused with the dual ("parity of overlap") functional
$\varphi_{T}=\sum_{u_{j}\in T}\varphi_{u_{j}}:\wp\left(  U\right)  \rightarrow%
\mathbb{Z}
_{2}$ where $\varphi_{u_{j}}\left(  \left\{  u_{k}\right\}  \right)
=\delta_{jk}$ for $U=\left\{  u_{1},...,u_{n}\right\}  $.}

\begin{center}
$\left\langle T|_{U}S\right\rangle =\left\vert T\cap S\right\vert $.
\end{center}

\noindent This is the QM/Sets-version of the Dirac brackets in the mathematics
of QM.

As noted above, this treatment of the brackets is motivated by the general
heuristic for transporting basis-set-defined structures between vector spaces
over different fields, e.g., from $%
\mathbb{C}
^{n}$ to $%
\mathbb{Z}
_{2}^{n}$. In both cases, the bracket gives a measure of the overlap or
indistinctness of the two vectors.\footnote{One possible misinterpretation of
QM/Sets is to misinterpret the transporting method as an embedding $%
\mathbb{Z}
_{2}^{n}\rightarrow%
\mathbb{C}
^{n}$ defined by $\left\{  u_{j}\right\}  \longmapsto\left\vert u_{j}%
\right\rangle $ using a basis for each space. But such an embedding from a
vector space over a field of finite characteristic to a vector space of
characteristic zero cannot be linear. The repeated sum of a nonzero element in
the domain space will eventually be $0$ but its repeated nonzero image in the
codomain space can never be $0$. Indeed in QM/Sets, the brackets $\left\langle
T|_{U}S\right\rangle =\left\vert T\cap S\right\vert $ for $T,T^{\prime
},S\subseteq U$ should be thought of \textit{only} as a measure of the overlap
since they are not even linear, e.g., $\left\langle T+T^{\prime}%
|_{U}S\right\rangle \neq\left\langle T|_{U}S\right\rangle +\left\langle
T^{\prime}|_{U}S\right\rangle $ whenever $T\cap T^{\prime}\neq\emptyset$.} The
ket $\left\vert S\right\rangle $ is the same as the ket $\left\vert S^{\prime
}\right\rangle $ for some subset $S^{\prime}\subseteq U^{\prime}$ in another
$U^{\prime}$-basis, but when the bra $\left\langle \left\{  u_{j}\right\}
\right\vert _{U}$ is applied to the ket $\left\vert S\right\rangle =\left\vert
S^{\prime}\right\rangle $, then it is the subset $S\subseteq U$, not
$S^{\prime}\subseteq U^{\prime}$, that comes outside the ket symbol
$\left\vert \ \right\rangle $\noindent\ in $\left\langle \left\{
u_{j}\right\}  |_{U}S\right\rangle =\left\vert \left\{  u_{j}\right\}  \cap
S\right\vert $.\footnote{The term "$\left\{  u_{j}\right\}  \cap S^{\prime}$"
is not even defined in general since it is the intersection of subsets
$\left\{  u_{j}\right\}  \subseteq U$ and $S^{\prime}\subseteq U^{\prime}$ of
two different universe sets $U$ and $U^{\prime}$.} Heuristically, the bra
$\left\langle T\right\vert _{U}$ can be thought of as a row-vector of zeros
and ones expressed in the $U$-basis, and then the ket $\left\vert
S\right\rangle $ is expressed as a column vector in the $U$-basis, and
$\left\langle T|_{U}S\right\rangle $ is their dot product \textit{computed in
the reals}.

The $U$\textit{-norm} $\left\Vert S\right\Vert _{U}:\wp\left(  U\right)
\rightarrow%
\mathbb{R}
$ is defined, as usual, as the square root of the bracket:\footnote{We use the
double-line notation $\left\Vert S\right\Vert _{U}$ for the $U$-norm of a set
to distinguish it from the single-line notation $\left\vert S\right\vert $ for
the cardinality of a set. We also use the double-line notation $\left\Vert
\psi\right\Vert $ for the norm in QM although sometimes the single line
notation $\left\vert \psi\right\vert $ is used elsewhere.}

\begin{center}
$\left\Vert S\right\Vert _{U}=\sqrt{\left\langle S|_{U}S\right\rangle }%
=\sqrt{\left\vert S\cap S\right\vert }=\sqrt{|S|}$
\end{center}

\noindent for $S\in\wp\left(  U\right)  $ which is the QM/Sets-version of the
norm $\left\Vert \psi\right\Vert =\sqrt{\left\langle \psi|\psi\right\rangle }$
in ordinary QM. Hence $\left\Vert S\right\Vert _{U}^{2}=\left\vert
S\right\vert $ is the counting measure on $\wp\left(  U\right)  $. Note that a
ket has to be expressed in the $U$-basis to apply the $U$-norm definition so,
for example, $\left\Vert \left\{  a^{\prime}\right\}  \right\Vert _{U}%
=\sqrt{2}$ since $\left\vert \left\{  a^{\prime}\right\}  \right\rangle
=\left\vert \left\{  a,b\right\}  \right\rangle $.

\subsection{Numerical attributes, linear operators, and DSDs}

In classical physics, the observables are numerical attributes, e.g., the
assignment of a position and momentum to particles in phase space. One of the
differences between classical and quantum physics is the replacement of these
observable numerical attributes by linear operators associated with the
observables where the values of the observables appear as eigenvalues of the
operators. But this difference may be smaller than it would seem at first
since a numerical attribute $f:U\rightarrow%
\mathbb{R}
$ can be recast into an operator-like format in QM/Sets where it determines
direct-sum decomposition of "eigenspaces," and there is even a
QM/Sets-analogue of spectral decomposition.

An observable, i.e., a self-adjoint operator, on a finite-dimensional Hilbert
space $V$ has a "home" basis set of orthonormal eigenvectors. Using the
transport method, a real-valued attribute $f:U\rightarrow%
\mathbb{R}
$ defined on $U=\left\{  u_{1},...,u_{n}\right\}  $ has the $U$-basis for
$\wp\left(  U\right)  \cong%
\mathbb{Z}
_{2}^{n}$ as a "home" basis set. The connection between the numerical
attributes $f:U\rightarrow%
\mathbb{R}
$ of QM/Sets and the self-adjoint operators of full QM can also be established
by seeing the function $f$ as being \textit{like} an "operator"
$f\upharpoonright()$ on $\wp\left(  U\right)  $ in that it is used to define a
sets-version of an "eigenvalue" equation [where $f\upharpoonright S$ is the
\textit{restriction} of $f$ to $S\in\wp\left(  U\right)  $]. For any subset
$S\in\wp\left(  U\right)  $, the definition of the equation is:

\begin{center}
$f\upharpoonright S=\phi_{i}S$ holds $\equiv_{df}$ $f$ is constant on the
subset $S$ with the value $\phi_{i}$.
\end{center}

\noindent This is the QM/Sets-version of an \textit{eigenvalue equation} for
arbitrary functions on a set $f:U\rightarrow%
\mathbb{R}
$. Whenever $S$ satisfies $f\upharpoonright S=\phi_{i}S$ for some $\phi_{i}$,
then $S$ is said to be an \textit{eigenvector} (= "level set") in the vector
space $\wp\left(  U\right)  $ of the numerical attribute $f:U\rightarrow%
\mathbb{R}
$, and $\phi_{i}\in%
\mathbb{R}
$ is the associated \textit{eigenvalue }(= constant value on a level set).
Each eigenvalue $\phi_{i}$ determines as usual an \textit{eigenspace}
$\wp\left(  f^{-1}\left(  \phi_{i}\right)  \right)  $ of its eigenvectors
which is a subspace of the vector space $\wp\left(  U\right)  \cong%
\mathbb{Z}
_{2}^{n}$. The whole space $\wp\left(  U\right)  $ can be expressed as usual
as the direct sum of the eigenspaces: $\wp\left(  U\right)  =\oplus_{\phi
_{i}\in f\left(  U\right)  }\wp\left(  f^{-1}\left(  \phi_{i}\right)  \right)
$ so $\left\{  \wp\left(  f^{-1}\left(  \phi_{i}\right)  \right)  \right\}
_{\phi_{i}\in f\left(  U\right)  }$ is a DSD in $DSD\left(
\mathbb{Z}
_{2}^{n}\right)  $. Since $f:U\rightarrow%
\mathbb{R}
$ does not define an operator $%
\mathbb{Z}
_{2}^{n}\rightarrow%
\mathbb{Z}
_{2}^{n}$ (unless it is a characteristic function), we see one of the reasons
for developing quantum partition logic using DSDs.

Moreover, for distinct eigenvalues $\phi_{i}\neq\phi_{i^{\prime}}$, any
corresponding eigenvectors $S\in\wp\left(  f^{-1}\left(  \phi_{i}\right)
\right)  $ and $T\in\wp\left(  f^{-1}\left(  \phi_{i^{\prime}}\right)
\right)  $ are \textit{orthogonal} in the sense that $\left\langle
T|_{U}S\right\rangle =0$. In general, for vectors $S,T\in\wp\left(  U\right)
$, orthogonality means zero overlap, i.e., disjointness.

The characteristic function $\chi_{S}:U\rightarrow%
\mathbb{R}
$ for $S\subseteq U$ has the eigenvalues of $0$ and $1$ so it is a numerical
attribute that \textit{can} be "internalized" as a linear operator
$S\cap():\wp\left(  U\right)  \rightarrow\wp\left(  U\right)  $. Hence in this
case, the "eigenvalue equation" $f\upharpoonright T=\phi_{i}T$ for $f=\chi
_{S}$ internalizes as an actual eigenvalue equation $S\cap T=\phi_{i}T$ for a
linear\footnote{It should be noted that the projection operator $S\cap
():\wp\left(  U\right)  \rightarrow\wp\left(  U\right)  $ is not only
idempotent but linear, i.e., $\left(  S\cap T_{1}\right)  +(S\cap T_{2}%
)=S\cap\left(  T_{1}+T_{2}\right)  $. Indeed, this is the distributive law
when $\wp\left(  U\right)  $ is interpreted as a Boolean ring with
intersection as multiplication.} operator $S\cap()$ with the resulting
eigenvalues of $1$ and $0$, and with the resulting eigenspaces $\wp\left(
S\right)  $ and $\wp\left(  S^{c}\right)  $ (where $S^{c}$ is the complement
of $S$) and atomic DSD $\left\{  \wp\left(  S\right)  ,\wp\left(
S^{c}\right)  \right\}  $.

The characteristic attributes $\chi_{S}:U\rightarrow%
\mathbb{R}
$ are characterized by the property that their value-wise product, i.e.,
$\left(  \chi_{S}\bullet\chi_{S}\right)  \left(  u_{j}\right)  =\chi
_{S}\left(  u_{j}\right)  \chi_{S}\left(  u_{j}\right)  $, is equal to the
attribute value $\chi_{S}\left(  u_{j}\right)  $, and that is reflected in the
idempotency of the corresponding operators:

\begin{center}
$\wp\left(  U\right)  \overset{S\cap()}{\longrightarrow}\wp\left(  U\right)
\overset{S\cap()}{\longrightarrow}\wp\left(  U\right)  =\wp\left(  U\right)
\overset{S\cap()}{\longrightarrow}\wp\left(  U\right)  $.
\end{center}

\noindent Thus the operators $S\cap()$ corresponding to the characteristic
functions $\chi_{S}$ are \textit{projection operators}.

The (maximum) eigenvectors $f^{-1}\left(  r\right)  $ for $f$, with $\phi_{i}$
in the \textit{image} or \textit{spectrum} $f\left(  U\right)  \subseteq%
\mathbb{R}
$, span the set $U$, i.e., $U=\cup_{\phi_{i}\in f\left(  U\right)  }%
f^{-1}\left(  \phi_{i}\right)  $. Hence the attribute $f:U\rightarrow%
\mathbb{R}
$ has a spectral decomposition in terms of its (projection-defining)
characteristic functions:

\begin{center}
$f=\sum_{i=1}^{m}\phi_{i}\chi_{f^{-1}\left(  \phi_{i}\right)  }:U\rightarrow%
\mathbb{R}
$

\textit{Spectral decomposition} of real-valued function $f:U\rightarrow%
\mathbb{R}
$
\end{center}

\noindent which is the QM/Sets-version of the spectral decomposition
$F=\sum_{i=1}^{m}\phi_{i}P_{i}$ of a self-adjoint operator $F$ in terms of the
projection operators $P_{i}$ for its eigenvalues $\phi_{i}$.

\subsection{The Born Rule for measurement in QM and QM/Sets}

An orthogonal decomposition of a finite set $U$ is just a partition
$\pi=\left\{  B,...\right\}  $ of $U$ since the blocks $B,B^{\prime},...$ are
orthogonal (i.e., disjoint) and their union, which is a disjoint union, is
$U$. Given such a disjoint-union decomposition of $U$, we have the:

\begin{center}
$\left\Vert U\right\Vert _{U}^{2}=\sum_{B\in\pi}\left\Vert B\right\Vert
_{U}^{2}$

Pythagorean Theorem

for disjoint-union decompositions of sets.
\end{center}

An old question is: "why the squaring in the Born rule of QM?" A superposition
state between certain definite orthogonal alternatives $A$ and $B$, where the
latter are represented by vectors $\overrightarrow{A}$ and $\overrightarrow{B}%
$, is represented by the vector sum $\overrightarrow{C}=\overrightarrow{A}%
+\overrightarrow{B}$. But what is the "strength," "intensity," or relative
importance of the vectors $\overrightarrow{A}$ and $\overrightarrow{B}$ in the
vector sum $\overrightarrow{C}$? That question requires a \textit{scalar}
measure of strength or intensity. The magnitude or "length" given by the norm
$\left\Vert {}\right\Vert $ does not answer the question since $\left\Vert
\overrightarrow{A}\right\Vert +\left\Vert \overrightarrow{B}\right\Vert
\neq\left\Vert \overrightarrow{C}\right\Vert $. But the Pythagorean Theorem
shows that the norm-squared gives the scalar measure of "intensity" that
answers the question: $\left\Vert \overrightarrow{A}\right\Vert ^{2}%
+\left\Vert \overrightarrow{B}\right\Vert ^{2}=\left\Vert \overrightarrow{C}%
\right\Vert ^{2}$ in vector spaces over $%
\mathbb{Z}
_{2}$ or over $%
\mathbb{C}
$. And when the superposition state is reduced by a measurement, then
the\textit{ probability} that the indefinite state will reduce to one of the
definite alternatives is given by that relative scalar measure of the
eigen-alternative's "strength" or "intensity"--and that is the Born Rule. In a
slogan, Born is the off-spring of Pythagoras.

Given an observable-operator $F$ in ordinary QM/$%
\mathbb{C}
$ and a numerical attribute in QM/Sets, the corresponding Pythagorean Theorems
for the complete sets of orthogonal projection operators are:

\begin{center}
$\left\Vert \psi\right\Vert ^{2}=\sum_{i}\left\Vert P_{i}\left(  \psi\right)
\right\Vert ^{2}$ and

$\left\Vert S\right\Vert _{U}^{2}=\sum_{i}\left\Vert f^{-1}\left(  \phi
_{i}\right)  \cap S\right\Vert _{U}^{2}=\sum_{i}\left\vert f^{-1}\left(
\phi_{i}\right)  \cap S\right\vert =\left\vert S\right\vert $.
\end{center}

\noindent Normalizing gives:

\begin{center}
$\sum_{i}\frac{\left\Vert P_{i}\left(  \psi\right)  \right\Vert ^{2}%
}{\left\Vert \psi\right\Vert ^{2}}=1$ and

$\sum_{i}\frac{\left\Vert f^{-1}\left(  \phi_{i}\right)  \cap S\right\Vert
_{U}^{2}}{\left\Vert S\right\Vert _{U}^{2}}=\sum_{i}\frac{\left\vert
f^{-1}\left(  \phi_{i}\right)  \cap S\right\vert }{\left\vert S\right\vert
}=1$
\end{center}

\noindent so the non-negative summands can be interpreted as
probabilities--which is the Born rule in QM and in QM/Sets.\footnote{Note that
there is no notion of a normalized vector in a vector space over $%
\mathbb{Z}
_{2}$ (another consequence of the lack of an inner product). The normalization
is, as it were, postponed to the probability algorithm which is computed in
the reals. This "external" probability algorithm is "internalized" when $%
\mathbb{Z}
_{2}$ is strengthened to $%
\mathbb{C}
$ in going from QM/sets to full QM.}

Here $\frac{\left\Vert P_{i}\left(  \psi\right)  \right\Vert ^{2}}{\left\Vert
\psi\right\Vert ^{2}}$ is the quantum probability of getting $\phi_{i}$ in an
$F$-measurement of $\psi$, while $\frac{\left\Vert f^{-1}\left(  \phi
_{i}\right)  \cap S\right\Vert _{U}^{2}}{\left\Vert S\right\Vert _{U}^{2}%
}=\frac{\left\vert f^{-1}\left(  \phi_{i}\right)  \cap S\right\vert
}{\left\vert S\right\vert }$ has the classical interpretation as the
probability $\Pr\left(  \phi_{i}|S\right)  $ of the numerical attribute
$f:U\rightarrow%
\mathbb{R}
$ having the eigenvalue $\phi_{i}$ when "measuring" $S\subseteq U$. Thus the
QM/Sets-version of the Born Rule is the perfectly ordinary Laplace-Boole rule
for the conditional probability $\Pr\left(  \phi_{i}|S\right)  =\frac
{\left\vert f^{-1}\left(  \phi_{i}\right)  \cap S\right\vert }{\left\vert
S\right\vert }$, that given an event $S$ on the sample space $U$, a random
variable $f:U\rightarrow%
\mathbb{R}
$ takes the value $\phi_{i}$.

In QM/Sets, when the state $S$ is being "measured" using the observable $f$
where the probability $\Pr\left(  \phi_{i}|S\right)  $ of getting the
eigenvalue $\phi_{i}$ is $\frac{\left\Vert f^{-1}\left(  \phi_{i}r\right)
\cap S\right\Vert _{U}^{2}}{\left\Vert S\right\Vert _{U}^{2}}=\frac{\left\vert
f^{-1}\left(  \phi_{i}\right)  \cap S\right\vert }{\left\vert S\right\vert }$,
the "damned quantum jump" (Schr\"{o}dinger) goes from $S$ by the projection
operator $f^{-1}\left(  \phi_{i}\right)  \cap()$ to the projected resultant
state $f^{-1}\left(  \phi_{i}\right)  \cap S$ which is in the eigenspace
$\wp\left(  f^{-1}\left(  \phi_{i}\right)  \right)  $ for that eigenvalue
$\phi_{i}$. The state resulting from the measurement represents a
more-definite state $f^{-1}\left(  \phi_{i}\right)  \cap S$ that now has the
definite $f$-value of $\phi_{i}$--so a second measurement would yield the same
eigenvalue $\phi_{i}$ with probability:

\begin{center}
$\Pr\left(  \phi_{i}|f^{-1}\left(  \phi_{i}\right)  \cap S\right)
=\frac{\left\vert f^{-1}\left(  \phi_{i}\right)  \cap\left[  f^{-1}\left(
\phi_{i}\right)  \cap S\right]  \right\vert }{\left\vert f^{-1}\left(
\phi_{i}\right)  \cap S\right\vert }=\frac{\left\vert f^{-1}\left(  \phi
_{i}\right)  \cap S\right\vert }{\left\vert f^{-1}\left(  \phi_{i}\right)
\cap S\right\vert }=1$
\end{center}

\noindent and the same resulting vector $f^{-1}\left(  \phi_{i}\right)
\cap\left[  f^{-1}\left(  \phi_{i}\right)  \cap S\right]  =f^{-1}\left(
\phi_{i}\right)  \cap S$ using the idempotency of the projection operators.

This treatment of measurement in QM/Sets is just the set-version of the
treatment of measurement in standard Dirac-von-Neumann QM.

\subsection{Summary of QM/Sets and QM}

The QM/set-versions of the corresponding QM notions are summarized in the
following table for the finite $U$-basis of the $%
\mathbb{Z}
_{2}$-vector space $\wp\left(  U\right)  \cong%
\mathbb{Z}
_{2}^{n}$ and for a finite dimensional Hilbert space $V$.

\begin{center}%
\begin{tabular}
[c]{|c|c|}\hline
QM/Sets over $%
\mathbb{Z}
_{2}$ & Standard QM over $%
\mathbb{C}
$\\\hline\hline
Projections: $S\cap():\wp\left(  U\right)  \rightarrow\wp\left(  U\right)  $ &
$P:V\rightarrow V$ where $P^{2}=P$\\\hline
Spectral Decomposition: $f=\sum_{i}\phi_{i}\chi_{f^{-1}\left(  \phi
_{i}\right)  }$ & $F=\sum_{i}\phi_{i}P_{i}$\\\hline
Brackets: $\left\langle S|_{U}T\right\rangle =\left\vert S\cap T\right\vert $
= overlap of $S,T\subseteq U$ & $\left\langle \psi|\varphi\right\rangle =$
"overlap" of $\psi$ and $\varphi$\\\hline
Norm: $\left\Vert S\right\Vert _{U}=\sqrt{\left\langle S|_{U}S\right\rangle
}=\sqrt{\left\vert S\right\vert }$ where $S\subseteq U$ & $\left\Vert
\psi\right\Vert =\sqrt{\left\langle \psi|\psi\right\rangle }$\\\hline
Pythagoras: $\left\Vert S\right\Vert _{U}^{2}=\sum_{i}\left\Vert f^{-1}\left(
\phi_{i}\right)  \cap S\right\Vert _{U}^{2}$ & $\left\Vert \psi\right\Vert
^{2}=\sum_{i}\left\Vert P_{i}\left(  \psi\right)  \right\Vert ^{2}$\\\hline
Normalized: $\sum_{i}\frac{\left\Vert f^{-1}\left(  \phi_{i}\right)  \cap
S\right\Vert _{U}^{2}}{\left\Vert S\right\Vert _{U}^{2}}=\sum_{i}%
\frac{\left\vert f^{-1}\left(  \phi_{i}\right)  \cap S\right\vert }{\left\vert
S\right\vert }=1$ & $\sum_{i}\frac{\left\Vert P_{i}\left(  \psi\right)
\right\Vert ^{2}}{\left\Vert \psi\right\Vert ^{2}}=1$\\\hline
Born rule: $\Pr(\phi_{i}|S)=\frac{\left\Vert f^{-1}\left(  \phi_{i}\right)
\cap S\right\Vert _{U}^{2}}{\left\Vert S\right\Vert _{U}^{2}}=\frac{\left\vert
f^{-1}\left(  \phi_{i}\right)  \cap S\right\vert }{\left\vert S\right\vert }$
& $\Pr\left(  \phi_{i}|\psi\right)  =\frac{\left\Vert P_{i}\left(
\psi\right)  \right\Vert ^{2}}{\left\Vert \psi\right\Vert ^{2}}$\\\hline
\end{tabular}

Probability calculus for QM/Sets over $%
\mathbb{Z}
_{2}$ and for standard QM over $%
\mathbb{C}
$
\end{center}

The word "logic" is thrown around quite loosely as in constant references to
"the logic" of this and that. But there is a clear sense in which $%
\mathbb{Z}
_{2}^{n}$ is "logical" since given a basis set $U$, it is isomorphic to the
powerset $\wp\left(  U\right)  $ with the symmetric difference as the vector
sum, and each ket in $%
\mathbb{Z}
_{2}^{n}$ represented in the $U$-coordinates is just a subset of $U$. Hence
the transporting of the structures of the observables and the probability
algorithm from a Hilbert space $%
\mathbb{C}
^{n}$ to $%
\mathbb{Z}
_{2}^{n}$ can be seen as extracting the \textit{logical skeleton} of QM in
QM/Sets, and, in that sense, QM/Sets can also be seen as a "logic of QM."

\section{Measurement in QM/Sets}

\subsection{Set partitions and DSDs in vector spaces over $%
\mathbb{Z}
_{2}$}

The language of sets, e.g., $S\subseteq U$, can be translated into the
language of vector spaces over $%
\mathbb{Z}
_{2}$, e.g. $S\in\wp\left(  U\right)  \cong%
\mathbb{Z}
_{2}^{\left\vert U\right\vert }$. QM/Sets uses either language depending on
the context.

\begin{itemize}
\item A set partition $\pi=\left\{  B,...\right\}  $ on set $U$ is a
disjoint-union decomposition of $U=$ $\cup_{B\in\pi}B$, and;

\item the "vector space partition" $\left\{  \wp\left(  B\right)  \right\}
_{B\in\pi}$ is a direct-sum decomposition of the vector space $\wp\left(
U\right)  =\oplus_{B\in\pi}\wp\left(  B\right)  $.
\end{itemize}

In particular, a real-valued numerical attribute $f:U\rightarrow%
\mathbb{Z}
$ defines both:

\begin{itemize}
\item a set partition of constant sets $\left\{  f^{-1}\left(  r\right)
\right\}  _{r\in f\left(  U\right)  }$, and

\item a DSD of eigenspaces $\phi=\left\{  \wp\left(  f^{-1}\left(  r\right)
\right)  \right\}  _{r\in f\left(  U\right)  }$.
\end{itemize}

Given an attribute $f:U\rightarrow%
\mathbb{R}
$ on a basis set $U$ and an attribute $g:U^{\prime}\rightarrow%
\mathbb{R}
$ on a different (equicardinal) basis set $U^{\prime}$ for $%
\mathbb{Z}
_{2}^{n}$, the two attributes (or "observables") are \textit{compatible} if
their corresponding DSDs $\phi=\left\{  \wp\left(  f^{-1}\left(  r\right)
\right)  \right\}  _{r\in f\left(  U\right)  }$ and $\gamma=\left\{
\wp\left(  g^{-1}\left(  s\right)  \right)  \right\}  _{s\in g\left(
U^{\prime}\right)  }$ are compatible, i.e., if the proto-join $\left\{
\wp\left(  f^{-1}\left(  r\right)  \right)  \cap\wp\left(  g^{-1}\left(
s\right)  \right)  \right\}  _{r\in f\left(  U\right)  ,s\in g\left(
U^{\prime}\right)  }$ spans the whole space. Then choosing a basis set for the
nonzero subspaces $\wp\left(  f^{-1}\left(  r\right)  \right)  \cap\wp\left(
g^{-1}\left(  s\right)  \right)  $ yields a basis set $U^{\prime\prime}$ of
"simultaneous eigenvectors" so that the two attributes can each be redefined
on $U^{\prime\prime}$ so as to determine the same DSDs $\phi$ and $\gamma$.
Hence we can always consider compatible attributes as being defined on the
same basis set, which we can take as $U$.

Measurement will later be treated using density matrices, but we can begin
with a simpler framework. In the correspondences between QM/Sets and QM, a
block $S$ in a partition on $U$ [i.e., a vector $S\in\wp\left(  U\right)  $]
corresponds to \textit{pure} state in QM, and a partition $\pi=\left\{
B,...\right\}  $ on $U$ is the \textit{mixed state} of orthogonal pure states
$B$ with the probabilities $p_{B}=\frac{\left\vert B\right\vert }{\left\vert
U\right\vert }$. In QM, a measurement makes distinctions, i.e., makes
alternatives distinguishable, and that turns a pure state into a mixture of
probabilistic outcomes.

A measurement of an attribute $f:U\rightarrow%
\mathbb{R}
$ in QM/Sets is the distinction-creating process that uses the set partition
$\left\{  f^{-1}\left(  r\right)  \right\}  _{r\in f\left(  U\right)  }$ of
$U$ to decompose (or "decohere") a "pure state" $S\subseteq U$ into a "mixed
state" set partition $\left\{  f^{-1}\left(  r\right)  \cap S\right\}  _{r\in
f\left(  U\right)  }$ on $S$ obtained by applying the decomposing operations
$f^{-1}\left(  r\right)  \cap()$ to $S$ with the block $f^{-1}\left(
r\right)  \cap S$ having the probability $\Pr\left(  r|S\right)
=\frac{\left\vert f^{-1}\left(  r\right)  \cap S\right\vert }{\left\vert
S\right\vert }$.

Alternatively this could be described in vector space terms. A measurement of
an attribute $f:U\rightarrow%
\mathbb{R}
$ in QM/Sets is the distinction-creating process that uses the DSD
$\phi=\left\{  \wp\left(  f^{-1}\left(  r\right)  \right)  \right\}  _{r\in
f\left(  U\right)  }$ to decompose a vector $S\in\wp\left(  U\right)  $ into
the parts obtained by applying the projection operators $f^{-1}\left(
r\right)  \cap():\wp\left(  U\right)  \rightarrow\wp\left(  U\right)  $ to the
vector $S$ with each part having the probability $\Pr(r|S)=\frac{\left\Vert
f^{-1}\left(  r\right)  \cap S\right\Vert _{U}^{2}}{\left\Vert S\right\Vert
_{U}^{2}}=\frac{\left\vert f^{-1}\left(  r\right)  \cap S\right\vert
}{\left\vert S\right\vert }$.

Either way, this is just the pedagogical QM/Sets version of the full QM
measurement of a given state $\psi$ using an observable operator
$F=\sum_{\lambda}\lambda P_{\lambda}$ where the state $\psi$ projects to the
parts $P_{\lambda}\left(  \psi\right)  $ which occur with the probabilities
$\Pr\left(  \lambda|\psi\right)  =\frac{\left\Vert P_{\lambda}\left(
\psi\right)  \right\Vert ^{2}}{\left\Vert \psi\right\Vert ^{2}}$.

\subsection{Weyl's anticipation of QM/Sets}

The pedagogical model, QM/Sets, could be seen as a development of some of the
hints in Hermann Weyl's expository writings about quantum mechanics. He called
a partition a "grating" or "sieve"\footnote{Arthur Eddington made a very early
use of the sieve idea:
\par
\begin{quotation}
\noindent In Einstein's theory of relativity the observer is a man who sets
out in quest of truth armed with a measuring-rod. In quantum theory he sets
out armed with a sieve. \cite[p. 267]{edd:pathways}
\end{quotation}
\par
\noindent This passage was quoted by Weyl \cite[p. 255]{weyl:phil} in his
treatment of gratings.}, and then considered \textit{both} set partitions and
vector space partitions (direct sum decompositions) as the respective types of
gratings.\cite[pp. 255-257]{weyl:phil} He started with a numerical attribute
on a set, e.g., $f:U\rightarrow%
\mathbb{R}
$ (in my notation), which defined the set partition or "grating" \cite[p.
255]{weyl:phil} with blocks having the same attribute-value, e.g., $\left\{
f^{-1}\left(  r\right)  \right\}  _{r\in f\left(  U\right)  }$. Then he moved
to the QM case where the universe set, e.g., $U=\left\{  u_{1},...,u_{n}%
\right\}  $, or "aggregate of $n$ states has to be replaced by an
$n$-dimensional Euclidean vector space" \cite[p. 256]{weyl:phil}.\footnote{In
his expository popular writings, Weyl used a Euclidean space instead of a
Hilbert space.} The appropriate notion of a vector space partition or
"grating" is a "splitting of the total vector space into mutually orthogonal
subspaces" so that "each vector $\overrightarrow{x}$ splits into $r$ component
vectors lying in the several subspaces" \cite[p. 256]{weyl:phil}, i.e., a
direct sum decomposition of the space. After referring to a partition as a
"grating" or "sieve," Weyl notes that "Measurement means application of a
sieve or grating" \cite[p. 259]{weyl:phil}. In QM/Sets, this "application" of
the set-grating or set partition $\left\{  f^{-1}\left(  r\right)  \right\}
_{r\in f\left(  U\right)  }$ to the "pure" state $S$ gives the "mixed state"
set partition $\left\{  f^{-1}\left(  r\right)  \cap S\right\}  _{r\in
f\left(  U\right)  }$of $S$. In terms of the DSD $\phi=\left\{  \wp\left(
f^{-1}\left(  r\right)  \right)  \right\}  _{r\in f\left(  U\right)  }$
determined by the attribute $f:U\rightarrow%
\mathbb{R}
$, this is just decomposing the vector $S$ into the disjoint parts
$S=\sum_{r\in f\left(  U\right)  }f^{-1}\left(  r\right)  \cap S$ where
$f^{-1}\left(  r\right)  \cap S\in\wp\left(  f^{-1}\left(  r\right)  \right)
$. By the "projection postulate," the state $S$ projects to one of its parts
$f^{-1}\left(  r\right)  \cap S$ with the probability $\Pr(r|S)=\left\vert
f^{-1}\left(  r\right)  \cap S\right\vert /\left\vert S\right\vert $.

For some visual imagery of measurement, we might think of a grating or sieve
as a series of regular-polygonal-shaped holes that might shape an indefinite
blob of dough. For illustrative purposes, we say the blob of dough is the sum
of the more definite shapes: $\bigcirc=\triangle+\square+...$. In a
measurement, the blob of dough falls through one of the polygonal holes in the
grating with equal probability and then takes on or "projects to" that shape.%

\begin{center}
\includegraphics[
height=1.9873in,
width=2.284in
]%
{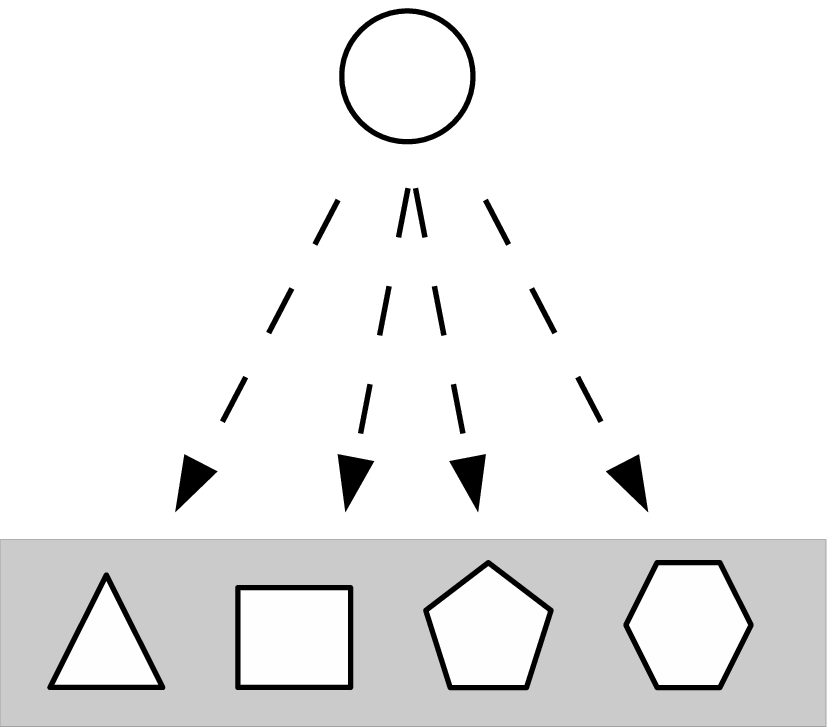}%
\end{center}

\begin{center}
Figure 4: Measurement as randomly giving an indefinite blob of dough a
definite polygonal shape.
\end{center}

\subsection{Example of measurements}

In the simple example illustrated below, we start at the one block or state of
the indiscrete partition or blob which is $\left\{  a,b,c\right\}  $. A
measurement uses some attribute that defines an inverse-image partition on
$U=\left\{  a,b,c\right\}  $. In the case at hand, there are "essentially"
four possible attributes that could be used to "measure" the state $\left\{
a,b,c\right\}  $ (since there are four partitions that refine the indiscrete partition).

For an example of a degenerate measurement, we choose an attribute with a
non-discrete inverse-image partition such as the partition $\pi=\left\{
\left\{  a\right\}  ,\left\{  b,c\right\}  \right\}  $ which determines a
non-maximal DSD $\left\{  \wp\left(  \left\{  a\right\}  \right)  ,\wp\left(
\left\{  b,c\right\}  \right)  \right\}  $. Hence the attribute could just be
the characteristic function $\chi_{\left\{  b,c\right\}  }$ with the two
eigenspaces $\wp(\left\{  a\right\}  )$ and $\wp(\left\{  b,c\right\}  )$ and
the two eigenvalues $0$ and $1$ respectively. Since the eigenspace $\wp\left(
\chi_{\left\{  b,c\right\}  }^{-1}\left(  1\right)  \right)  =\wp\left(
\left\{  b,c\right\}  \right)  $ is not one dimensional, the eigenvalue of $1$
is a QM/Sets-version of a \textit{degenerate} eigenvalue. This attribute
$\chi_{\left\{  b,c\right\}  }$ has four (non-zero) eigenvectors:

\begin{center}
$\chi_{\left\{  b,c\right\}  }\upharpoonright\left\{  b,c\right\}  =1\left\{
b,c\right\}  $, $\chi_{\left\{  b,c\right\}  }\upharpoonright\left\{
b\right\}  =1\left\{  b\right\}  $, $\chi_{\left\{  b,c\right\}
}\upharpoonright\left\{  c\right\}  =1\left\{  c\right\}  $, and
$\chi_{\left\{  b,c\right\}  }\upharpoonright\left\{  a\right\}  =0\left\{
a\right\}  $.
\end{center}

The "measuring apparatus" makes distinctions by joining the attribute's
inverse-image partition

\begin{center}
$\chi_{\left\{  b,c\right\}  }^{-1}=\left\{  \chi_{\left\{  b,c\right\}
}^{-1}\left(  1\right)  ,\chi_{\left\{  b,c\right\}  }^{-1}\left(  0\right)
\right\}  =\left\{  \left\{  b,c\right\}  ,\left\{  a\right\}  \right\}  $
\end{center}

\noindent with the pure state representing the indefinite entity $U=\left\{
a,b,c\right\}  $. The action on the pure state is:

\begin{center}
$U\rightarrow\left\{  U\right\}  \vee\chi_{\left\{  b,c\right\}  }^{-1}%
=\chi_{\left\{  b,c\right\}  }^{-1}=\left\{  \left\{  b,c\right\}  ,\left\{
a\right\}  \right\}  $.
\end{center}

The measurement of that attribute returns one of the eigenvalues with the probabilities:

\begin{center}
$\Pr(0|U)=\frac{\left\vert \left\{  a\right\}  \cap\left\{  a,b,c\right\}
\right\vert }{\left\vert \left\{  a,b,c\right\}  \right\vert }=\frac{1}{3}$
and $\Pr\left(  1|U\right)  =\frac{\left\vert \left\{  b,c\right\}
\cap\left\{  a,b,c\right\}  \right\vert }{\left\vert \left\{  a,b,c\right\}
\right\vert }=\frac{2}{3}$.
\end{center}

\noindent Suppose it returns the eigenvalue $1$. Then the indefinite entity
$\left\{  a,b,c\right\}  $ reduces to the projected eigenstate $\chi_{\left\{
b,c\right\}  }^{-1}\left(  1\right)  \cap\left\{  a,b,c\right\}  =\left\{
b,c\right\}  $ for that eigenvalue \cite[p. 221]{cohen-t:QM1}.

Since this is a degenerate result (i.e., the eigenspace $\wp\left(
\chi_{\left\{  b,c\right\}  }^{-1}\left(  1\right)  \right)  =\wp\left(
\left\{  b,c\right\}  \right)  $ doesn't have dimension one), another
measurement is needed to make more distinctions. Measurements by attributes,
such as $\chi_{\left\{  a,b\right\}  }$ or $\chi_{\left\{  a,c\right\}  }$,
that give either of the other two partitions, $\left\{  \left\{  a,b\right\}
,\{c\right\}  \}$ or $\left\{  \left\{  b\right\}  ,\left\{  a,c\right\}
\right\}  $ as inverse images, would suffice to distinguish $\left\{
b,c\right\}  $ into $\left\{  b\right\}  $ or $\left\{  c\right\}  $. Then
either attribute together with the attribute $\chi_{\left\{  b,c\right\}  }$
would form a \textit{Complete Set of Compatible Attributes} or CSCA (i.e., the
QM/Sets-version of Dirac's Complete Set of Commuting Operators or CSCO), where
\textit{complete} means that the join of the attributes' inverse-image
partitions gives the discrete partition and where \textit{compatible} means
that all the attributes can be taken as defined on the same set of
(simultaneous) basis eigenvectors, e.g., the $U$-basis.

Taking, for example, the other attribute as $\chi_{\left\{  a,b\right\}  }$,
the join of the two attributes' partitions is discrete:

\begin{center}
$\mathbf{\chi}_{\left\{  b,c\right\}  }^{-1}\vee\mathbf{\chi}_{\left\{
a,b\right\}  }^{-1}=\left\{  \left\{  a\right\}  ,\left\{  b,c\right\}
\right\}  \vee\left\{  \left\{  a,b\right\}  ,\{c\right\}  \}=\left\{
\left\{  a\right\}  ,\left\{  b\right\}  ,\left\{  c\right\}  \right\}
=\mathbf{1}$.
\end{center}

\noindent Hence all the eigenstate singletons can be characterized by the
ordered pairs of the eigenvalues of these two attributes: $\left\{  a\right\}
=\left\vert 0,1\right\rangle $, $\left\{  b\right\}  =\left\vert
1,1\right\rangle $, and $\left\{  c\right\}  =\left\vert 1,0\right\rangle $
(using Dirac's ket-notation to give the ordered pairs and listing the
eigenvalues of $\chi_{\left\{  b,c\right\}  }$ first on the left).

The second projective measurement of the indefinite entity $\left\{
b,c\right\}  $ using the attribute $\chi_{\left\{  a,b\right\}  }$ with the
inverse-image partition $\chi_{\left\{  a,b\right\}  }^{-1}=\left\{  \left\{
a,b\right\}  ,\{c\right\}  \}$ would have the pure-to-mixed state action:

\begin{center}
$\left\{  b,c\right\}  \rightarrow\left\{  \left\{  b,c\right\}  \cap
\chi_{\left\{  a,b\right\}  }(1),\left\{  b,c\right\}  \cap\chi_{\left\{
a,b\right\}  }\left(  0\right)  \right\}  =\left\{  \left\{  b\right\}
,\left\{  c\right\}  \right\}  $.
\end{center}

The distinction-making measurement would cause the indefinite entity $\left\{
b,c\right\}  $ to turn into one of the definite entities of $\left\{
b\right\}  $ or $\left\{  c\right\}  $ with the probabilities:

\begin{center}
$\Pr\left(  1|\left\{  b,c\right\}  \right)  =\frac{\left\vert \left\{
a,b\right\}  \cap\left\{  b,c\right\}  \right\vert }{\left\vert \left\{
b,c\right\}  \right\vert }=\frac{1}{2}$ and $\Pr\left(  0|\left\{
b,c\right\}  \right)  =\frac{\left\vert \left\{  c\right\}  \cap\left\{
b,c\right\}  \right\vert }{\left\vert \left\{  b,c\right\}  \right\vert
}=\frac{1}{2}$.
\end{center}

\noindent If the measured eigenvalue is $0$, then the state $\left\{
b,c\right\}  $ projects to $\chi_{\left\{  a,b\right\}  }^{-1}\left(
0\right)  \cap\left\{  b,c\right\}  =\left\{  c\right\}  $ as pictured below.%

\begin{center}
\includegraphics[
height=1.471in,
width=2.258in
]%
{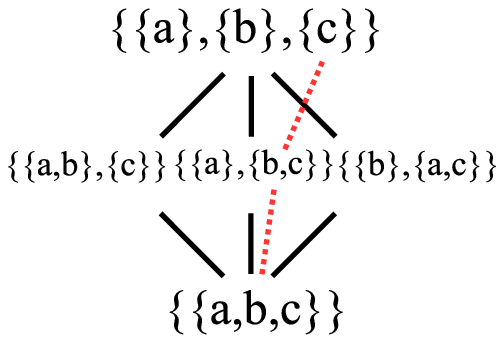}%
\end{center}

\begin{center}
Figure 5: Degenerate measurement
\end{center}

\noindent The two projective measurements of $\left\{  a,b,c\right\}  $ using
the complete set of compatible (e.g., both defined on $U$) attributes
$\chi_{\left\{  b,c\right\}  }$ and $\chi_{\left\{  a,b\right\}  }$ produced
the respective eigenvalues $1$ and $0$ so the resulting eigenstate was
characterized by the eigenket $\left\vert 1,0\right\rangle =\{c\}$.

Again, this is all analogous to standard Dirac-von-Neumann quantum mechanics.

\subsection{Density matrices in QM/Sets}

The previous treatment of the role of partitions in measurement can be
restated using density matrices over the reals. Given a partition
$\pi=\left\{  B,...\right\}  $ on $U=\left\{  u_{1},...,u_{n}\right\}  $, the
blocks $B\in\pi$ can be thought of as (nonoverlapping or "orthogonal") "pure
states" where the "state" $B$ occurs with the probability $p_{B}%
=\frac{\left\vert B\right\vert }{\left\vert U\right\vert }$. Then we can
transport the usual procedure for forming the density matrix $\rho\left(
\pi\right)  $ for the "orthogonal pure states" $B$ with the probabilities
$p_{B}$. The "pure state" $B$ normalized in the reals to length $1$ is
represented by the column vector $\left\vert B\right\rangle _{1}=\frac
{1}{\sqrt{\left\vert B\right\vert }}\left[  \chi_{B}\left(  u_{1}\right)
,...,\chi_{B}\left(  u_{n}\right)  \right]  ^{t}$(where $\left[  \ \right]
^{t}$ indicates the transpose). Then the \textit{density matrix }$\rho\left(
B\right)  $\textit{\ for the pure state }$B\subseteq U$ is then (calculating
in the reals):

\begin{center}
$\rho\left(  B\right)  =\left\vert B\right\rangle _{1}\left(  \left\vert
B\right\rangle _{1}\right)  ^{t}=\frac{1}{\left\vert B\right\vert }%
\begin{bmatrix}
\chi_{B}\left(  u_{1}\right) \\
\chi_{B}\left(  u_{2}\right) \\
\vdots\\
\chi_{B}\left(  u_{n}\right)
\end{bmatrix}
\left[  \chi_{B}\left(  u_{1}\right)  ,...,\chi_{B}\left(  u_{n}\right)
\right]  $

$=\frac{1}{\left\vert B\right\vert }%
\begin{bmatrix}
\chi_{B}\left(  u_{1}\right)  & \chi_{B}\left(  u_{1}\right)  \chi_{B}\left(
u_{2}\right)  & \cdots & \chi_{B}\left(  u_{1}\right)  \chi_{B}\left(
u_{n}\right) \\
\chi_{B}\left(  u_{2}\right)  \chi_{B}\left(  u_{1}\right)  & \chi_{B}\left(
u_{2}\right)  & \cdots & \chi_{B}\left(  u_{2}\right)  \chi_{B}\left(
u_{n}\right) \\
\vdots & \vdots & \ddots & \vdots\\
\chi_{B}\left(  u_{n}\right)  \chi_{B}\left(  u_{1}\right)  & \chi_{B}\left(
u_{n}\right)  \chi_{B}\left(  u_{2}\right)  & \cdots & \chi_{B}\left(
u_{n}\right)
\end{bmatrix}
$.
\end{center}

For instance if $U=\left\{  u_{1},u_{2},u_{3}\right\}  $, then for the blocks
in the partition $\pi=\left\{  \left\{  u_{1},u_{2}\right\}  ,\left\{
u_{3}\right\}  \right\}  $:

\begin{center}
$\rho\left(  \left\{  u_{1},u_{2}\right\}  \right)  =%
\begin{bmatrix}
\frac{1}{2} & \frac{1}{2} & 0\\
\frac{1}{2} & \frac{1}{2} & 0\\
0 & 0 & 0
\end{bmatrix}
$ and $\rho\left(  \left\{  u_{3}\right\}  \right)  =%
\begin{bmatrix}
0 & 0 & 0\\
0 & 0 & 0\\
0 & 0 & 1
\end{bmatrix}
$.
\end{center}

\noindent Then the "mixed state" \textit{density matrix }$\rho\left(
\pi\right)  $\textit{\ of the partition} $\pi$ is the weighted sum:

\begin{center}
$\rho\left(  \pi\right)  =\sum_{B\in\pi}p_{B}\rho\left(  B\right)  $.
\end{center}

In the example, this is:

\begin{center}
$\rho\left(  \pi\right)  =\frac{2}{3}%
\begin{bmatrix}
\frac{1}{2} & \frac{1}{2} & 0\\
\frac{1}{2} & \frac{1}{2} & 0\\
0 & 0 & 0
\end{bmatrix}
+\frac{1}{3}%
\begin{bmatrix}
0 & 0 & 0\\
0 & 0 & 0\\
0 & 0 & 1
\end{bmatrix}
=%
\begin{bmatrix}
\frac{1}{3} & \frac{1}{3} & 0\\
\frac{1}{3} & \frac{1}{3} & 0\\
0 & 0 & \frac{1}{3}%
\end{bmatrix}
$.
\end{center}

In partition logic \cite{ell:intropartlogic}, given a set partition
$\pi=\left\{  B,B^{\prime},...\right\}  $ on a universe set $U$, an ordered
pair $\left(  u,u^{\prime}\right)  \in U\times U$ is called a
\textit{distinction} or \textit{dit} of $\pi$ if the elements are in different
blocks of $\pi$, and the set of all distinctions is the \textit{dit set}
$\operatorname*{dit}\left(  \pi\right)  $. An ordered pair $\left(
u,u^{\prime}\right)  $ is called an \textit{indistinction} or \textit{indit}
of $\pi$ if the two elements are in the same block of $\pi$, and the set of
all indistinctions is the indit set $\operatorname*{indit}\left(  \pi\right)
$. A partition $\pi$ has an associated binary equivalence relation which is
its indit set $\operatorname*{indit}\left(  \pi\right)  \subseteq U\times U$,
and an associated partition relation or apartness relation which is the
complementary dit set $\operatorname*{dit}\left(  \pi\right)  =U\times
U-\operatorname*{indit}\left(  \pi\right)  $. The density matrix $\rho\left(
\pi\right)  $ of the partition can then be directly interpreted in terms of
its indit set:

\begin{center}
$\rho_{jk}\left(  \pi\right)  =\left\{
\begin{array}
[c]{c}%
\frac{1}{\left\vert U\right\vert }\text{ if }\left(  u_{j},u_{k}\right)
\in\operatorname*{indit}\left(  \pi\right) \\
0\text{ if }\left(  u_{j},u_{k}\right)  \notin\operatorname*{indit}\left(
\pi\right)
\end{array}
\right.  $.
\end{center}

\noindent All the entries are real "amplitudes" whose squares are the two-draw
probabilities of drawing a pair of elements from $U$ (with replacement) that
is an indistinction of $\pi$. Like in the full quantum case, the non-zero
entries of the density matrix $\rho_{jk}\left(  \pi\right)  =\sqrt{\frac
{1}{\left\vert U\right\vert }\frac{1}{\left\vert U\right\vert }}=\frac
{1}{\left\vert U\right\vert }$ are the "coherences" \cite[p. 302]{cohen-t:QM1}
which indicate that $u_{j}$ and $u_{k}$ "cohere" together in a block or "pure
state" of the partition, i.e., for some block $B\in\pi$, $u_{j},u_{k}\in B$.
Since the ordered pairs $\left(  u_{j},u_{j}\right)  $ in the diagonal
$\Delta\subseteq U\times U$ are always indits of any partition, the diagonal
entries in $\rho\left(  \pi\right)  $ are always $\frac{1}{\left\vert
U\right\vert }$.

Combinatorial theory gives a natural way to define the same density matrix
$\rho\left(  \pi\right)  $ of a partition $\pi$. A binary relation $R\subseteq
U\times U$ on $U=\left\{  u_{1},...,u_{n}\right\}  $ can be represented by an
$n\times n$ \textit{incidence matrix} $I(R)$ where

\begin{center}
$I\left(  R\right)  _{jk}=\left\{
\begin{array}
[c]{c}%
1\text{ if }\left(  u_{j},u_{k}\right)  \in R\\
0\text{ if }\left(  u_{j},u_{k}\right)  \notin R\text{.}%
\end{array}
\right.  $
\end{center}

\noindent Taking $R$ as the equivalence relation $\operatorname*{indit}\left(
\pi\right)  $ associated with a partition $\pi$, the density matrix
$\rho\left(  \pi\right)  $ defined above is just the incidence matrix
$I\left(  \operatorname*{indit}\left(  \pi\right)  \right)  $ rescaled to be
of trace $1$ (i.e., sum of diagonal entries is $1$):

\begin{center}
$\rho\left(  \pi\right)  =\frac{1}{\left\vert U\right\vert }I\left(
\operatorname*{indit}\left(  \pi\right)  \right)  $.
\end{center}

\subsection{Measurement in QM/Sets using density matrices}

If the subsets $T\in\wp\left(  U\right)  $ are represented by the $n$-ary
column vectors $\left[  \chi_{T}\left(  u_{1}\right)  ,...,\chi_{T}\left(
u_{n}\right)  \right]  ^{t}$, then the action of the projection operator
$B\cap():\wp\left(  U\right)  \rightarrow\wp\left(  U\right)  $ is represented
in the $U$-basis by the $n\times n$ diagonal matrix $P_{B}$ where the diagonal
entries are:

\begin{center}
$\left(  P_{B}\right)  _{jj}=\left\{
\begin{array}
[c]{c}%
1\text{ if }u_{j}\in B\\
0\text{ if }u_{j}\notin B
\end{array}
\right.  =\chi_{B}\left(  u_{j}\right)  $
\end{center}

\noindent which is idempotent, $P_{B}^{2}=P_{B}$, and symmetric, $P_{B}%
^{t}=P_{B}$. For any state $S\in\wp\left(  U\right)  $, the trace (sum of
diagonal entries) of $P_{B}\rho\left(  S\right)  $ is:

\begin{center}
$\operatorname*{tr}\left[  P_{B}\rho\left(  S\right)  \right]  =\frac
{1}{\left\vert S\right\vert }\sum_{j=1}^{n}\chi_{S}\left(  u_{j}\right)
\chi_{B}\left(  u_{j}\right)  =\frac{\left\vert B\cap S\right\vert
}{\left\vert S\right\vert }=\Pr\left(  B|S\right)  $
\end{center}

\noindent so given $f:U\rightarrow%
\mathbb{R}
$,

\begin{center}
$\Pr\left(  r|S\right)  =\frac{\left\vert f^{-1}\left(  r\right)  \cap
S\right\vert }{\left\vert S\right\vert }=\operatorname*{tr}\left[
P_{f^{-1}\left(  r\right)  }\rho\left(  S\right)  \right]  $
\end{center}

\noindent This is the QM/Sets version of the usual result: $\Pr\left(
\lambda|\psi\right)  =\frac{\left\Vert P_{\lambda}\left(  \psi\right)
\right\Vert ^{2}}{\left\Vert \psi\right\Vert ^{2}}=\operatorname*{tr}\left[
P_{\lambda}\rho\left(  \psi\right)  \right]  $.

Given a state $S$, the measurement by the $f$-attribute DSD $\left\{
\wp\left(  f^{-1}\left(  r\right)  \right)  \right\}  _{r\in f\left(
U\right)  }$ projects $S$ to the state $f^{-1}\left(  r\right)  \cap S$ with
the probability $\operatorname*{tr}[P_{f^{-1}\left(  r\right)  }\rho\left(
S\right)  ]=\frac{|f^{-1}\left(  r\right)  \cap S|}{\left\vert S\right\vert
}=\Pr\left(  r|S\right)  $. We need to convert this into the language of
density matrices. Starting with the pure state $S$ as a normalized column
vector $\left\vert S\right\rangle _{1}$, the subset $f^{-1}\left(  r\right)
\cap S$ resulting from that projection is the column vector $P_{f^{-1}\left(
r\right)  }\left\vert S\right\rangle $. To calculate the corresponding density
matrix we must first normalize the column vector $P_{f^{-1}\left(  r\right)
}\left\vert S\right\rangle $ by dividing through by $\sqrt{\left\vert
f^{-1}\left(  r\right)  \cap S\right\vert }$ (where nonzero). But the
normalizing factor to compute $\rho\left(  S\right)  $ was $\sqrt{\left\vert
S\right\vert }$, i.e., $\left\vert S\right\rangle _{1}=\frac{1}{\sqrt
{\left\vert S\right\vert }}\left\vert S\right\rangle $. Since
$\operatorname*{tr}\left[  P_{f^{-1}\left(  r\right)  }\rho\left(  S\right)
\right]  =\frac{\left\vert f^{-1}\left(  r\right)  \cap S\right\vert
}{\left\vert S\right\vert }$, the normalized version of $P_{f^{-1}\left(
r\right)  }\left\vert S\right\rangle $ is:

\begin{center}
$\frac{1}{\sqrt{\left\vert f^{-1}\left(  r\right)  \cap S\right\vert }%
}P_{f^{-1}\left(  r\right)  }\left\vert S\right\rangle =\frac{1}%
{\sqrt{\left\vert f^{-1}\left(  r\right)  \cap S\right\vert }}P_{f^{-1}\left(
r\right)  }\sqrt{\left\vert S\right\vert }\left\vert S\right\rangle _{1}%
=\frac{1}{\sqrt{\operatorname*{tr}\left[  P_{f^{-1}\left(  r\right)  }%
\rho\left(  S\right)  \right]  }}P_{f^{-1}\left(  r\right)  }\left\vert
S\right\rangle _{1}$.
\end{center}

Hence the density matrix corresponding to the projected state $P_{f^{-1}%
\left(  r\right)  }\left\vert S\right\rangle $ is:

\begin{center}
$\frac{1}{\operatorname*{tr}\left[  P_{f^{-1}\left(  r\right)  }\rho\left(
S\right)  \right]  }\left(  P_{f^{-1}\left(  r\right)  }\left\vert
S\right\rangle _{1}\right)  \left(  P_{f^{-1}\left(  r\right)  }\left\vert
S\right\rangle _{1}\right)  ^{t}$

$=\frac{1}{\operatorname*{tr}\left[  P_{f^{-1}\left(  r\right)  }\rho\left(
S\right)  \right]  }P_{f^{-1}\left(  r\right)  }\left\vert S\right\rangle
_{1}\left(  \left\vert S\right\rangle _{1}\right)  ^{t}\left(  P_{f^{-1}%
\left(  r\right)  }\right)  ^{t}=\frac{P_{f^{-1}\left(  r\right)  }\rho\left(
S\right)  P_{f^{-1}\left(  r\right)  }}{\operatorname*{tr}\left[
P_{f^{-1}\left(  r\right)  }\rho\left(  S\right)  \right]  }$.
\end{center}

This might be illustrated by using the second part of the above degenerate
measurement where $f=\chi_{\left\{  a,b\right\}  }$ and $S=\left\{
b,c\right\}  $. Then the density matrix is:

\begin{center}
$\rho\left(  \left\{  b,c\right\}  \right)  =%
\begin{bmatrix}
0 & 0 & 0\\
0 & \frac{1}{2} & \frac{1}{2}\\
0 & \frac{1}{2} & \frac{1}{2}%
\end{bmatrix}
$ and $\chi_{\left\{  a,b\right\}  }^{-1}\left(  1\right)  =f^{-1}\left(
1\right)  =\left\{  a,b\right\}  $ so $P_{f^{-1}\left(  1\right)  }=%
\begin{bmatrix}
1 & 0 & 0\\
0 & 1 & 0\\
0 & 0 & 0
\end{bmatrix}
$.

$P_{f^{-1}\left(  1\right)  }\rho\left(  \left\{  b,c\right\}  \right)  =%
\begin{bmatrix}
1 & 0 & 0\\
0 & 1 & 0\\
0 & 0 & 0
\end{bmatrix}%
\begin{bmatrix}
0 & 0 & 0\\
0 & \frac{1}{2} & \frac{1}{2}\\
0 & \frac{1}{2} & \frac{1}{2}%
\end{bmatrix}
=\allowbreak%
\begin{bmatrix}
0 & 0 & 0\\
0 & \frac{1}{2} & \frac{1}{2}\\
0 & 0 & 0
\end{bmatrix}
$

$P_{f^{-1}\left(  1\right)  }\rho\left(  \left\{  b,c\right\}  \right)
P_{f^{-1}\left(  1\right)  }=\allowbreak%
\begin{bmatrix}
0 & 0 & 0\\
0 & \frac{1}{2} & \frac{1}{2}\\
0 & 0 & 0
\end{bmatrix}%
\begin{bmatrix}
1 & 0 & 0\\
0 & 1 & 0\\
0 & 0 & 0
\end{bmatrix}
=\allowbreak%
\begin{bmatrix}
0 & 0 & 0\\
0 & \frac{1}{2} & 0\\
0 & 0 & 0
\end{bmatrix}
$
\end{center}

Since $\operatorname*{tr}\left[  P_{f^{-1}\left(  1\right)  }\rho\left(
\left\{  b,c\right\}  \right)  \right]  =\frac{1}{2}$, the resultant state
from that projection is:

\begin{center}
$\frac{P_{f^{-1}\left(  1\right)  }\rho\left(  \left\{  b,c\right\}  \right)
P_{f^{-1}\left(  1\right)  }}{\operatorname*{tr}\left[  P_{f^{-1}\left(
1\right)  }\rho\left(  \left\{  b,c\right\}  \right)  \right]  }=\frac{1}%
{1/2}\allowbreak%
\begin{bmatrix}
0 & 0 & 0\\
0 & \frac{1}{2} & 0\\
0 & 0 & 0
\end{bmatrix}
=\allowbreak%
\begin{bmatrix}
0 & 0 & 0\\
0 & 1 & 0\\
0 & 0 & 0
\end{bmatrix}
$
\end{center}

\noindent with the density matrix $\rho\left(  \left\{  b\right\}  \right)  $
where $\left\{  b\right\}  =f^{-1}\left(  1\right)  \cap\left\{  b,c\right\}
=\left\{  a,b\right\}  \cap\left\{  b,c\right\}  $. For the other eigenvalue
of $0$, we have

\begin{center}
$P_{f^{-1}\left(  0\right)  }\rho\left(  \left\{  b,c\right\}  \right)
P_{f^{-1}\left(  0\right)  }=%
\begin{bmatrix}
0 & 0 & 0\\
0 & 0 & 0\\
0 & 0 & 1
\end{bmatrix}%
\begin{bmatrix}
0 & 0 & 0\\
0 & \frac{1}{2} & \frac{1}{2}\\
0 & \frac{1}{2} & \frac{1}{2}%
\end{bmatrix}%
\begin{bmatrix}
0 & 0 & 0\\
0 & 0 & 0\\
0 & 0 & 1
\end{bmatrix}
$

$=\allowbreak%
\begin{bmatrix}
0 & 0 & 0\\
0 & 0 & 0\\
0 & \frac{1}{2} & \frac{1}{2}%
\end{bmatrix}%
\begin{bmatrix}
0 & 0 & 0\\
0 & 0 & 0\\
0 & 0 & 1
\end{bmatrix}
=\allowbreak%
\begin{bmatrix}
0 & 0 & 0\\
0 & 0 & 0\\
0 & 0 & \frac{1}{2}%
\end{bmatrix}
$
\end{center}

\noindent and $\operatorname*{tr}\left[  P_{f^{-1}\left(  0\right)  }%
\rho\left(  \left\{  b,c\right\}  \right)  \right]  =\frac{1}{2}$ so

\begin{center}
$\frac{P_{f^{-1}\left(  0\right)  }\rho\left(  \left\{  b,c\right\}  \right)
P_{f^{-1}\left(  0\right)  }}{\operatorname*{tr}\left[  P_{f^{-1}\left(
0\right)  }\rho\left(  \left\{  b,c\right\}  \right)  \right]  }=\frac{1}{1/2}%
\begin{bmatrix}
0 & 0 & 0\\
0 & 0 & 0\\
0 & 0 & \frac{1}{2}%
\end{bmatrix}
=%
\begin{bmatrix}
0 & 0 & 0\\
0 & 0 & 0\\
0 & 0 & 1
\end{bmatrix}
$
\end{center}

\noindent which is the density matrix for the pure state $\left\{  c\right\}
=f^{-1}\left(  0\right)  \cap\left\{  b,c\right\}  =\left\{  c\right\}
\cap\left\{  b,c\right\}  $.

The final formula for the post-measurement mixed state $\hat{\rho}\left(
S\right)  $ would weigh the projected states by their probability, so we have:

\begin{center}
$\hat{\rho}\left(  S\right)  =\sum_{r\in f\left(  U\right)  }\Pr\left(
r|S\right)  \frac{P_{f^{-1}\left(  r\right)  }\rho\left(  S\right)
P_{f^{-1}\left(  r\right)  }}{\operatorname*{tr}\left[  P_{f^{-1}\left(
r\right)  }\rho\left(  S\right)  \right]  }$

$=\sum_{r\in f\left(  U\right)  }\operatorname*{tr}\left[  P_{f^{-1}\left(
r\right)  }\rho\left(  S\right)  \right]  \frac{P_{f^{-1}\left(  r\right)
}\rho\left(  S\right)  P_{f^{-1}\left(  r\right)  }}{\operatorname*{tr}\left[
P_{f^{-1}\left(  r\right)  }\rho\left(  S\right)  \right]  }=\sum_{r\in
f\left(  U\right)  }P_{f^{-1}\left(  r\right)  }\rho\left(  S\right)
P_{f^{-1}\left(  r\right)  }$.
\end{center}

\noindent Thus the action of the measurement is:

\begin{center}
$\rho\left(  S\right)  \longrightarrow\hat{\rho}\left(  S\right)  =\sum_{r\in
f\left(  U\right)  }P_{f^{-1}\left(  r\right)  }\rho\left(  S\right)
P_{f^{-1}\left(  r\right)  }$

Measurement of $S$ using $f$-attribute in density matrix form.
\end{center}

This result is just the "transported" QM/Sets version of the description of
measurement in full QM. Consider the projective measurement using a
self-adjoint operator $F$ on $V$ with the DSD $\left\{  V_{\lambda}\right\}  $
of eigenspaces and the projections to the eigenspaces $P_{\lambda
}:V\rightarrow V_{\lambda}$. The measurement of a normalized pure state
$\left\vert \psi\right\rangle $ results in the state $P_{\lambda}\left\vert
\psi\right\rangle $ with the probability $p_{\lambda}=\operatorname*{tr}%
\left[  P_{\lambda}\rho\left(  \psi\right)  \right]  =\Pr\left(  \lambda
|\psi\right)  $ where $\rho\left(  \psi\right)  =\left\vert \psi\right\rangle
\left\langle \psi\right\vert $. The projected resultant state $P_{\lambda
}\left\vert \psi\right\rangle $ has the density matrix $\frac{P_{\lambda
}\left\vert \psi\right\rangle \left\langle \psi\right\vert P_{\lambda}%
}{\operatorname*{tr}\left[  P_{\lambda}\rho\left(  \psi\right)  \right]
}=\frac{P_{\lambda}\rho\left(  \psi\right)  P_{\lambda}}{\operatorname*{tr}%
\left[  P_{\lambda}\rho\left(  \psi\right)  \right]  }$ so the mixed state
describing the probabilistic results of the measurement is \cite[p. 101 or p.
515]{nielsen-chuang:bible}:

\begin{center}
$\hat{\rho}\left(  \psi\right)  =\sum_{\lambda}p_{\lambda}\frac{P_{\lambda
}\rho\left(  \psi\right)  P_{\lambda}}{\operatorname*{tr}\left[  P_{\lambda
}\rho\left(  \psi\right)  \right]  }=\sum_{\lambda}\operatorname*{tr}\left[
P_{\lambda}\rho\left(  \psi\right)  \right]  \frac{P_{\lambda}\rho\left(
\psi\right)  P_{\lambda}}{\operatorname*{tr}\left[  P_{\lambda}\rho\left(
\psi\right)  \right]  }=\sum_{\lambda}P_{\lambda}\rho\left(  \psi\right)
P_{\lambda}$.
\end{center}

Thus we see how the density matrix treatment of measurement in QM/Sets is just
a sets-version of the density matrix treatment of projective measurement in
standard Dirac-von-Neumann QM:

\begin{center}
$\rho\left(  \psi\right)  \longmapsto\hat{\rho}\left(  \psi\right)
=\sum_{\lambda}P_{\lambda}\rho\left(  \psi\right)  P_{\lambda}$.
\end{center}

\section{Final remarks}

The usual version of quantum logic can be viewed as the extension of the
Boolean logic of subsets to the logic of subspaces of a vector space
(specifically, closed subspaces of a Hilbert space). Since the notion of a set
partition (or equivalence relation or quotient set) is the category-theoretic
dual to the notion of a subset, the logic of set partitions is, in that sense,
dual to the Boolean logic of subsets. Hence there is a dual form of quantum
logic that can be viewed as the extension of the logic of set partitions to
the logic of direct-sum decompositions of a vector space (specifically, a
Hilbert space).

The usual quantum logic of subspaces can be viewed as focusing on
propositions, i.e., the proposition that a given state vector is in a
subspace, and projection operators. Since a self-adjoint operator (observable)
determines a direct-sum decomposition (losing only the specific numerical
eigenvalues), the quantum logic of DSDs can be viewed as focusing on
observables or self-adjoint operators (abstracted from specific
eigenvalues)--with the two projection operators associated with a proposition
and its negation included in the form of the atomic DSDs (and the blob).
Unlike the quantum logic of subspaces, the logic of DSDs (vector space
partitions) provides the natural setting to model measurement since, as Weyl
put it: "Measurement means application of a sieve or grating" \cite[p.
259]{weyl:phil}.

In this introductory treatment, we have focused on first developing that
quantum logic of direct-sum decompositions of a general finite-dimensional
vector space $V$ over a field $\mathbb{K}$. Then we turned to the special case
of the quantum logic of direct-sum decompositions of a finite vector space
over $%
\mathbb{Z}
_{2}$ which applies to the pedagogical model of quantum mechanics over sets,
QM/Sets, the logical skeleton of QM. In the Appendix, we give an elementary
treatment of combinatorics of DSDs of finite vector spaces over finite fields
with $q$ elements, with some numerical calculations and examples for the
special case of QM/Sets where $q=2$.

\section{Appendix: Counting DSDs of finite vector spaces}

\subsection{Reviewing q-analogs: From sets to vector spaces}

The theory of $q$-analogs shows how many "classical" combinatorial formulas
for finite sets can be extended to finite vector spaces where $q$ is the
cardinality of the finite base field $GF(q)$, i.e., $q=p^{n}$, a power of a prime.

The natural number $n$ is replaced by:

\begin{center}
$\left[  n\right]  _{q}=\frac{q^{n}-1}{q-1}=1+q+q^{2}+...+q^{n-1}$
\end{center}

\noindent so as $q\rightarrow1$, then $\left[  n\right]  _{q}\rightarrow n$ in
the passage from vector spaces to sets. The factorial $n!$ is replaced, in the
$q$-analog

\begin{center}
$\left[  n\right]  _{q}!=\left[  n\right]  _{q}\left[  n-1\right]
_{q}...\left[  1\right]  _{q}$
\end{center}

\noindent where $\left[  1\right]  _{q}=\left[  0\right]  _{q}=1$.

To obtain the Gaussian binomial coefficients we calculate with ordered bases
of a $k$-dimensional subspace of an $n$-dimensional vector space over the
finite field $GF\left(  q\right)  $ with $q$ elements. There are $q^{n}$
elements in the space so the first choice for a basis vector has $\left(
q^{n}-1\right)  $ (excluding $0$) possibilities, and since that vector
generated a subspace of dimension $q$, the choice of the second basis vector
is limited to $\left(  q^{n}-q\right)  $ elements, and so forth. Thus:

\begin{center}
$\left(  q^{n}-1\right)  \left(  q^{n}-q\right)  \left(  q^{n}-q^{2}\right)
...\left(  q^{n}-q^{k-1}\right)  $

$=\left(  q^{n}-1\right)  q^{1}\left(  q^{n-1}-1\right)  q^{2}\left(
q^{n-1}-1\right)  ...q^{k-1}\left(  q^{n-k+1}-1\right)  $

$=\frac{\left[  n\right]  _{q}!}{\left[  n-k\right]  _{q}!}q^{\left(
1+2+...+\left(  k-1\right)  \right)  }=\frac{\left[  n\right]  _{q}!}{\left[
n-k\right]  _{q}!}q^{k\left(  k-1\right)  /2}=\frac{\left[  n\right]  _{q}%
!}{\left[  n-k\right]  _{q}!}q^{\binom{k}{2}}$.

Number of ordered bases for a $k$-dimensional subspace in an $n$-dimensional space.
\end{center}

\noindent But for a space of dimension $k$, the number of ordered bases are:

\begin{center}
$\left(  q^{k}-1\right)  \left(  q^{k}-q\right)  \left(  q^{k}-q^{2}\right)
...\left(  q^{k}-q^{k-1}\right)  $

$=\left(  q^{k}-1\right)  q^{1}\left(  q^{k-1}-1\right)  q^{2}\left(
q^{k-1}-1\right)  ...q^{k-1}\left(  q^{k-k+1}-1\right)  $

$=\left[  k\right]  _{q}!q^{k\left(  k-1\right)  /2}=\left[  k\right]
_{q}!q^{\binom{k}{2}}$

Number of ordered bases for a $k$-dimensional space.
\end{center}

Thus the number of subspaces of dimension $k$ is the ratio:

\begin{center}
$\binom{n}{k}_{q}=\frac{\left[  n\right]  _{q}!q^{k\left(  k-1\right)  /2}%
}{\left[  n-k\right]  _{q}!\left[  k\right]  _{q}!q^{k\left(  k-1\right)  /2}%
}=\frac{\left[  n\right]  _{q}!}{\left[  n-k\right]  _{q}!\left[  k\right]
_{q}!}$

Gaussian binomial coefficient
\end{center}

\noindent where $\binom{n}{k}_{q}\rightarrow\binom{n}{k}$ as $q\rightarrow1$,
i.e., the number of $k$-dimensional subspaces $\rightarrow$ number of
$k$-element subsets. Many classical identities for binomial coefficients
generalize to Gaussian binomial coefficients \cite{goldman-rota:foundations4}.

\subsection{Counting partitions of finite sets and vector spaces}

\subsubsection{The direct formulas for counting partitions of finite sets}

Using sophisticated techniques, the direct-sum decompositions of a finite
vector space over $GF\left(  q\right)  $ have been enumerated in the sense of
giving the exponential generating function for the numbers
(\cite{bender-goldman}; \cite{stanley:expstructures}). Our goal is to derive,
by elementary methods, the formulas to enumerate these and some related
direct-sum decompositions.

Two subspaces of a vector space are said to be \textit{disjoint} if their
intersection is the zero subspace $0$. A \textit{direct-sum decomposition}
(DSD) of a finite-dimensional vector space $V$ over a base field $\mathbb{K}$
is a set of (nonzero) pair-wise disjoint subspaces, called \textit{blocks} (as
with partitions), $\left\{  V_{i}\right\}  _{i=1,...,m}$ that span the space.
Then each vector $v\in V$ has a unique expression $v=\sum_{i=1}^{m}v_{i}$ with
each $v_{i}\in V_{i}$. Since a direct-sum decomposition can be seen as the
vector-space version of a set partition, we begin with counting the number of
partitions on a set.

Each set partition $\left\{  B_{1},...,B_{m}\right\}  $ of an $n$-element set
has a "type" or "signature" number partition giving the cardinality of the
blocks where they might be presented in nondecreasing order which we can
assume to be: $\left(  \left\vert B_{1}\right\vert ,\left\vert B_{2}%
\right\vert ,...,\left\vert B_{m}\right\vert \right)  $ which is a number
partition of $n$. For our purposes, there is another way to present number
partitions, the \textit{part-count representation}, where $a_{k}$ is the
number of times the integer $k$ occurs in the number partition (and $a_{k}=0$
if $k$ does not appear) so that:

\begin{center}
$a_{1}1+a_{2}2+...+a_{n}n=\sum_{k=1}^{n}a_{k}k=n$.

Part-count representation of number partitions keeping track of repetitions.
\end{center}

Each set partition $\left\{  B_{1},...,B_{m}\right\}  $ of an $n$-element set
has a part-count signature $a_{1},...,a_{n}$, and then there is a "classical"
formula for the number of partitions with that signature (\cite[p.
215]{andrews:partitions}; \cite[p. 427]{knuth:vol4a}).

\begin{proposition}
The number of set partitions for the given signature: $a_{1},...,a_{n}$ where
$\sum_{k=1}^{n}a_{k}k=n$ is:
\end{proposition}

\begin{center}
$\frac{n!}{a_{1}!a_{2}!...a_{n}!\left(  1!\right)  ^{a_{1}}\left(  2!\right)
^{a_{2}}...\left(  n!\right)  ^{a_{n}}}$.

.
\end{center}

\noindent Proof: Suppose we count the number of set partitions $\left\{
B_{1},...,B_{m}\right\}  $ of an $n$-element set when the blocks have the
given cardinalities: $n_{j}=\left\vert B_{j}\right\vert $ for $j=1,...,m$ so
$\sum_{j=1}^{m}n_{j}=n$. The first block $B_{1}$ can be chosen in $\binom
{n}{n_{1}}$ ways, the second block in $\binom{n-n_{1}}{n_{2}}$ ways and so
forth, so the total number of ways is:

\begin{center}
$\binom{n}{n_{1}}\binom{n-n_{1}}{n_{2}}...\binom{n-n_{1}-...-n_{m-1}}{n_{m}%
}=\frac{n!}{n_{1}!\left(  n-n_{1}\right)  !}\frac{\left(  n-n_{1}\right)
!}{n_{2}!\left(  n-n_{1}-n_{2}\right)  !}...\frac{\left(  n-n_{1}%
-...-n_{m-1}\right)  !}{n_{m}!\left(  n-n_{1}-...-n_{m}\right)  !}$

$=\frac{n!}{n_{1}!...n_{m}!}=\binom{n}{n_{1},...,n_{m}}$
\end{center}

\noindent the multinomial coefficient. This formula can then be restated in
terms of the part-count signature $a_{1},...,a_{n}$ where $\sum_{k=1}^{n}%
a_{k}k=n$ as: $\frac{n!}{\left(  1!\right)  ^{a_{1}}\left(  2!\right)
^{a_{2}}...\left(  n!\right)  ^{a_{n}}}$. But that overcounts since the
$a_{k}$ blocks of size $k$ can be permuted without changing the partition's
signature so one needs to divide by $a_{k}!$ for $k=1,...,n$ which yields the
formula for the number of partitions with that signature. $\square$

The \textit{Stirling numbers} $S\left(  n,m\right)  $ \textit{of the second
kind} are the number of partitions of an $n$-element set with $m$ blocks.
Since $\sum_{k=1}^{n}a_{k}=m$ is the number of blocks, the direct formula (as
opposed to a recurrence formula) is:

\begin{center}
$S\left(  n,m\right)  =%
{\textstyle\sum\limits_{\substack{1a_{1}+2a_{2}+...+na_{n}=n\\a_{1}%
+a_{2}+...+a_{n}=m}}}
\frac{n!}{a_{1}!a_{2}!...a_{n}!\left(  1!\right)  ^{a_{1}}\left(  2!\right)
^{a_{2}}...\left(  n!\right)  ^{a_{n}}}$

Direct formula for Stirling numbers of the second kind.
\end{center}

The \textit{Bell numbers} $B\left(  n\right)  $ are the total number of
partitions on an $n$-element set so the direct formula is:

\begin{center}
$B\left(  n\right)  =\sum_{m=1}^{n}S\left(  n,m\right)  =%
{\textstyle\sum\limits_{1a_{1}+2a_{2}+...+na_{n}=n}}
\frac{n!}{a_{1}!a_{2}!...a_{n}!\left(  1!\right)  ^{a_{1}}\left(  2!\right)
^{a_{2}}...\left(  n!\right)  ^{a_{n}}}$

Direct formula for total number of partitions of an $n$-element set.
\end{center}

\subsubsection{The direct formulas for counting DSDs of finite vector spaces}

Each DSD $\pi=\left\{  V_{i}\right\}  _{i=1,...,m}$ of a finite vector space
of dimension $n$ also determines a number partition of $n$ using the
dimensions $n_{i}=\dim\left(  V_{i}\right)  $ in place of the set
cardinalities, and thus each DSD also has a signature $a_{1},...,a_{n}$ where
the subspaces are ordered by nondecreasing dimension and where $\sum_{k=1}%
^{n}a_{k}k=n$ and $\sum_{k=1}^{n}a_{k}=m$.

\begin{proposition}
The number of DSDs of a vector space $V$ of dimension $n$ over $GF\left(
q\right)  $ with the part-count signature $a_{1},...,a_{n}$ is:
\end{proposition}

\begin{center}
$\frac{1}{a_{1}!a_{2}!...a_{n}!}\frac{\left[  n\right]  _{q}!}{\left(  \left[
1\right]  _{q}!\right)  ^{a_{1}}...\left(  \left[  n\right]  _{q}!\right)
^{a_{n}}}q^{\frac{1}{2}\left(  n^{2}-\sum_{k}a_{k}k^{2}\right)  }$

Number of DSDs for the given signature $a_{1},...,a_{n}$ where $\sum_{k=1}%
^{n}a_{k}k=n$.
\end{center}

\noindent Proof: Reasoning first in terms of the dimensions $n_{i}$, we
calculate the number of ordered bases in a subspace of dimension $n_{1}$ of a
vector space of dimension $n$ over the finite field $GF\left(  q\right)  $
with $q$ elements. There are $q^{n}$ elements in the space so the first choice
for a basis vector is $\left(  q^{n}-1\right)  $ (excluding $0$), and since
that vector generated a subspace of dimension $q$, the choice of the second
basis vector is limited to $\left(  q^{n}-q\right)  $ elements, and so forth. Thus:

\begin{center}
$\left(  q^{n}-1\right)  \left(  q^{n}-q\right)  \left(  q^{n}-q^{2}\right)
...\left(  q^{n}-q^{n_{1}-1}\right)  $

$=\left(  q^{n}-1\right)  q^{1}\left(  q^{n-1}-1\right)  q^{2}\left(
q^{n-1}-1\right)  ...q^{n_{1}-1}\left(  q^{n-n_{1}+1}-1\right)  $

$=\left(  q^{n}-1\right)  \left(  q^{n-1}-1\right)  ...\left(  q^{n-n_{1}%
-1}-1\right)  q^{\left(  1+2+...+\left(  n_{1}-1\right)  \right)  }$

$=\left(  q^{n}-1\right)  \left(  q^{n-1}-1\right)  ...\left(  q^{n-n_{1}%
-1}-1\right)  q^{\binom{n_{1}}{2}}$

Number of ordered bases for an $n_{1}$-dimensional subspace of an
$n$-dimensional space.
\end{center}

\noindent If we then divide by the number of ordered bases for an $n_{1}%
$-dimension space:

\begin{center}
$\left(  q^{n_{1}}-1\right)  \left(  q^{n_{1}}-q\right)  ...\left(  q^{n_{1}%
}-q^{n_{1}-1}\right)  =\left(  q^{n_{1}}-1\right)  \left(  q^{n_{1}%
-1}-1\right)  ...\left(  q-1\right)  q^{\left(  1+2+...+\left(  n_{1}%
-1\right)  \right)  }$
\end{center}

\noindent we could cancel the $q^{n_{1}\left(  n_{1}-1\right)  /2}%
=q^{\binom{n_{1}}{2}}$ terms to obtain the Gaussian binomial coefficient

\begin{center}
$\frac{\left(  q^{n}-1\right)  \left(  q^{n-1}-1\right)  ...\left(
q^{n-n_{1}-1}-1\right)  q^{\left(  1+2+...+\left(  n_{1}-1\right)  \right)  }%
}{\left(  q^{n_{1}}-1\right)  \left(  q^{n_{1}-1}-1\right)  ...\left(
q-1\right)  q^{\left(  1+2+...+\left(  n_{1}-1\right)  \right)  }}=\binom
{n}{n_{1}}_{q}=\frac{\left[  n\right]  _{q}!}{\left[  n-n_{1}\right]
_{q}!\left[  n_{1}\right]  _{q}!}$

Number of different $n_{1}$-dimensional subspaces of an $n$-dimensional space.
\end{center}

\noindent If instead we continue to develop the numerator by multiplying by
the number of ordered bases for an $n_{2}$-dimensional space that could be
chosen from the remaining space of dimension $n-n_{1}$ to obtain:

\begin{center}
$\left(  q^{n}-1\right)  \left(  q^{n}-q\right)  \left(  q^{n}-q^{2}\right)
...\left(  q^{n}-q^{n_{1}-1}\right)  \times\left(  q^{n}-q^{n_{1}}\right)
\left(  q^{n}-q^{n_{1}+1}\right)  ...\left(  q^{n}-q^{n_{1}+n_{2}-1}\right)  $

$=\left(  q^{n}-1\right)  \left(  q^{n-1}-1\right)  ...\left(  q^{n-n_{1}%
-n_{2}+1}-1\right)  q^{\left(  1+2+...+\left(  n_{1}+n_{2}-1\right)  \right)
}$.
\end{center}

\noindent Then dividing by the number of ordered bases of an $n_{1}%
$-dimensional space times the number of ordered bases of an $n_{2}%
$-dimensional space gives the number of different "disjoint" (i.e., only
overlap is zero subspace) subspaces of $n_{1}$-dimensional and $n_{2}%
$-dimensional subspaces.

\begin{center}
$=\frac{\left(  q^{n}-1\right)  \left(  q^{n-1}-1\right)  ...\left(
q^{n-n_{1}-n_{2}+1}-1\right)  q^{\left(  1+2+...+\left(  n_{1}+n_{2}-1\right)
\right)  }}{\left(  q^{n_{1}}-1\right)  \left(  q^{n_{1}-1}-1\right)
...\left(  q-1\right)  q^{\left(  1+2+...+\left(  n_{1}-1\right)  \right)
}\times\left(  q^{n_{2}}-1\right)  \left(  q^{n_{2}-1}-1\right)  ...\left(
q-1\right)  q^{1+2+...+\left(  n_{2}-1\right)  }}$.
\end{center}

Continuing in this fashion we arrive at the number of disjoint subspaces of
dimensions $n_{1},n_{2},...,n_{m}$ where $\sum_{i=1}^{m}n_{i}=n$:

\begin{center}
$\frac{\left(  q^{n}-1\right)  \left(  q^{n-1}-1\right)  ...\left(
q-1\right)  q^{\left(  1+2+...+\left(  n-1\right)  \right)  }}{\prod
\limits_{i=1,...,m}\left(  q^{n_{i}}-1\right)  \left(  q^{n_{i}-1}-1\right)
...\left(  q-1\right)  q^{\left(  1+2+...+\left(  n_{i}-1\right)  \right)  }%
}=\frac{\left[  n\right]  _{q}!q^{n\left(  n-1\right)  /2}}{\left[
n_{1}\right]  _{q}!q^{n_{1}\left(  n_{1}-1\right)  /2}\times...\times\left[
n_{m}\right]  _{q}!q^{n_{m}\left(  n_{m}-1\right)  /2}}$

$=\frac{\left[  n\right]  _{q}!}{\left[  n_{1}\right]  _{q}!...\left[
n_{m}\right]  _{q}!}q^{\frac{1}{2}\left[  n\left(  n-1\right)  -\sum_{i=1}%
^{m}n_{i}\left(  n_{i}-1\right)  \right]  }$.
\end{center}

\noindent There may be a number $a_{k}$ of subspaces with the same dimension,
e.g., if $n_{j}=n_{j+1}=k$, then $a_{k}=2$ so the term $\left[  n_{j}\right]
_{q}!q^{n_{j}\left(  n_{j}-1\right)  /2}\times\left[  n_{j+1}\right]
_{q}!q^{n_{j+1}\left(  n_{j+1}-1\right)  /2}$ in the denominator could be
replaced by $\left(  \left[  k\right]  _{q}!\right)  ^{a_{k}}q^{a_{k}k\left(
k-1\right)  /2}$. Hence the previous result could be rewritten in the
part-count representation:

\begin{center}
$\frac{\left[  n\right]  _{q}!}{\left(  \left[  1\right]  _{q}!\right)
^{a_{1}}...\left(  \left[  n\right]  _{q}!\right)  ^{a_{n}}}q^{\frac{1}%
{2}\left[  n\left(  n-1\right)  -\sum_{k}a_{k}k\left(  k-1\right)  \right]  }$.
\end{center}

\noindent And permuting subspaces of the same dimension $k$ yields a DSD with
the same signature, so we need to divide by $a_{k}!$ to obtain the formula:

\begin{center}
$\frac{\left[  n\right]  _{q}!}{a_{1}!...a_{n}!\left(  \left[  1\right]
_{q}!\right)  ^{a_{1}}...\left(  \left[  n\right]  _{q}!\right)  ^{a_{n}}%
}q^{\frac{1}{2}\left[  n\left(  n-1\right)  -\sum_{k}a_{k}k\left(  k-1\right)
\right]  }$.
\end{center}

\noindent The exponent on the $q$ term can be simplified since $\sum_{k}%
a_{k}k=n$:

\begin{center}
$\frac{1}{2}\left[  n\left(  n-1\right)  -\left(  \sum_{k}a_{k}k\left(
k-1\right)  \right)  \right]  =\frac{1}{2}\left[  n^{2}-n-\left(  \sum
_{k}a_{k}k^{2}-\sum_{k}a_{k}k\right)  \right]  $

$=\frac{1}{2}\left[  n^{2}-n-\left(  \sum_{k}a_{k}k^{2}-n\right)  \right]
=\frac{1}{2}\left(  n^{2}-\sum_{k}a_{k}k^{2}\right)  $.
\end{center}

This yields the final formula for the number of DSDs with the part-count
signature $a_{1},...,a_{n}$:

\begin{center}
$\frac{\left[  n\right]  _{q}!}{a_{1}!...a_{n}!\left(  \left[  1\right]
_{q}!\right)  ^{a_{1}}...\left(  \left[  n\right]  _{q}!\right)  ^{a_{n}}%
}q^{\frac{1}{2}\left(  n^{2}-\sum_{k}a_{k}k^{2}\right)  }$. $\square$
\end{center}

Note that the formula is not obtained by a simple substitution of $\left[
k\right]  _{q}!$ for $k!$ in the set partition formula due to the extra term
$q^{\frac{1}{2}\left(  n^{2}-\sum_{k}a_{k}k^{2}\right)  }$, but that it still
reduces to the classical formula for set partitions with that signature as
$q\rightarrow1$. This formula leads directly to the vector space version of
the Stirling numbers of the second kind to count the DSDs with $m$ parts and
to the vector space version of the Bell numbers to count the total number of DSDs.

Before giving those formulas, it should be noted that there is another
$q$-analog formula called "generalized Stirling numbers" (of the second
kind)--but it generalizes only one of the recurrence formulas for $S\left(
n,m\right)  $. It does not generalize the \textit{interpretation} "number of
set partitions on an $n$-element set with $m$ parts" to count the vector space
partitions (DSDs) of finite vector spaces of dimension $n$ with $m$ parts. The
Stirling numbers satisfy the recurrence formula:

\begin{center}
$S\left(  n+1,m\right)  =mS\left(  n,m\right)  +S\left(  n-1,m\right)  $ with
$S\left(  0,m\right)  =\delta_{0m}$.
\end{center}

\noindent Donald Knuth uses the braces notation for the Stirling numbers, $%
\genfrac{\{}{\}}{0pt}{}{n}{m}%
=S\left(  n,m\right)  $, and then he defines the "generalized Stirling number"
\cite[p. 436]{knuth:vol4a} $%
\genfrac{\{}{\}}{0pt}{}{n}{m}%
_{q}$ by the $q$-analog recurrence relation:

\begin{center}
$%
\genfrac{\{}{\}}{0pt}{}{n+1}{m}%
_{q}=\left(  1+q+...+q^{m-1}\right)
\genfrac{\{}{\}}{0pt}{}{n}{m}%
_{q}+%
\genfrac{\{}{\}}{0pt}{}{n}{m-1}%
_{q}$; $%
\genfrac{\{}{\}}{0pt}{}{0}{m}%
_{q}=\delta_{0m}$.
\end{center}

It is easy to generalize the direct formula for the Stirling numbers and it
generalizes the partition interpretation:

\begin{center}
$D_{q}\left(  n,m\right)  =%
{\textstyle\sum\limits_{\substack{1a_{1}+2a_{2}+...+na_{n}=n\\a_{1}%
+a_{2}+...+a_{n}=m}}}
\frac{\left[  n\right]  _{q}!}{a_{1}!...a_{n}!\left(  \left[  1\right]
_{q}!\right)  ^{a_{1}}...\left(  \left[  n\right]  _{q}!\right)  ^{a_{n}}%
}q^{\frac{1}{2}\left(  n^{2}-\sum_{k}a_{k}k^{2}\right)  }$

Number of DSDs of a finite vector space of dimension $n$ over $GF\left(
q\right)  $ with $m$ parts.
\end{center}

\noindent The number $D_{q}\left(  n,m\right)  $ is $S_{nm}$ in
\cite{stanley:expstructures}. Taking $q\rightarrow1$ yields the Stirling
numbers of the second kind, i.e., $D_{1}\left(  n,m\right)  =S\left(
n,m\right)  $. Knuth's generalized Stirling numbers $%
\genfrac{\{}{\}}{0pt}{}{n}{m}%
_{q}$ and $D_{q}\left(  n,m\right)  $ start off the same, e.g., $%
\genfrac{\{}{\}}{0pt}{}{0}{0}%
_{q}=1=D_{q}\left(  0,0\right)  $ and $%
\genfrac{\{}{\}}{0pt}{}{1}{1}%
_{q}=1=D_{q}\left(  1,1\right)  $, but then quickly diverge. For instance, all
$%
\genfrac{\{}{\}}{0pt}{}{n}{n}%
_{q}=1$ for all $n$, whereas the special case of $D_{q}(n,n)$ is the number of
DSDs of $1$-dimensional subspaces in a finite vector space of dimension $n$
over $GF\left(  q\right)  $ (see table below for $q=2$). The formula
$D_{q}\left(  n,n\right)  $ is $M\left(  n\right)  $ in \cite[Example
5.5.2(b), pp. 45-6]{stanley:ec-vol2} or \cite[Example 2.2, p. 75]%
{stanley:expstructures}.

The number $D_{q}(n,n)$ of DSDs of $1$-dimensional subspaces is closely
related to the number of basis sets. The old formula for that number of bases
is \cite[p. 71]{lidl:finite-fields}:

\begin{center}
$\frac{1}{n!}\left(  q^{n}-1\right)  \left(  q^{n}-q\right)  ...\left(
q^{n}-q^{n-1}\right)  $

$=\frac{1}{n!}\left(  q^{n}-1\right)  \left(  q^{n-1}-1\right)  ...(q^{1}%
-1)q^{\left(  1+2+...+\left(  n-1\right)  \right)  }$

$=\frac{1}{n!}\left[  n\right]  _{q}!q^{\binom{n}{2}}\left(  q-1\right)  ^{n}$
\end{center}

\noindent\ since $\left[  k\right]  _{q}=\frac{q^{k}-1}{q-1}$ for $k=1,...,n$.

In the formula for $D_{q}\left(  n,n\right)  $, there is only one signature
$a_{1}=n$ and $a_{k}=0$ for $k=2,...,n$ which immediately gives the formula
for the number of DSDs with $n$ $1$-dimensional blocks and each $1$%
-dimensional block has $q-1$ choices for a basis vector so the total number of
sets of basis vectors is given by the same formula:

\begin{center}
$D_{q}\left(  n,n\right)  \left(  q-1\right)  ^{n}=\frac{\left[  n\right]
_{q}!}{a_{1}!}q^{\frac{1}{2}\left(  n^{2}-a_{1}1^{2}\right)  }\left(
q-1\right)  ^{n}=\frac{1}{n!}\left[  n\right]  _{q}!q^{\binom{n}{2}}\left(
q-1\right)  ^{n}$.
\end{center}

\noindent Note that for $q=2$, $\left(  q-1\right)  ^{n}=1$ so $D_{2}\left(
n,n\right)  $ is the number of different basis sets.

Summing the $D_{q}\left(  n,m\right)  $ for all $m$ gives the vector space
version of the Bell numbers $B\left(  n\right)  $:

\begin{center}
$D_{q}\left(  n\right)  =\sum_{m=1}^{n}D_{q}\left(  n,m\right)  =%
{\textstyle\sum\limits_{1a_{1}+2a_{2}+...+na_{n}=n}}
\frac{1}{a_{1}!a_{2}!...a_{n}!}\frac{\left[  n\right]  _{q}!}{\left(  \left[
1\right]  _{q}!\right)  ^{a_{1}}...\left(  \left[  n\right]  _{q}!\right)
^{a_{n}}}q^{\frac{1}{2}\left(  n^{2}-\sum_{k}a_{k}k^{2}\right)  }$

Number of DSDs of a vector space of dimension $n$ over $GF\left(  q\right)  $.
\end{center}

\noindent Our notation $D_{q}\left(  n\right)  $ is $D_{n}\left(  q\right)  $
in Bender and Goldman \cite{bender-goldman} and $\left\vert Q_{n}\right\vert $
in Stanley (\cite{stanley:expstructures}, \cite{stanley:ec-vol2}). Setting
$q=1$ gives the Bell numbers, i.e., $D_{1}\left(  n\right)  =B\left(
n\right)  $.

\subsection{Counting DSDs with a block containing a designated vector
$v^{\ast}$}

Set partitions have a property not shared by vector space partitions, i.e.,
DSDs. Given a designated element $u^{\ast}$ of the universe set $U$, the
element is contained in some block of every partition on $U$. But given a
nonzero vector $v^{\ast}$ in a space $V$, it is not necessarily contained in a
block of any given DSD of $V$. Some proofs of formulas use this property of
set partitions so the proofs do not generalize to DSDs.

Consider one of the formulas for the Stirling numbers of the second kind:

\begin{center}
$S\left(  n,m\right)  =\sum_{k=0}^{n-1}\binom{n-1}{k}S\left(  k,m-1\right)  $

Summation formula for $S\left(  n,m\right)  $.
\end{center}

The proof using the designated element $u^{\ast}$ reasoning starts with the
fact that any partition of $U$ with $\left\vert U\right\vert =n$ with $m$
blocks will have one block containing $u^{\ast}$ so we then only need to count
the number of $m-1$ blocks on the subset disjoint from the block containing
$u^{\ast}$. If the block containing $u^{\ast}$ had $n-k$ elements, there are
$\binom{n-1}{k}$ blocks that could be complementary to an $\left(  n-k\right)
$-element block containing $u^{\ast}$ and each of those $k$-element blocks had
$S\left(  k,m-1\right)  $ partitions on it with $m-1$ blocks. Hence the total
number of partitions on an $n$-element set with $m$ blocks is that sum.

This reasoning can be extended to DSDs over finite vector spaces, but it only
counts the number of DSDs with a block containing a designated nonzero vector
$v^{\ast}$ (it doesn't matter which one), not all DSDs. Furthermore, it is not
a simple matter of substituting $\binom{n-1}{k}_{q}$ for $\binom{n-1}{k}$.
Each $\left(  n-k\right)  $-element subset has a unique $k$-element subset
disjoint from it (its complement), but the same does not hold in general
vector spaces. Thus given a subspace with $\left(  n-k\right)  $-dimensions,
we must compute the number of $k$-dimensional subspaces disjoint from it.

Let $V$ be an $n$-dimensional vector space over $GF\left(  q\right)  $ and let
$v^{\ast}$ be a specific nonzero vector in $V$. In a DSD with an $\left(
n-k\right)  $-dimensional block containing $v^{\ast}$, how many $k$%
-dimensional subspaces are there disjoint from the $\left(  n-k\right)
$-dimensional subspace containing $v^{\ast}$? The number of ordered basis sets
for a $k$-dimensional subspace disjoint from the given $\left(  n-k\right)
$-dimensional space is:

\begin{center}
$\left(  q^{n}-q^{n-k}\right)  \left(  q^{n}-q^{n-k+1}\right)  ...\left(
q^{n}-q^{n-1}\right)  =\left(  q^{k}-1\right)  q^{n-k}\left(  q^{k-1}%
-1\right)  q^{n-k+1}...\left(  q-1\right)  q^{n-1}$

$=\left(  q^{k}-1\right)  \left(  q^{k-1}-1\right)  ...\left(  q-1\right)
q^{\left(  n-k\right)  +\left(  n-k+1\right)  +...+\left(  n-1\right)  }$

$=\left(  q^{k}-1\right)  \left(  q^{k-1}-1\right)  ...\left(  q-1\right)
q^{k\left(  n-k\right)  +\frac{1}{2}k\left(  k-1\right)  }$
\end{center}

\noindent since we use the usual trick to evaluate twice the exponent:

\begin{center}
$\left(  n-k\right)  +\left(  n-k+1\right)  +...+\left(  n-1\right)  $

$+$\underline{$\left(  n-1\right)  +\left(  n-2\right)  +...+\left(
n-k\right)  $}

$=\left(  2n-k-1\right)  +...+\left(  2n-k-1\right)  $

$=k\left(  2n-k-1\right)  =2k\left(  k+\left(  n-k\right)  \right)
-k^{2}-k=2k\left(  n-k\right)  +k^{2}-k$.
\end{center}

Now the number of ordered basis set of a $k$-dimensional space is:

\begin{center}
$\left(  q^{k}-1\right)  \left(  q^{k-1}-1\right)  ...\left(  q-1\right)
q^{\frac{1}{2}k\left(  k-1\right)  }$
\end{center}

\noindent so dividing by that gives:

\begin{center}
$q^{k\left(  n-k\right)  }$

The number of $k$-dimensional subspaces disjoint from any $\left(  n-k\right)
$-dimensional subspace.\footnote{This was proven using M\"{o}bius inversion on
the lattice of subspaces by Crapo \cite{crapo:mobius-lattice}.}
\end{center}

\noindent Note that taking $q\rightarrow1$ yields the fact that an $\left(
n-k\right)  $-element subset of an $n$-element set has a unique $k$-element
subset disjoint from it.

Hence in the $q$-analog formula, the binomial coefficient $\binom{n-1}{k}$ is
replaced by the Gaussian binomial coefficient $\binom{n-1}{k}_{q}$ times
$q^{k\left(  n-k\right)  }$. Then the rest of the proof proceeds as usual. Let
$D_{q}^{\ast}\left(  n,m\right)  $ denote the number of DSDs of $V$ with $m$
blocks with one block containing a designated $v^{\ast}$. Then we can mimic
the proof of the formula $S\left(  n,m\right)  =\sum_{k=0}^{n-1}\binom{n-1}%
{k}S\left(  k,m-1\right)  $ to derive the following:

\begin{proposition}
Given a designated nonzero vector $v^{\ast}\in V$, the number of DSDs of $V$
with $m$ blocks one of which contains $v^{\ast}$ is:
\end{proposition}

\begin{center}
$D_{q}^{\ast}\left(  n,m\right)  =\sum_{k=0}^{n-1}$ $\binom{n-1}{k}%
_{q}q^{k\left(  n-k\right)  }D_{q}\left(  k,m-1\right)  $. $\square$
\end{center}

\noindent Note that taking $q=1$ gives the right-hand side of: $\sum
_{k=0}^{n-1}$ $\binom{n-1}{k}S\left(  k,m-1\right)  $ since $D_{1}\left(
k,m-1\right)  =S\left(  k,m-1\right)  $, and the left-hand side is the same as
$S\left(  n,m\right)  $ since \textit{every} set partition of an $n$-element
with $m$ blocks has to have a block containing some designated element
$u^{\ast}$.

Since the Bell numbers can be obtained from the Stirling numbers of the second
time as: $B\left(  n\right)  =\sum_{m=1}^{n}S\left(  n,m\right)  $, there is
clearly a similar formula for the Bell numbers:

\begin{center}
$B\left(  n\right)  =\sum_{k=0}^{n-1}\binom{n-1}{k}B\left(  k\right)  $

Summation formula for $B\left(  n\right)  $.
\end{center}

This formula can also be directly proven using the designated element
$u^{\ast}$ reasoning, so it can be similarly be extended to computing
$D_{q}^{\ast}\left(  n\right)  $, the number of DSDs of $V$ with a block
containing a designated nonzero vector $v^{\ast}$.

\begin{proposition}
Given an designated nonzero vector $v^{\ast}\in V$, the number of DSDs of $V$
with a block containing $v^{\ast}$ is:
\end{proposition}

\begin{center}
$D_{q}^{\ast}\left(  n\right)  =\sum_{k=0}^{n-1}$ $\binom{n-1}{k}%
_{q}q^{k\left(  n-k\right)  }D_{q}\left(  k\right)  $. $\square$
\end{center}

\noindent In the same manner, taking $q=1$ yields the classical summation
formula for $B\left(  n\right)  $ since $D_{1}\left(  k\right)  =B\left(
k\right)  $, and every partition has to have a block containing a designated
element $u^{\ast}$.

Furthermore the $D^{\ast}$ numbers have the expected relation:

\begin{corollary}
$D_{q}^{\ast}(n)=\sum_{m=1}^{n}D_{q}^{\ast}\left(  n,m\right)  $. $\square$
\end{corollary}

Note that both $D_{q}\left(  n,m\right)  $ and $D_{q}^{\ast}\left(
n,m\right)  $ are $q$-analogs of the Stirling numbers of the second kind
$S\left(  n,m\right)  $, and that both $D_{q}\left(  n\right)  $ and
$D_{q}^{\ast}\left(  n\right)  $ are $q$-analogs of the Bell numbers $B\left(
n\right)  $.

In QM/Sets, the "observables" or attributes are defined by real-valued
functions on basis sets. Given a basis set $U=\left\{  u_{1},...,u_{n}%
\right\}  $ for $V=%
\mathbb{Z}
_{2}^{n}\cong\wp\left(  U\right)  $, a real-valued attribute $f:U\rightarrow%
\mathbb{R}
$ determines a set partition $\left\{  f^{-1}\left(  r\right)  \right\}
_{r\in f\left(  U\right)  }$ on $U$ and a DSD $\left\{  \wp\left(
f^{-1}\left(  r\right)  \right)  \right\}  _{r\in f\left(  U\right)  }$ on
$\wp\left(  U\right)  $. In full QM, the important thing about an "observable"
is not the specific numerical eigenvalues, but its eigenspaces for distinct
eigenvalues, and that information is in the DSD of its eigenspaces. The
attribute $f:U\rightarrow%
\mathbb{R}
$ cannot be internalized as an operator on $\wp\left(  U\right)  \cong%
\mathbb{Z}
_{2}^{n}$ (unless its values are $0,1$), but it nevertheless determines the
DSD $\left\{  \wp\left(  f^{-1}\left(  r\right)  \right)  \right\}  _{r\in
f\left(  U\right)  }$ which is sufficient to pedagogically model many quantum
results. Hence a DSD can be thought of an "abstract attribute" (without the
eigenvalues) with its blocks serving as "eigenspaces." Then a natural question
to ask is given any nonzero vector $v^{\ast}\in V=%
\mathbb{Z}
_{2}^{n}$, how many "abstract attributes" are there where $v^{\ast}$ is an
"eigenvector"--and the answer is $D_{2}^{\ast}\left(  n\right)  $. And
$D_{2}^{\ast}\left(  n,m\right)  $ is the number of "abstract attributes" with
$m$ distinct "eigenvalues" where $v^{\ast}$ is an "eigenvector."

\subsection{Atoms, maximal DSDs, and segments}

For a finite $n$-dimensional vector space $V$ over $GF\left(  q\right)  $, the
partially ordered set $DSD\left(  V\right)  $ of DSDs is denoted as $Q_{n}$ in
Stanley \cite[Example 5.5.2(b), p. 45]{stanley:ec-vol2} where, as usual in the
combinatorial theory literature, the "unrefinement" ordering is used on
partitions and DSDs--although there are some exceptions as in Andrews \cite[p.
217]{andrews:partitions}. Hence $DSD\left(  V\right)  $ is the opposite
partial order $Q_{n}^{op}$ of Stanley's $Q_{n}$ which reverses maximal and
minimal DSDs so our number of maximal DSDs $D_{q}\left(  n,n)\right)  $ is
Stanley's number of minimal DSDs $M(n)$.

To compute the number of atoms below a given maximal element $\omega=\left\{
U_{k}\right\}  _{k=1,...,n}$ of $DSD\left(  V\right)  $, note that there are
$\sum_{m=1}^{n-1}\binom{n}{m}=2^{n}-2$ proper subsets of the set $\omega$ and
each determines (by direct sum) a subspace of dimension $1$, $2$,..., or $n-1$
and the proper subsets occur in complementary pairs in the atoms or binary
DSDs so there are:

\begin{center}
$2^{n-1}-1$

Number of atoms below each maximal DSD $\omega$.
\end{center}

\noindent Each atom defines two projection operators and the blob $\mathbf{0}$
determines two more projection operators for a total of $2\left(
2^{n-1}-1\right)  +2=2^{n}$ projection operators which gives the Boolean
algebra of subsets of the given $n$-element set $\omega=\left\{
U_{k}\right\}  _{k=1,...,n}$. Note that once a maximal DSD $\omega$ is picked
in $DSD\left(  V\right)  $ and the DSDs are restricted to those below $\omega
$, i.e., to $\prod\left(  \omega\right)  $, then that is a "classical"
partition lattice on an $n$-element set where $2^{n-1}-1$ is indeed the number
of atoms.

Conversely the number of maximal elements $\omega$ above a given atom
$\left\{  V_{1},V_{2}\right\}  $ in $DSD\left(  V\right)  $ will depend on the
positive dimensions $k=\dim\left(  V_{1}\right)  $ and $n-k-\dim\left(
V_{2}\right)  $ of the blocks $V_{1}$and $V_{2}$ in the atom. There are
$D_{q}\left(  k,k\right)  $ maximal DSDs in $DSD\left(  V_{1}\right)  $ and
$D_{q}\left(  n-k,n-k\right)  $ maximal DSDs in $DSD\left(  V_{2}\right)  $.
Picking one maximal DSD of each set will give a maximal DSD for the whole
space that is above the given atom and a maximal DSD above the atom can be
partitioned into two such subsets, so we have:

\begin{center}
$D_{q}\left(  k,k\right)  D_{q}\left(  n-k,n-k\right)  =\frac{\left[
k\right]  _{q}!}{k!}q^{\frac{1}{2}\left(  k^{2}-k\right)  }\frac{\left[
n-k\right]  _{q}!}{\left(  n-k\right)  !}q^{\frac{1}{2}\left(  \left(
n-k\right)  ^{2}-\left(  n-k\right)  \right)  }$

$\frac{\left[  k\right]  _{q}!\left[  n-k\right]  _{q}!}{k!\left(  n-k\right)
!}q^{\frac{1}{2}\left[  \left(  k^{2}-k\right)  +\left(  \left(  n-k\right)
^{2}-\left(  n-k\right)  \right)  \right]  }=\frac{\left[  k\right]
_{q}!\left[  n-k\right]  _{q}!}{k!\left(  n-k\right)  !}q^{\binom{k}{2}%
+\binom{n-k}{2}}$.

Number of maximal DSDs above a given atom.
\end{center}

The reasoning clearly generalizes so for any given DSD $\sigma=\left\{
W_{j}\right\}  _{j=1,...,m}$ where $\dim\left(  W_{j}\right)  =n_{j}$, we have:

\begin{center}
$%
{\textstyle\prod\limits_{j=1}^{m}}
D_{q}\left(  n_{j},n_{j}\right)  $

Number of maximal DSDs in $DSD\left(  V\right)  $ refining $\sigma$.
\end{center}

Moreover, the reasoning can be generalized to arbitrary DSDs above $\sigma$,
i.e., to the upper segment $\left\{  \pi\in DSD\left(  V\right)
:\sigma\preceq\pi\right\}  $:

\begin{center}
$%
{\textstyle\prod\limits_{j=1}^{m}}
D_{q}\left(  \dim\left(  W_{j}\right)  \right)  $

Number of DSDs in $DSD\left(  V\right)  $ refining $\sigma$.
\end{center}

As $q\rightarrow1$, $D_{q}\left(  n\right)  \rightarrow B\left(  n\right)  $,
the Bell numbers, and thus we get the classical formula for the number of set
partitions $\prod_{i=1}^{m}B\left(  \left\vert C_{i}\right\vert \right)  $
that refine a given set partition $\sigma=\left\{  C_{1},...,C_{m}\right\}  $.

\subsection{Computing initial values for $q=2$}

In the case of $n=1,2,3$, the DSDs can be enumerated "by hand" to check the
formulas, and then the formulas can be used to compute higher values of
$D_{2}\left(  n,m\right)  $ or $D_{2}\left(  n\right)  $.

Since all subspaces contain the zero element which is the empty set
$\emptyset$, it will be usually suppressed when listing the elements of a
subspace. And subsets like $\left\{  a\right\}  $ or $\left\{  a,b\right\}  $
will be denoted as just $a$ and $ab$. Thus the subspace $\left\{
\emptyset,\left\{  a\right\}  ,\left\{  b\right\}  ,\left\{  a,b\right\}
\right\}  $ is denoted for brevity as $\left\{  a,b,ab\right\}  $. A
$k$-dimensional subspace has $2^{k}$ elements so only $2^{k}-1$ are listed.

For $n=1$, there is only one nonzero subspace $\left\{  a\right\}  $, i.e.,
$\left\{  \emptyset,\left\{  a\right\}  \right\}  $, and $D_{2}\left(
1,1\right)  =D_{2}\left(  1\right)  =1$.

For $n=2$, the whole subspace is $\left\{  a,b,ab\right\}  $ and it has three
bases $\left\{  a,b\right\}  $, $\left\{  a,ab\right\}  $, and $\left\{
b,ab\right\}  $. The formula for the number of bases gives $D_{2}\left(
2,2\right)  =3$. The only $D_{2}\left(  2,1\right)  =1$ DSD is the whole space.

For $n=3$, the whole space $\left\{  a,b,c,ab,ac,bc,abc\right\}  $ is the only
$D_{2}\left(  3,1\right)  =1$ and indeed for any $n$ and $q$, $D_{q}\left(
n,1\right)  =1$. For $n=3$ and $m=3$, $D_{2}\left(  3,3\right)  $ is the
number of maximal DSDs of $DSD(%
\mathbb{Z}
_{2}^{3})$ which is the number (unordered) bases of $%
\mathbb{Z}
_{2}^{3}$ (recall $%
\genfrac{\{}{\}}{0pt}{}{n}{n}%
_{q}=1$ for all $q$). Since we know the signature, i.e., $a_{1}=3$ and
otherwise $a_{k}=0$, we can easily compute $D_{2}\left(  3,3\right)  $:

\begin{center}
$\frac{1}{a_{1}!a_{2}!...a_{n}!}\frac{\left[  n\right]  _{q}!}{\left(  \left[
1\right]  _{q}!\right)  ^{a_{1}}...\left(  \left[  n\right]  _{q}!\right)
^{a_{n}}}q^{\frac{1}{2}\left(  n^{2}-\sum_{k}a_{k}k^{2}\right)  }$

$=\frac{1}{3!}\frac{\left[  3\right]  _{2}!}{\left(  \left[  1\right]
_{2}\right)  ^{3}}2^{\frac{1}{2}\left(  3^{2}-3\right)  }=\frac{1}{6}%
\frac{7\times3}{1}2^{\frac{1}{2}\left(  6\right)  }=28=D_{2}(3,3)$.
\end{center}

\noindent And here they are.

\begin{center}

\begin{tabular}
[c]{|c|c|c|c|}\hline
$\left\{  a,b,c\right\}  $ & $\left\{  a,b,ac\right\}  $ & $\left\{
a,b,bc\right\}  $ & $\left\{  a,b,abc\right\}  $\\\hline
$\left\{  a,c,ab\right\}  $ & $\left\{  a,c,bc\right\}  $ & $\left\{
a,c,abc\right\}  $ & $\left\{  a,ab,ac\right\}  $\\\hline
$\left\{  a,ab,bc\right\}  $ & $\left\{  a,ab,abc\right\}  $ & $\left\{
a,ac,bc\right\}  $ & $\left\{  a,ac,abc\right\}  $\\\hline
$\left\{  b,c,ab\right\}  $ & $\left\{  b,c,ac\right\}  $ & $\left\{
b,c,abc\right\}  $ & $\left\{  b,ab,ac\right\}  $\\\hline
$\left\{  b,ab,bc\right\}  $ & $\left\{  b,ab,abc\right\}  $ & $\left\{
b,ac,bc\right\}  $ & $\left\{  b,bc,abc\right\}  $\\\hline
$\left\{  c,ab,ac\right\}  $ & $\left\{  c,ab,bc\right\}  $ & $\left\{
c,ac,bc\right\}  $ & $\left\{  c,ac,abc\right\}  $\\\hline
$\left\{  ab,ac,abc\right\}  $ & $\left\{  ab,bc,abc\right\}  $ & $\left\{
ac,bc,abc\right\}  $ & $\left\{  bc,ab,abc\right\}  $\\\hline
\end{tabular}

All maximal DSDs in $DSD\left(
\mathbb{Z}
_{2}^{3}\right)  $ = all bases of $%
\mathbb{Z}
_{2}^{3}$.
\end{center}

\noindent For $n=3$ and $m=2$, $D_{2}\left(  3,2\right)  $ is the number of
atomic (i.e., binary) DSDs, each of which has the signature $a_{1}=a_{2}=1$ so
the total number of atomic DSDs is:

\begin{center}
$D_{2}\left(  3,2\right)  =\frac{1}{1!1!}\frac{\left[  3\right]  _{2}%
!}{\left(  \left[  1\right]  !\right)  ^{1}\left(  \left[  2\right]  !\right)
^{1}}2^{\frac{1}{2}\left(  3^{2}-1-2^{2}\right)  }=\frac{7\times3}{3}%
2^{\frac{1}{2}\left(  4\right)  }=7\times4=28$.
\end{center}

\noindent And here they are:

\begin{center}%
\begin{tabular}
[c]{|c|c|c|c|}\hline
$\left\{  \left\{  a\right\}  ,\left\{  b,c,bc\right\}  \right\}  $ &
$\left\{  \left\{  a\right\}  ,\left\{  ab,ac,bc\right\}  \right\}  $ &
$\left\{  \left\{  a\right\}  ,\left\{  c,ab,abc\right\}  \right\}  $ &
$\left\{  \left\{  a\right\}  ,\left\{  b,ac,abc\right\}  \right\}  $\\\hline
$\left\{  \left\{  b\right\}  ,\left\{  a,c,ac\right\}  \right\}  $ &
$\left\{  \left\{  b\right\}  ,\left\{  ab,ac,bc\right\}  \right\}  $ &
$\left\{  \left\{  b\right\}  ,\left\{  c,ab,abc\right\}  \right\}  $ &
$\left\{  \left\{  b\right\}  ,\left\{  a,bc,abc\right\}  \right\}  $\\\hline
$\left\{  \left\{  ab\right\}  ,\left\{  b,c,bc\right\}  \right\}  $ &
$\left\{  \left\{  ab\right\}  ,\left\{  a,bc,abc\right\}  \right\}  $ &
$\left\{  \left\{  ab\right\}  ,\left\{  b,ac,abc\right\}  \right\}  $ &
$\left\{  \left\{  ab\right\}  ,\left\{  a,c,ac\right\}  \right\}  $\\\hline
$\left\{  \left\{  c\right\}  ,\left\{  a,b,ab\right\}  \right\}  $ &
$\left\{  \left\{  c\right\}  ,\left\{  ab,ac,bc\right\}  \right\}  $ &
$\left\{  \left\{  c\right\}  ,\left\{  a,bc,abc\right\}  \right\}  $ &
$\left\{  \left\{  c\right\}  ,\left\{  b,ac,abc\right\}  \right\}  $\\\hline
$\left\{  \left\{  ac\right\}  ,\left\{  a,b,ab\right\}  \right\}  $ &
$\left\{  \left\{  ac\right\}  ,\left\{  a,bc,abc\right\}  \right\}  $ &
$\left\{  \left\{  ac\right\}  ,\left\{  c,ab,abc\right\}  \right\}  $ &
$\left\{  \left\{  ac\right\}  ,\left\{  b,c,bc\right\}  \right\}  $\\\hline
$\left\{  \left\{  bc\right\}  ,\left\{  a,b,ab\right\}  \right\}  $ &
$\left\{  \left\{  bc\right\}  ,\left\{  b,ac,abc\right\}  \right\}  $ &
$\left\{  \left\{  bc\right\}  ,\left\{  c,ab,abc\right\}  \right\}  $ &
$\left\{  \left\{  bc\right\}  ,\left\{  a,c,ac\right\}  \right\}  $\\\hline
$\left\{  \left\{  abc\right\}  ,\left\{  a,b,ab\right\}  \right\}  $ &
$\left\{  \left\{  abc\right\}  ,\left\{  b,c,bc\right\}  \right\}  $ &
$\left\{  \left\{  abc\right\}  ,\left\{  a,c,ac\right\}  \right\}  $ &
$\left\{  \left\{  abc\right\}  ,\left\{  ab,ac,bc\right\}  \right\}
$\\\hline
\end{tabular}

All atomic DSDs in $DSD\left(
\mathbb{Z}
_{2}^{3}\right)  $ = all binary DSDs for $%
\mathbb{Z}
_{2}^{3}$.
\end{center}

The above table has been arranged to illustrate the result that any given
$k$-dimensional subspaces has $q^{k\left(  n-k\right)  }$ subspaces disjoint
from it. For $n=3$ and $k=1$, each row gives the $2^{2}=4$ subspaces disjoint
from any given $1$-dimensional subspace represented by $\left\{  a\right\}  $,
$\left\{  b\right\}  $,..., $\left\{  abc\right\}  $. For instance, the four
subspaces disjoint from the subspace $\left\{  ab\right\}  $ (shorthand for
$\left\{  \emptyset,\left\{  a,b\right\}  \right\}  $) are given in the third
row since those are the "complementary" subspaces that together with $\left\{
ab\right\}  $ form a DSD.

To illustrate the number of atoms below a maximal element $\omega$, recall
that a maximal DSD and a basis set are the "same thing" for $q=2$. The basis
set $\left\{  a,ac,bc\right\}  $ for $%
\mathbb{Z}
_{2}^{3}$ has $2^{2}-1=3$ atoms below it, namely $\left\{  \left\{  a\right\}
,\left\{  ab,ac,bc\right\}  \right\}  $, $\left\{  \left\{  bc\right\}
,\left\{  a,c,ac\right\}  \right\}  $, and $\left\{  \left\{  ac\right\}
,\left\{  a,bc,abc\right\}  \right\}  $, and those three atoms determine the
eight element BA $\wp\left(  \left\{  a,ac,bc\right\}  \right)  $ of subsets
of the basis set.

To illustrate the number of maximal DSDs above a given atom for $q=2$ (where
maximal DSD = basis set) and $n=3$, each atom has $k=1$ and $n-k=2$, so the
formula gives $3$ which is correct. For instance, the atom $\left\{  \left\{
ac\right\}  ,\left\{  a,bc,abc\right\}  \right\}  $ has three maximal DSDs
above it, namely $\left\{  a,ac,bc\right\}  $, $\left\{  a,ac,abc\right\}  $,
and $\left\{  ac,bc,abc\right\}  $.

Note that for $q=1$, the formula gives the number of maximal partitions above
a given binary partition on a set, namely $1$, the discrete partition on the set.

For $q=2$, the initial values up to $n=6$ of $D_{2}\left(  n,m\right)  $are
given the following table.

\begin{center}%
\begin{tabular}
[c]{|c||c|c|c|c|c|c|c|}\hline
$n\backslash m$ & $0$ & $1$ & $2$ & $3$ & $4$ & $5$ & $6$\\\hline\hline
$0$ & $1$ &  &  &  &  &  & \\\hline
$1$ & $0$ & $1$ &  &  &  &  & \\\hline
$2$ & $0$ & $1$ & $3$ &  &  &  & \\\hline
$3$ & $0$ & $1$ & $28$ & $28$ &  &  & \\\hline
$4$ & $0$ & $1$ & $400$ & $1,680$ & $840$ &  & \\\hline
$5$ & $0$ & $1$ & $10,416$ & $168,640$ & $277,760$ & $83,328$ & \\\hline
$6$ & $0$ & $1$ & $525,792$ & $36,053,248$ & $159,989,760$ & $139,991,040$ &
$27,998,208$\\\hline
\end{tabular}

$D_{2}\left(  n,m\right)  $ with $n,m=1,2,...,6$.
\end{center}

The seventh row $D_{2}\left(  7,m\right)  $ for $m=0,1,...,7$ is: $0$, $1$,
$51116992$, $17811244032$, $209056841728$, $419919790080$, $227569434624$, and
$32509919232$ which sum to $D_{2}\left(  7\right)  $.

The row sums give the values of $D_{2}\left(  n\right)  $ for $n=0,1,2,...,7$.

\begin{center}%
\begin{tabular}
[c]{|c||c|}\hline
$n$ & $D_{2}\left(  n\right)  $\\\hline\hline
$0$ & $1$\\\hline
$1$ & $1$\\\hline
$2$ & $4$\\\hline
$3$ & $57$\\\hline
$4$ & $2,921$\\\hline
$5$ & $540,145$\\\hline
$6$ & $364,558,049$\\\hline
$7$ & $906,918,346,689$\\\hline
\end{tabular}

$D_{2}\left(  n\right)  $ for $n=0,1,...,7$.
\end{center}

We can also compute the $D^{\ast}$ examples of DSDs with a block containing a
designated element. For $q=2$, the $D_{2}^{\ast}\left(  n,m\right)  $ numbers
for $n,m=0,1,...,7$ are given in the following table.

\begin{center}%
\begin{tabular}
[c]{|c||c|c|c|c|c|c|c|c|}\hline
$n\backslash m$ & $0$ & $1$ & $2$ & $3$ & $4$ & $5$ & $6$ & $7$\\\hline\hline
$0$ & $1$ &  &  &  &  &  &  & \\\hline
$1$ & $0$ & $1$ &  &  &  &  &  & \\\hline
$2$ & $0$ & $1$ & $2$ &  &  &  &  & \\\hline
$3$ & $0$ & $1$ & $16$ & $12$ &  &  &  & \\\hline
$4$ & $0$ & $1$ & $176$ & $560$ & $224$ &  &  & \\\hline
$5$ & $0$ & $1$ & $3456$ & $40000$ & $53760$ & $13440$ &  & \\\hline
$6$ & $0$ & $1$ & $128000$ & $5848832$ & $20951040$ & $15554560$ & $2666496$ &
\\\hline
$7$ & $0$ & $1$ & $9115648$ & $1934195712$ & $17826414592$ & $30398054400$ &
$14335082496$ & $1791885312$\\\hline
\end{tabular}

Number of DSDs $D_{2}^{\ast}\left(  n,m\right)  $ containing any given nonzero
vector $v^{\ast}$
\end{center}

For $n=3$ and $m=2$, the table says there are $D_{2}^{\ast}\left(  3,2\right)
=16$ DSDs with $2$ blocks one of which contains a given vector, say $v^{\ast
}=ab$ which represents $\left\{  a,b\right\}  $, and here they are.

\begin{center}%
\begin{tabular}
[c]{|c|c|c|c|}\hline
$\left\{  \left\{  ab\right\}  ,\left\{  b,c,bc\right\}  \right\}  $ &
$\left\{  \left\{  ab\right\}  ,\left\{  a,bc,abc\right\}  \right\}  $ &
$\left\{  \left\{  ab\right\}  ,\left\{  b,ac,abc\right\}  \right\}  $ &
$\left\{  \left\{  ab\right\}  ,\left\{  a,c,ac\right\}  \right\}  $\\\hline
$\left\{  \left\{  c\right\}  ,\left\{  a,b,ab\right\}  \right\}  $ &
$\left\{  \left\{  c\right\}  ,\left\{  ab,ac,bc\right\}  \right\}  $ &
$\left\{  \left\{  ac\right\}  ,\left\{  c,ab,abc\right\}  \right\}  $ &
$\left\{  \left\{  bc\right\}  ,\left\{  c,ab,abc\right\}  \right\}  $\\\hline
$\left\{  \left\{  ac\right\}  ,\left\{  a,b,ab\right\}  \right\}  $ &
$\left\{  \left\{  a\right\}  ,\left\{  ab,ac,bc\right\}  \right\}  $ &
$\left\{  \left\{  a\right\}  ,\left\{  c,ab,abc\right\}  \right\}  $ &
$\left\{  \left\{  abc\right\}  ,\left\{  a,b,ab\right\}  \right\}  $\\\hline
$\left\{  \left\{  bc\right\}  ,\left\{  a,b,ab\right\}  \right\}  $ &
$\left\{  \left\{  b\right\}  ,\left\{  ab,ac,bc\right\}  \right\}  $ &
$\left\{  \left\{  b\right\}  ,\left\{  c,ab,abc\right\}  \right\}  $ &
$\left\{  \left\{  abc\right\}  ,\left\{  ab,ac,bc\right\}  \right\}
$\\\hline
\end{tabular}

Two-block DSDs of $\wp\left(  \left\{  a,b,c\right\}  \right)  $ with a block
containing $ab=\left\{  a,b\right\}  $.
\end{center}

\noindent The table also says there are $D_{2}^{\ast}\left(  3,3\right)  =12$
basis sets containing \textit{any} given element which we could take to be
$v^{\ast}=abc=\left\{  a,b,c\right\}  $, and here they are.

\begin{center}%
\begin{tabular}
[c]{|c|c|c|c|}\hline
$\left\{  a,b,abc\right\}  $ & $\left\{  b,ab,abc\right\}  $ & $\left\{
a,c,abc\right\}  $ & $\left\{  b,bc,abc\right\}  $\\\hline
$\left\{  a,ab,abc\right\}  $ & $\left\{  a,ac,abc\right\}  $ & $\left\{
b,c,abc\right\}  $ & $\left\{  c,ac,abc\right\}  $\\\hline
$\left\{  ab,ac,abc\right\}  $ & $\left\{  ab,bc,abc\right\}  $ & $\left\{
ac,bc,abc\right\}  $ & $\left\{  bc,ab,abc\right\}  $\\\hline
\end{tabular}

Three-block DSDs (basis sets) of $\wp\left(  \left\{  a,b,c\right\}  \right)
$ with a basis element $abc=\left\{  a,b,c\right\}  $.
\end{center}

Summing the rows in the $D_{2}^{\ast}\left(  n,m\right)  $ table gives the
values for $D_{2}^{\ast}\left(  n\right)  $ for $n=0,1,...,7$.

\begin{center}%
\begin{tabular}
[c]{|c||c|}\hline
$n$ & $D_{2}^{\ast}\left(  n\right)  $\\\hline\hline
$0$ & $1$\\\hline
$1$ & $1$\\\hline
$2$ & $3$\\\hline
$3$ & $29$\\\hline
$4$ & $961$\\\hline
$5$ & $110,657$\\\hline
$6$ & $45,148,929$\\\hline
$7$ & $66,294,748,161$\\\hline
\end{tabular}

$D_{2}^{\ast}\left(  n\right)  $ for $n=0,1,...,7$.
\end{center}

In QM/Sets, a DSD on $%
\mathbb{Z}
_{2}^{n}$ is an "abstract observable" that abstracts from the specific
eigenvalues and gives only the DSD of eigenspaces. $D_{2}^{\ast}\left(
n\right)  $ counts the number of abstract observables that will have any given
nonzero vector $v^{\ast}\in%
\mathbb{Z}
_{2}^{n}$ as an eigenvector. For instance, for $n=3$, any given nonzero vector
$v^{\ast}$ will be an eigenvector for $D_{2}^{\ast}\left(  3\right)  =29$
abstract observables, i.e., $12$ three-block DSDs, $16$ two-block DSDs, and
$1$ single-block DSD (the blob).

The integer sequence $D_{2}\left(  n,n\right)  $ for $n=0,1,2,...$ is known
as: $A053601$"Number of bases of an $n$-dimensional vector space over $GF(2)$"
in the \textit{On-Line Encyclopedia of Integer Sequences} (https://oeis.org/).
The sequences defined and tabulated here for $q=2$ have been added to the
\textit{Encyclopedia} as: $A270880$ [$D_{2}(n,m)$], $A270881$ [$D_{2}\left(
n\right)  $], $A270882$ [$D_{2}^{\ast}\left(  n,m\right)  $], $A270883$
[$D_{2}^{\ast}\left(  n\right)  $].


\begin{thebibliography}{99}                                                                                               %


\bibitem {abram-coecke:catqm}Abramsky, Samson and Coecke, Bob 2004. A
categorical semantics of quantum protocols. in: \textit{Proceedings of the
19th IEEE Symposium on Logic in Computer Science (LiCS'04)}, IEEE Computer
Science Press, 415--425.

\bibitem {andrews:partitions}Andrews, George E. 1998. \textit{The Theory of
Partitions}. Cambridge UK: Cambridge University Press.

\bibitem {awodey:cat-theory}Awodey, Steve. 2006. \textit{Category Theory.}
Oxford: Clarendon Press.

\bibitem {bender-goldman}Bender, Edward A., and Jay R. Goldman. 1971.
Enumerative Uses of Generating Functions. \textit{Indiana University
Mathematics Journal,} 20 (8): 753--65.

\bibitem {birkhoff:lt}Birkhoff, Garrett 1948. \textit{Lattice Theory}. New
York: American Mathematical Society.

\bibitem {boole:lot}Boole, George 1854. \textit{An Investigation of the Laws
of Thought on which are founded the Mathematical Theories of Logic and
Probabilities}. Cambridge: Macmillan and Co.

\bibitem {cohen-t:QM1}Cohen-Tannoudji, Claude, Bernard Diu and Franck
Lalo\"{e} 2005. \textit{Quantum Mechanics Vol. 1}. New York: John Wiley \& Sons.

\bibitem {crapo:mobius-lattice}Crapo, Henry. 1966. The M\"{o}bius Function of
a Lattice. \textit{Journal of Combinatorial Theory,} I (1 June): 126--31.

\bibitem {dirac:principles}Dirac, Paul A. M. 1958. \textit{The Principles of
Quantum Mechanics (4th ed.)}. Oxford: Clarendon.

\bibitem {edd:pathways}Eddington, Arthur S. 1947. \textit{New Pathways in
Science (Messenger Lectures 1934)}. Cambridge UK: Cambridge University Press.

\bibitem {ell:partitions}Ellerman, David 2010. The Logic of Partitions:
Introduction to the Dual of the Logic of Subsets. \textit{Review of Symbolic
Logic}. 3 (2 June): 287-350.

\bibitem {ell:qmoversets}Ellerman, David 2013. \textit{Quantum mechanics over
sets}. arXiv:1310.8221 [quant-ph].

\bibitem {ell:intropartlogic}Ellerman, David 2014. An Introduction of
Partition Logic. \textit{Logic Journal of the IGPL.} 22, no. 1: 94--125.

\bibitem {ell:mammoth}Ellerman, David. On the Objective Indefiniteness
Interpretation of Quantum Mechanics. In \textit{The Mammoth Book of Quantum
Mechanics Interpretations}, edited by Ulf Edvinsson. Open Academic Press, forthcoming.

\bibitem {goldman-rota:foundations4}Goldman, Jay R., and Gian-Carlo Rota.
1970. On the Foundations of Combinatorial Theory IV: Finite Vector Spaces and
Eulerian Generating Functions. \textit{Studies in Applied Mathematics.} XLIX
(3 Sept.): 239--58.

\bibitem {grat:lattice-theory}Gr\"{a}tzer, George. 2003. \textit{General
Lattice Theory (2nd Ed.)}. Boston: Birkh\"{a}user Verlag.

\bibitem {hansonsabry:dqt}Hanson, Andrew J., Gerardo Ortiz, Amr Sabry, and
Yu-Tsung Tai 2013. \textit{Discrete Quantum Theories}. arXiv:1305.3292v1.

\bibitem {knuth:vol4a}Knuth, Donald E. 2011. \textit{The Art of Computer
Programming: Vol. 4A Combinatorial Algorithms Part 1}. Boston: Pearson Education.

\bibitem {kochen:hv-in-qm}Kochen, Simon, and E. P. Specker. 1967. The Problem
of Hidden Variables in Quantum Mechanics. \textit{Journal of Mathematics and
Mechanics.} 17 (1): 59--87.

\bibitem {kolmogorov:foundations}Kolmogorov, A. N. 1956. \textit{Foundations
of the Theory of Probability}. Translated by Nathan Morrison. Second English
ed. New York: Chelsea.

\bibitem {kung;rota}Kung, Joseph P.S., Gian-Carlo Rota and Catherine H. Yan
2009. \textit{Combinatorics: The Rota Way}. New York: Cambridge University Press.

\bibitem {laplace:probs}Laplace, Pierre-Simon. 1995 (1825).
\textit{Philosophical Essay on Probabilities}. Translated by A. I. Dale. New
York: Springer-Verlag.

\bibitem {lidl:finite-fields}Lidl, Rudolf, and Harald Niederreiter. 1986.
\textit{Introduction to Finite Fields and Their Applications}. Cambridge UK:
Cambridge University Press.

\bibitem {maclane:cwm2}Mac Lane, Saunders. 1998. \textit{Categories for the
Working Mathematician}. 2nd Ed. New York: Springer Science+Business Media.

\bibitem {mceliece:coding}McEliece, Robert J. 1977. \textit{The Theory of
Information and Coding: A Mathematical Framework for Communication
(Encyclopedia of Mathematics and its Applications, Vol. 3)}. Reading MA: Addison-Wesley.

\bibitem {nielsen-chuang:bible}Nielsen, M., and I. Chuang. 2000.
\textit{Quantum Computation and Quantum Information}. Cambridge: Cambridge
University Press.

\bibitem {ore:eq-rel}Ore, Oystein. 1942. Theory of Equivalence Relations.
\textit{Duke Mathematical Journal,} 9: 573--627.

\bibitem {schum:modal}Schumacher, Benjamin and Michael Westmoreland 2012.
Modal Quantum Theory. \textit{Foundations of Physics}, 42, 918-925.

\bibitem {stanley:expstructures}Stanley, Richard P. 1978. Exponential
Structures. \textit{Studies in Applied Mathematics,} 59 (1 July): 73--82.

\bibitem {stanley:ec-vol2}Stanley, Richard P. 1999. \textit{Enumerative
Combinatorics Vol. 2}. New York: Cambridge University Press.

\bibitem {tak:mutant}Takeuchi, T., L.N. Chang, Z. Lewis, and D. Minic. 2012.
\textit{Some Mutant Forms of Quantum Mechanics}. [quant-Ph] arXiv:1208.5544v1.

\bibitem {weyl:phil}Weyl, Hermann 1949. \textit{Philosophy of Mathematics and
Natural Science}. Princeton NJ: Princeton University Press.
\end{thebibliography}
\end{document}